\documentclass[aps, prd, twocolumn, showpacs, showkeys, superscriptaddress, preprintnumbers, floatfix, reprint]{revtex4}
\usepackage{multirow}
\usepackage{slashed}
\usepackage{graphicx}
\usepackage{amsmath}
\usepackage{bm}
\usepackage{yhmath}
\usepackage{hyperref}
\usepackage{subfigure}
\usepackage{color}
\usepackage{cases}
\usepackage{subfigure}
\usepackage{dcolumn, booktabs, bm}
\usepackage{amsfonts, amssymb, stmaryrd, latexsym, amsmath}
\usepackage{textcomp}
\usepackage{amsthm}
\usepackage{float}
\usepackage{mathrsfs}

\usepackage{array}
\newcounter{RomanNumber}

\newcommand{\lyxmathsym}[1]{\ifmmode\begingroup\def\b@ld{bold}
	\text{\ifx\math@version\b@ld\bfseries\fi#1}\endgroup\else#1\fi}

\renewcommand{\arraystretch}{1.8}
\allowdisplaybreaks
\begin{document}
	
	\title{Analysis  of the  hidden-charm pentaquark states  based on magnetic moment and transition magnetic moment }
	\author{Fei Guo  }\email{guofei@stumail.nwu.edu.cn}\affiliation{School of Physics, Northwest University, Xian 710127, China}
	\author{Hao-Song Li}\email{haosongli@nwu.edu.cn}\affiliation{School of Physics, Northwest University, Xian 710127, China}\affiliation{Institute of Modern Physics, Northwest University, Xian 710127, China}\affiliation{Shaanxi Key Laboratory for Theoretical Physics Frontiers, Xian 710127, China}\affiliation{Peng Huanwu Center for Fundamental Theory, Xian 710127, China}
	\begin{abstract}
	 In this work, we  calculate the magnetic moments of the  $P^{N^{0}}_{\psi}$ states  and $P^{\Delta^{0}}_{\psi}$  states with valence quark content $\bar{c}cudd  $ in  molecular model, diquark-diquark-antiquark model and diquark-triquark model, as well as the transition magnetic moments in the molecular model.  At the same time, we also calculate  magnetic moments and transition magnetic moments of  $P^{\Delta^{++}}_{\psi}$ states and  $P^{\Delta^{-}}_{\psi}$ states in the molecular model as additional products. Our results show that in the diquark-diquark-antiquark model, the magnetic moments of  $ \lambda $ excitation state are  usually larger than the magnetic moments of $ \rho $ excitation state. We find some interesting proportional relationships between the expressions of  transition magnetic moments. The results  provide important insights for future experimental observation of hidden-charm pentaquark  states and help to distinguish their inner structures. With these efforts, our understanding of the
		properties for the hidden-charm pentaquark states will become more abundant.
	\end{abstract}
	\maketitle
	\thispagestyle{empty}
	
	\section{Introduction}
	In the early 1960s,  M. Gell-Mann and G. Zweig 	 proposed the multiquark states beyond the conventional meson and baryon in the quark model\cite{Gell-Mann:1964ewy,Zweig:1964ruk}. With the development of QCD, more complex quark structures are allowed theoretically. These  unconventional hadrons are called exotic hadrons, such as tetraquark $ (\bar{q}\bar{q}qq) $ and pentaquark $ (\bar{q}qqqq) $. In recent years, more and more exotic hadrons states beyond  the  conventional quark model have been observed experimentally\cite{Belle:2003nnu,Belle:2004lle,Belle:2008qeq,CDF:2009jgo,LHCb:2014zfx,LHCb:2016axx,LHCb:2018oeg,BaBar:2005hhc,Belle:2007dxy,Belle:2007umv,Belle:2008xmh,BESIII:2016adj,LHCb:2015yax,Gershon:2022xnn,LHCb:2019rmd,LHCb:2020jpq,LHCb:2021chn,LHCb:2022jad}. The exploration of multiquark states has entered upon
	a new era.
	
	The study of pentaquark states made important progress in 2015, the LHCb  collaboration  observed two hidden-charm pentaquark candidates $P^{N}_{\psi}(4380)^{^{+}}  $ and $P^{N}_{\psi}(4450)^{^{+}}  $ in the $J/\psi p  $ invariant mass distribution\cite{LHCb:2015yax}.   In 2019, $P^{N}_{\psi}(4450)^{^{+}}  $was updated to $P^{N}_{\psi}(4440)^{^{+}}  $and $P^{N}_{\psi}(4457)^{^{+}}  $, and a new structure $P^{N}_{\psi}(4312)^{^{+}}  $was observed\cite{LHCb:2019rmd}. In this work, we adopted a new naming convention for exotic hadrons\cite{Gershon:2022xnn}. The masses and widths for these states were reported as (in units of MeV):
	\begin{eqnarray}
&&P^{N}_{\psi}(4380)^{^{+}}:M =4380\pm8\pm29 ~~~~~ \Gamma =205\pm18\pm86\nonumber\\
&&P^{N}_{\psi}(4440)^{^{+}}:M=4440.3 \pm 1.3 ^{+ 4.1}_{-4.7}~~\Gamma= 20.6 \pm 4.9^{+8.7}_{-10.1}\nonumber\\
&&P^{N}_{\psi}(4457)^{^{+}}:M=4457.3 \pm 0.6 ^{+ 4.1}_{-1.7}~~\Gamma= 6.4 \pm 2.0^{+5.7}_{-1.9}\nonumber\\
&&P^{N}_{\psi}(4312)^{^{+}}:M=4311.9 \pm 0.7^{+ 6.8}_{-0.6}~~\Gamma= 9.8 \pm 2.7^{+3.7}_{-4.5}\nonumber\
	\end{eqnarray}

	Recently, the research on strange hidden-charm  pentaquark states ushered in new experimental progress. In 2020, the LHCb collaboration reported a  strange hidden-charmed pentaquark candidate $P^{\Lambda}_{\psi s}(4459)^{^{0}}  $ in the $ \Xi^{-}_{b}\to J/\psi\Lambda K^{-} $decay\cite{LHCb:2020jpq}.  After considering all systematic uncertainties, its significance exceeds $ 3\sigma $. In recent years, a hidden-charm  pentaquark candidates $P^{N}_{\psi }(4337)^{^{+}}  $\cite{LHCb:2021chn}  and a strange hidden-charm $P^{\Lambda}_{\psi s}(4338)^{^{0}}  $\cite{LHCb:2022jad} were reported in 2021 and 2022 respectively.
	The masses and widths for these states were reported as (in units of MeV):
	\begin{eqnarray}
		&&P^{\Lambda}_{\psi s}(4459)^{^{0}}:M=4458.8 \pm 2.9 ^{+ 4.7}_{-1.1} ~~~~ \Gamma =17.3\pm6.5^{+8.0}_{-5.7}\nonumber\\
		&&P^{N}_{\psi }(4337)^{^{+}}:M=4337 ^{+ 7+2}_{-4-2}~~~~~~~~~~~~~\Gamma= 29 ^{+26+14}_{-12-14}\nonumber\\
		&&P^{\Lambda}_{\psi s}(4338)^{^{0}}:M=4338.2 \pm 0.7\pm 0.4 ~~\Gamma= 7.0 \pm 1.2\pm 1.3\nonumber\
	\end{eqnarray}
	
	After these pentaquark states were observed, theorists proposed many models to explain their internal structures
	\cite{He:2019ify,Karliner:2015ina,Burns:2015dwa,Monemzadeh:2016eoj,Yang:2015bmv,Huang:2015uda,Shimizu:2019ptd,Chen:2015moa,Chen:2016heh,Nakamura:2021dix,Ling:2021lmq,Maiani:2015vwa,Li:2015gta,Wang:2015epa,Anisovich:2015cia,Shi:2021wyt,Ali:2020vee,Wang:2020eep,Lebed:2015tna,Zhu:2015bba,Giron:2021sla,	Eides:2019tgv,Fernandez-Ramirez:2019koa,Nakamura:2021qvy}.  Theorists have obtained a good description for the pentaquark mass spectrum and decay patterns in the three models including molecular model\cite{He:2019ify,Karliner:2015ina,Burns:2015dwa,Monemzadeh:2016eoj,Yang:2015bmv,Huang:2015uda,Shimizu:2019ptd,Chen:2015moa,Chen:2016heh,Nakamura:2021dix,Ling:2021lmq}, diquark-diquark-antiquark model\cite{Maiani:2015vwa,Li:2015gta,Wang:2015epa,Anisovich:2015cia,Shi:2021wyt,Ali:2020vee,Wang:2020eep} and diquark-triquark model\cite{Lebed:2015tna,Zhu:2015bba,Giron:2021sla}. In Ref. \cite{Yang:2022ezl},  the mass spectrum of pentaquark states in the molecular model are predicted with a phenomenological model. In Ref. \cite{Zhu:2022wpi}, the author performed the research on the coupled-channel of molecular states $\Xi^*_{c}\bar{D}^* $, $ \Xi^{'}_{c}\bar{D}^* $, $  \Xi^{*}_{c}\bar{D} $, $  \Xi_{c}\bar{D}^* $, $  \Xi^{'}_{c}\bar{D} $, $  \Lambda_{c}\bar{D}^*_{s} $, $  \Xi_{c}\bar{D} $ and $\Lambda J/\psi  $, and estimated the invariant mass spectrum by constructing potential kernel with effective Lagrangians.  The mass spectrum of  hidden charm pentaquarks were studied in  framework of the chromomagnetic model \cite{Weng:2019ynv}.  
	
	The discovery of these hidden-charm pentaquarks $P^{N^{+}}_{\psi} ( \bar{c}cuud) $  naturally leads us to speculate that  other  hidden-charm pentaquarks $P^{N^{0}}_{\psi} ( \bar{c}cudd) $ also exists. The $P^{N^{+}}_{\psi}  $states are predicted isospin doublets  with neutral partners $P^{N^{0}}_{\psi} $ states.
	The valence quark content of the $P^{N^{0}}_{\psi} $ states and $P^{\Delta^{0}}_{\psi}$  states is  $ \bar{c}cudd $, and they are closely related, so we consider these two hidden-charm pentaquark states. 
	  In order to explore more valuable information, we  study  the magnetic moment and transition magnetic moment of $ P^{N^{0}}_{\psi} $ states and $P^{\Delta^{0}}_{\psi}$  states. 
	  The magnetic moment and transition magnetic moment provide useful clues for studying the internal structure of these exotic hadrons	\cite{Wang:2022tib,	Zhou:2022gra,Zhang:2021yul,Wang:2016dzu,Li:2021ryu,Gao:2021hmv,Wang:2022ugk,Wang:2019mhm,Ortiz-Pacheco:2018ccl,Liu:2003ab,Ozdem:2022kei,Ozdem:2021ugy,Ozdem:2022iqk,Ozdem:2022vip}.
	  The magnetic moments, the transition magnetic moments, and the radiative decay behaviors of the of the $ S $-wave  isoscalar $\Xi_c^{(*)} \bar{D}^{(*)}$ molecular pentaquark states have been studied  in Ref. \cite{Wang:2022tib}. In Ref. \cite{Wang:2016dzu}, the author discussed the composition of the color-flavor configurations of the pentaquark states in the  molecular model, diquark-diquark-antiquark model and diquark-triquark model, and calculates the magnetic moments in these three models. In Ref. \cite{Li:2021ryu},the author considered $P^{N}_{\psi}  $ states as pure molecular states without flavor mixing, and it can be divided into $P^{N^{+}}_{\psi}  $ states and $P^{N^{0}}_{\psi}  $ states, and then calculated the magnetic moments and transition magnetic moments of the $P^{N}_{\psi}  $ states and $P^{\Lambda}_{\psi s}  $ states  in the molecular model. 
	
	The study of the magnetic moments and transition magnetic moments of the $P^{N^{0}}_{\psi}  $ states and $P^{\Delta^{0}}_{\psi}$  states will help us understand its inner structure and search for $P^{N^{0}}_{\psi}  $states and $P^{\Delta^{0}}_{\psi}$  states in the photoproduction process. We believe that with the continuous progress of the experimental and theoretical research on the  pentaquark states, the discovery of the $P^{N^{0}}_{\psi}  $ states and $P^{\Delta^{0}}_{\psi}$  states will become possible in the future, which will enrich the  exotic hadron family.
								
	The organizational structure of this paper is as follows. In Sec. \ref{sec2},  we construct the flavor wave functions in the molecular model, diquark-diquark-antiquark model and diquark-triquark model,  and discuss the color configurations in these three models. In Sec. \ref{sec3},  we calculate the magnetic moments and transition magnetic moments of the $P^{N^{0}}_{\psi}   $ states and $P^{\Delta^{0}}_{\psi}$  states in the molecular model. In Sec. \ref{sec4}, we calculate the magnetic moments  of the $P^{N^{0}}_{\psi}   $ states and $P^{\Delta^{0}}_{\psi}$  states in the diquark-diquark-antiquark model. In Sec.  \ref{sec5}, we calculate the magnetic moments  of the $P^{N^{0}}_{\psi}   $ states and $P^{\Delta^{0}}_{\psi}$  states in the diquark-triquark  model.   In Sec. \ref{sec6}, we analyze the numerical results of magnetic moments and transition magnetic moments.
In Sec. \ref{sec7},	we briefly summarize our work.
	\section{Flavor-color Wave functions of the hidden-charm pentaquark states }
	\label{sec2}
\renewcommand\tabcolsep{0.03cm}
\renewcommand{\arraystretch}{1.550}
\begin{table}[!htbp]
	\caption{The flavor wave functions of the $P^{N^{0}}_{\psi}  $ states and $P^{\Delta^{0}}_{\psi}$  states  in different models, where \{$ q_{1}q_{2} $\}=$\sqrt{\frac{1}{2}}(q_{1}q_{2}+q_{2}q_{1})$, [$q_{1}q_{2}$] =$\sqrt{\frac{1}{2}}(q_{1}q_{2}-q_{2}q_{1})$. $I$ and $I_3$ are the isospin and its third component, respectively. }
	\label{tab:k}
	\begin{tabular}{c|c|c|c}
		\toprule[1.0pt]
		\toprule[1.0pt]
		\multicolumn{4}{c}{Molecular model~($\Sigma_c^{(*)} \bar{D}^{(*)}$/$\Lambda_c^{} {D}^{(*)}$)}\\
		\hline
		States&Multiple& $(I,I_3)$ & Wave funtion \\
		\hline
$P^{\Delta^{0}}_{\psi}$	&$ 10_{f} $&$(\frac{3}{2},-\frac{1}{2})$ &$  \sqrt{\frac{2}{3}}\left|\Sigma_c^{(*)+} {D}^{(*)-}\right\rangle+\sqrt{\frac{1}{3}}\left|\Sigma_c^{(*)0}\bar{D}^{(*)0}\right\rangle$ \\
		\hline
		\multirow{2}{*}	{	$P^{N^{0}}_{\psi}   $}	&$ 8_{1f} $	&$(\frac{1}{2},-\frac{1}{2})$ & $ \sqrt{\frac{1}{3}}\left|\Sigma_c^{(*)+} {D}^{(*)-}\right\rangle-\sqrt{\frac{2}{3}}\left|\Sigma_c^{(*)0}\bar{D}^{(*)0}\right\rangle$ \\
	\cline{2-4}
		
	&	$ 8_{2f} $&$(\frac{1}{2},-\frac{1}{2})$ & $ \left|\Lambda_c^{+} {D}^{(*)-}\right\rangle$\\
		\hline
		\multicolumn{4}{c}{Diquark-diquark-antiquark model}\\
		\hline
$P^{\Delta^{0}}_{\psi}$		&	$10_{f}$
		& $(\frac{3}{2},-\frac{1}{2})$
		& $\sqrt{\frac{2}{3}}({c}d)\{ud\}{\bar c}+\sqrt{\frac{1}{3}}({c}u)\{dd\}{\bar c}$
		\\
		\hline
	\multirow{2}{*}	{	$P^{N^{0}}_{\psi}   $}		&	$8_{1f}$
		& $(\frac{1}{2},-\frac{1}{2})$
		& $\sqrt{\frac{1}{3}}({c}d)\{ud\}{\bar c}-\sqrt{\frac{2}{3}}({c}u)\{dd\}{\bar c}$
		\\
		\cline{2-4}
		&$8_{2f}$
		& $(\frac{1}{2},-\frac{1}{2})$
		& $({c}d)[ud]{\bar c}$
		\\
		\hline
		
		\multicolumn{4}{c}{Diquark-triquark model}\\
		\hline
	$P^{\Delta^{0}}_{\psi}$		&	$10_{f}$
		& $(\frac{3}{2},-\frac{1}{2})$
		& $\sqrt{\frac{2}{3}}({c}d)(\bar c\{ud\})+\sqrt{\frac{1}{3}}({c}u)(\bar c\{dd\})$
		\\
		\hline
	\multirow{2}{*}	{	$P^{N^{0}}_{\psi}   $}		&	$8_{1f}$
		& $(\frac{1}{2},-\frac{1}{2})$
		& $\sqrt{\frac{1}{3}}({c}d)(\bar c\{ud\})-\sqrt{\frac{2}{3}}({c}u)(\bar c\{dd\})$
		\\
		\cline{2-4}
	&	$8_{2f}$
		& $(\frac{1}{2},-\frac{1}{2})$
		& $({c}d)(\bar c[ud])$
		\\

		\bottomrule[1.0pt]
		\bottomrule[1.0pt]
	\end{tabular}
\end{table}
In this part, we  introduce the flavor-color wave functions of the  hidden-charm pentaquark states in  molecular model $(\bar{c}q_{3})(cq_{1}q_{2})  $,   diquark-diquark-antiquark model $(cq_{3})(q_{1}q_{2})\bar{c}  $ and  diquark-triquark model $(cq_{3})(\bar{c}q_{1}q_{2})  $, where $ q_{1,2,3} $ is the light quark.	We construct the flavor wave functions of the  pentaquark  states in the $ SU(3)_{f} $ frame. The $ q_{1}q_{2} $ forms the $ \bar{3}_{f} $ and $6 _{f} $ flavor representation with the total spin $S = 0 $ and $1$, respectively. The $ q_{3} $ combines with $ q_{1}q_{2} $ in $ \bar{3}_{f} $ and $6_{f}$ to form  the flavor representation $\bar{3}_{f} \otimes3_{f}=1_{f}\oplus8 _{2f}  $ and $6_{f} \otimes3_{f}=10_{f}\oplus8 _{1f}  $, respectively. After inserting c, $ \bar{c} $ and the Clebsch-Gordan coefficients, we obtained the flavor wave functions in the molecular model with the configuration $(\bar{c}q_{3})(cq_{1}q_{2})  $. The same method is applied to the other configurations $ (cq_{3})(\bar{c}q_{1}q_{2}) $ and
$(cq_{3})(q_{1}q_{2})\bar{c} $.  Here, the flavor wave function in the $ 1_{f} $ flavor representation is $\sqrt{\frac{1}{3}}\{({\bar{c}}d)( c[us])-(\bar{c}u)( c[ds])+(\bar{c}s)( c[ud])\}$, which includes charm, up,  down and  strange quarks, so we did not consider the $ 1_{f} $ flavor representation.
In our work, we study the hidden-charm pentaquark states with valence quark content $ \bar{c}cudd $,  including the $ P^{N^{0}}_{\psi} $ states with isospin $ I=\frac{1}{2} $ and $P^{\Delta^{0}}_{\psi}$  states with isospin $ I=\frac{3}{2} $. The $P^{N^{0}}_{\psi} $ states and $P^{\Delta^{0}}_{\psi}$  states correspond to $ 8_{f} $ and $ 10_{f} $ flavor representations, respectively.
Therefore, we list the flavor wave functions of the $P^{N^{0}}_{\psi}  $ states and $P^{\Delta^{0}}_{\psi}$  states in the molecular model,  diquark-diquark-antiquark model and diquark-triquark model in Table \ref{tab:k}. When calculating the magnetic moments in the three models, we simultaneously consider the $ S $-wave and $ P $-wave states.

     The color confinement hypothesis implies that all hadrons must be color singlets which means that they do not change in the color $ SU(3) $ space.  We  briefly introduce the color configurations of the  hidden-charm pentaquark states in  three models.

	1. Molecular model $(\bar{c}q_{3})(cq_{1}q_{2})  $: The  hidden-charm pentaquark states in the molecular model $(\bar{c}q_{3})(cq_{1}q_{2})  $ are composed of two clusters of meson and baryon.  As a molecular state, $ \bar{c}q_{3} $ and $cq_{1}q_{2}  $ are in the color singlet.   The color representation of $ (\bar{c}q_{3}) $ cluster  tends to be $\bar{3}_{c} \otimes3_{c}=1_{c}\oplus8 _{c}$, the color representation of $ (cq_{1}q_{2}) $ cluster  tends to be $3_{c} \otimes(3_{c}\otimes3_{c})=3_{c} \otimes(\bar3_{c}\oplus6_{c})=(1_{c}\oplus8_{2c})\oplus(8_{1c}\oplus10_{c}) $.
	
    2. Diquark-diquark-antiquark model  $(cq_{3})(q_{1}q_{2})\bar{c}  $: In this model, diquark $(cq_{3})$  and  diquark $(q_{1}q_{2}) $  are both antisymmetric $ \bar{3} _{c}$ color representation, and $(\bar{3}_{c}\otimes\bar{3}_{c})  $ prefers to form $ 3_{c} $.
    The $ \bar{c} $ is also on the $ \bar{3}_{c} $ color representation, using ($\bar{3}_{c} \otimes\bar{3}_{c})\otimes\bar{3}_{c}=3_{c}\otimes\bar{3}_{c}$ obtain the color singlet representation of the diquark-diquark-antiquark model.
	
	3. Diquark-triquark model $(cq_{3})(\bar{c}q_{1}q_{2})$: In the diquark-triquark model, triquark $(\bar{c}q_{1}q_{2})$ is symmetric $3_{c} $ color representation, and diquark $ (cq_{3}) $ is antisymmetric $\bar{3}_{c} $ color representation. So we have $3_{c}\otimes\bar{3}_{c}$ obtain a color singlet.

	\renewcommand\tabcolsep{0.525cm}
	\renewcommand{\arraystretch}{1.7}
	\begin{table*}[htbp]
		\caption{The magnetic moments of the $P^{N^{0}}_{\psi}  $ states in the molecular model with the $ 8_{1f} $ and $ 8_{2f} $ flavor representation.  The $J_{1}^{P_{1}}\otimes J_{2}^{P_{2}}\otimes J_{3}^{P_{3}}$ are corresponding to the angular momentum and parity of baryon, meson and orbital, respectively. The  $ \mu_{l(baryon/ meson)} $ represents the orbital excitation between corresponding hadrons. The unit is the nuclear magnetic moment $ \mu_{N} $.}
		\scriptsize
		\label{tab:mb} 
		\begin{tabular}{c|c|c|c|c}
			\toprule[1pt]
			\toprule[1pt]
			\multicolumn{5}{c}{$ \left|\Lambda_c^{+} {D}^{(*)-}\right\rangle$}\\
			\toprule[1pt]
			$J^P$	& $^{2s+1}L_J$ & $J_{1}^{P_{1}}\otimes J_{2}^{P_{2}}\otimes J_{3}^{P_{3}}$  & Expressions  &  Results \\
			\hline
			{$\frac{1}{2}^{-}$}	&{{${^{2}S_{\frac{1}{2}}}$}}  &${\frac{1}{2}}^{+}\otimes0^{-}\otimes0^{+}$  &$\mu _{c}$  &$0.403$ \\

			&& ${\frac{1}{2}}^{+}\otimes1^{-}\otimes0^{+}$ & $\frac{1}{3} \left(-4 \mu _{c}+2 \mu _{d}\right)$ & $-1.035$ \\
			\hline
			{$\frac{3}{2}^{-}$}  &{{${^{4}S_{\frac{3}{2}}}$}} & ${\frac{1}{2}}^{+}\otimes1^{-}\otimes0^{+}$ &$\mu_{d} $  & $-0.947$ \\
			\hline
			{$\frac{1}{2}^{+}$}	&{{${^{2}P_{\frac{1}{2}}}$}}  &${\frac{1}{2}}^{+}\otimes0^{-}\otimes1^{-}$  &$\frac{1}{3} \left(- \mu _{c}+2\mu_{  l(\Lambda_c^{+}/{D}^{-})  }\right)$  &$-0.196$  \\

			&& $({\frac{1}{2}}^{+}\otimes1^{-})_{\frac{1}{2}}\otimes1^{-}$ & $\frac{1}{9} \left(3 \mu_{c}-2 \mu _{d}+6\mu_{  l(\Lambda_c^{+}/{D}^{*-})  }\right)$  &$0.750 $ \\
			\cline{2-5}
			&{{${^{4}P_{\frac{1}{2}}}$}}  & $({\frac{1}{2}}^{+}\otimes1^{-})_{\frac{3}{2}}\otimes1^{-}$ &  $\frac{1}{9} \left(5 \mu _{d}-3\mu_{  l(\Lambda_c^{+}/{D}^{*-})  }\right)$ & $-0.507$ \\
			\hline
			{$\frac{3}{2}^{+}$}&{{${^{2}P_{\frac{3}{2}}}$}}  & ${\frac{1}{2}}^{+}\otimes0^{-}\otimes1^{-}$ &  $  \mu _{c}+\mu_{  l(\Lambda_c^{+}/{D}^{-})  }$ &$0.311 $ \\
			
			&&$({\frac{1}{2}}^{+}\otimes1^{-})_{\frac{1}{2}}\otimes1^{-}$  & $\frac{1}{3} \left(-3 \mu _{c}+2 \mu _{d}+3\mu_{  l(\Lambda_c^{+}/{D}^{*-})  }\right)$   &$ -1.092$ \\
			\cline{2-5}
			&{{${^{4}P_{\frac{3}{2}}}$}}  &$({\frac{1}{2}}^{+}\otimes1^{-})_{\frac{3}{2}}\otimes1^{-}$  &$\frac{1}{15} \left(11 \mu _{d}+6\mu_{  l(\Lambda_c^{+}/{D}^{*-})  }\right)$   &$ -0.718$ \\
			\hline
			{$\frac{5}{2}^{+}$}	&{{${^{4}P_{\frac{5}{2}}}$}}   &${\frac{1}{2}}^{+}\otimes1^{-}\otimes1^{-}$  &$  \mu _{d}+\mu_{  l(\Lambda_c^{+}/{D}^{*-})  }$  & $-1.004$ \\
			
			\toprule[1pt]
			\multicolumn{5}{c}{$ \sqrt{\frac{1}{3}}\left|\Sigma_c^{(*)+} {D}^{(*)-}\right\rangle-\sqrt{\frac{2}{3}}\left|\Sigma_c^{(*)0}\bar{D}^{(*)0}\right\rangle$}\\
			\toprule[1pt]
			$J^P$	& $^{2s+1}L_J$ &  $J_{1}^{P_{1}}\otimes J_{2}^{P_{2}}\otimes J_{3}^{P_{3}}$  &Expressions& Results \\
			\hline
			{$\frac{1}{2}^{-}$}	&{{${^{2}S_{\frac{1}{2}}}$}}  &${\frac{1}{2}}^{+}\otimes0^{-}\otimes0^{+}$  &$\frac{1}{9} \left(2 \mu _{u}+10\mu _{d}-3\mu _{c}\right)$  &$  -0.766$ \\

			&& ${\frac{1}{2}}^{+}\otimes1^{-}\otimes0^{+}$ & $\frac{1}{27} \left(10 \mu _{u}-4 \mu _{d}-15\mu _{c}\right)$ &$ 0.618$ \\
			
			&& ${\frac{3}{2}}^{+}\otimes1^{-}\otimes0^{+}$ & $\frac{1}{27} \left(-\mu _{u}+22 \mu _{d}+24\mu _{c}\right)$ & $-0.483$ \\
			\hline

			{$\frac{3}{2}^{-}$}  &{{${^{4}S_{\frac{3}{2}}}$}} & ${\frac{1}{2}}^{+}\otimes1^{-}\otimes0^{+}$ &$\frac{1}{9} \left(8\mu _{u}+13 \mu _{d}-12\mu _{c}\right)$  & $-0.222$ \\
			&& ${\frac{3}{2}}^{+}\otimes0^{-}\otimes0^{+}$ & $\frac{1}{3} \left(\mu _{u}+5\mu _{d}+3\mu _{c}\right)$ &$ -0.544$ \\
			&& ${\frac{3}{2}}^{+}\otimes1^{-}\otimes0^{+}$ & $\frac{1}{45} \left(23\mu _{u}+61\mu _{d}+15\mu _{c}\right)$ &$ -0.181$ \\
			\hline
			{$\frac{5}{2}^{-}$}  &{{${^{6}S_{\frac{5}{2}}}$}} & ${\frac{3}{2}}^{+}\otimes1^{-}\otimes0^{+}$ &$\mu _{u}+2\mu _{d}$  & 0 \\
			\hline
			{$\frac{1}{2}^{+}$}	&{{${^{2}P_{\frac{1}{2}}}$}}  &${\frac{1}{2}}^{+}\otimes0^{-}\otimes1^{-}$  &$\frac{1}{27} \left(-2\mu _{u}-10\mu _{d}+3\mu _{c}+6\mu_{  l(\Sigma_c^{+}/{D}^{-})  }+12\mu_{  l(\Sigma_c^{0}/\bar{D}^{0})  }\right)$  &$0.229$  \\
			
			&& $({\frac{1}{2}}^{+}\otimes1^{-})_{\frac{1}{2}}\otimes1^{-}$ & $\frac{1}{81} \left(- 10\mu _{u}+4\mu _{d}+15\mu _{c}+18\mu_{  l(\Sigma_c^{+}/{D}^{*-})  }+36\mu_{  l(\Sigma_c^{0}/\bar{D}^{*0})  }\right)$  &$-0.225$  \\
			\cline{2-5}
			&{{${^{4}P_{\frac{1}{2}}}$}}  & $({\frac{1}{2}}^{+}\otimes1^{-})_{\frac{3}{2}}\otimes1^{-}$ &$\frac{1}{81} \left(40\mu _{u}+65\mu _{d}-60\mu _{c}-9\mu_{  l(\Sigma_c^{+}/{D}^{*-})  }-18\mu_{  l(\Sigma_c^{0}/\bar{D}^{*0})  }\right)$ & $-0.114$ \\
			&& ${\frac{3}{2}}^{+}\otimes0^{-}\otimes1^{-}$ & $\frac{1}{27} \left(5\mu _{u}+25\mu _{d}+15\mu _{c}-3\mu_{  l(\Sigma_c^{*+}/{D}^{-})  }-6\mu_{  l(\Sigma_c^{*0}/\bar{D}^{0})  }\right)$  &$-0.289$  \\
			\cline{2-5}
			&{{${^{2}P_{\frac{1}{2}}}$}}  & $({\frac{3}{2}}^{+}\otimes1^{-})_{\frac{1}{2}}\otimes1^{-}$ &$\frac{1}{81} \left(\mu _{u}-22\mu _{d}-24\mu _{c}+18\mu_{  l(\Sigma_c^{*+}/{D}^{*-})  }+36\mu_{  l(\Sigma_c^{*0}/\bar{D}^{*0})  }\right)$ &$ 0.140 $\\
			&{{${^{4}P_{\frac{1}{2}}}$}}  & $({\frac{3}{2}}^{+}\otimes1^{-})_{\frac{3}{2}}\otimes1^{-}$ &$\frac{1}{81} \left(23\mu _{u}+61\mu _{d}+15\mu _{c}-9\mu_{  l(\Sigma_c^{*+}/{D}^{*-})  }-18\mu_{  l(\Sigma_c^{*0}/\bar{D}^{*0})  }\right)$ &$ -0.090$ \\
			\hline
			$ \frac{3}{2}^{+} $&{{${^{2}P_{\frac{3}{2}}}$}}  & ${\frac{1}{2}}^{+}\otimes0^{-}\otimes1^{-}$ &$\frac{1}{9} \left(2\mu _{u}+10\mu _{d}-3\mu _{c}+3\mu_{  l(\Sigma_c^{+}/{D}^{-})  }+6\mu_{  l(\Sigma_c^{0}/\bar{D}^{0})  }\right)$ & $-0.806$ \\
			& & $({\frac{1}{2}}^{+}\otimes1^{-})_{\frac{1}{2}}\otimes1^{-}$ &$\frac{1}{27} \left(10\mu _{u}-4\mu _{d}-15\mu _{c}+9\mu_{  l(\Sigma_c^{+}/{D}^{*-})  }+18\mu_{  l(\Sigma_c^{0}/\bar{D}^{*0})  }\right)$ &$ 0.590$ \\
			\cline{2-5}
			&{{${^{4}P_{\frac{3}{2}}}$}}  & $({\frac{1}{2}}^{+}\otimes1^{-})_{\frac{3}{2}}\otimes1^{-}$ &$\frac{1}{135} \left(88\mu _{u}+143\mu _{d}-132\mu _{c}+18\mu_{  l(\Sigma_c^{+}/{D}^{*-})  }+36\mu_{  l(\Sigma_c^{0}/\bar{D}^{*0})  }\right)$ &$ -0.174$ \\
			& & ${\frac{3}{2}}^{+}\otimes0^{-}\otimes1^{-}$ &$\frac{1}{45} \left(11\mu _{u}+55\mu _{d}+33\mu _{c}+6\mu_{  l(\Sigma_c^{*+}/{D}^{-})  }+12\mu_{  l(\Sigma_c^{*0}/\bar{D}^{0})  }\right)$ &$ -0.415$ \\
			\cline{2-5}
			&{{${^{2}P_{\frac{3}{2}}}$}}  & $({\frac{3}{2}}^{+}\otimes1^{-})_{\frac{1}{2}}\otimes1^{-}$ &$\frac{1}{27} \left(-\mu _{u}+22\mu _{d}+24\mu _{c}+9\mu_{  l(\Sigma_c^{*+}/{D}^{*-})  }+18\mu_{  l(\Sigma_c^{*0}/\bar{D}^{*0})  }\right)$ &$ -0.515$ \\
			&{{${^{4}P_{\frac{3}{2}}}$}}  & $({\frac{3}{2}}^{+}\otimes1^{-})_{\frac{3}{2}}\otimes1^{-}$ &$\frac{1}{675} \left(253\mu _{u}+671\mu _{d}+165\mu _{c}+90\mu_{  l(\Sigma_c^{*+}/{D}^{*-})  }+180\mu_{  l(\Sigma_c^{*0}/\bar{D}^{*0})  }\right)$ & $-0.146$ \\
			&{{${^{6}P_{\frac{3}{2}}}$}}  & $({\frac{3}{2}}^{+}\otimes1^{-})_{\frac{5}{2}}\otimes1^{-}$ &$\frac{1}{75} \left(63\mu _{u}+126\mu _{d}-15\mu_{  l(\Sigma_c^{*+}/{D}^{*-})  }-30\mu_{  l(\Sigma_c^{*0}/\bar{D}^{*0})  }\right)$   & $0.019$ \\
			\hline
			$ \frac{5}{2}^{+} $&{{${^{4}P_{\frac{5}{2}}}$}}  & ${\frac{1}{2}}^{+}\otimes1^{-}\otimes1^{-}$ &$\frac{1}{9} \left(8\mu _{u}+13\mu _{d}-12\mu _{c}+3\mu_{  l(\Sigma_c^{+}/{D}^{*-})  }+6\mu_{  l(\Sigma_c^{0}/\bar{D}^{*0})  }\right)$ & $-0.125$ \\
			& & ${\frac{3}{2}}^{+}\otimes0^{-}\otimes1^{-}$ &$\frac{1}{3} \left(\mu _{u}+5\mu _{d}+3\mu _{c}+\mu_{  l(\Sigma_c^{*+}/{D}^{-})  }+2\mu_{  l(\Sigma_c^{*0}/\bar{D}^{0})  }\right)$ &$ -0.584$ \\
			&& $({\frac{3}{2}}^{+}\otimes1^{-})_{\frac{3}{2}}\otimes1^{-}$ &$\frac{1}{45} \left(23\mu _{u}+61\mu _{d}+15\mu _{c}+15\mu_{  l(\Sigma_c^{*+}/{D}^{*-})  }+30\mu_{  l(\Sigma_c^{*0}/\bar{D}^{*0})  }\right)$   & $-0.213$ \\
			\cline{2-5}
			&{{${^{6}P_{\frac{5}{2}}}$}}  &$({\frac{3}{2}}^{+}\otimes1^{-})_{\frac{5}{2}}\otimes1^{-}$  &$\frac{1}{105} \left(93\mu _{u}+186\mu _{d}+10\mu_{  l(\Sigma_c^{*+}/{D}^{*-})  }+20\mu_{  l(\Sigma_c^{*0}/\bar{D}^{*0})  }\right)$ & $-0.009$ \\
			\hline
			$ \frac{7}{2}^{+} $&{{${^{6}P_{\frac{7}{2}}}$}}  & ${\frac{3}{2}}^{+}\otimes1^{-}\otimes1^{-}$ &$\frac{1}{3} \left(3\mu _{u}+6\mu _{d}+\mu_{  l(\Sigma_c^{*+}/{D}^{*-})  }+2\mu_{  l(\Sigma_c^{*0}/\bar{D}^{*0})  }\right)$ &$ -0.031$ \\
			\bottomrule[1pt]
			\bottomrule[1pt]
		\end{tabular}
	\end{table*}

	\renewcommand\tabcolsep{0.525cm}
	\renewcommand{\arraystretch}{1.7}
	\begin{table*}[htbp]
		\caption{ The magnetic moments of the  $P^{\Delta^{0}}_{\psi}$   states in the molecular model with the $ 10_{f} $  flavor representation.   The  $J_{1}^{P_{1}}\otimes J_{2}^{P_{2}}\otimes J_{3}^{P_{3}}$ are corresponding to the angular momentum and parity of baryon, meson and orbital, respectively.  The $ \mu_{l(baryon/ meson)} $ represents the orbital excitation between corresponding hadrons. The unit is the nuclear magnetic moment $ \mu_{N} $.}
		\scriptsize
		\label{tab:ms} 
		\begin{tabular}{c|c|c|c|c}
			\toprule[1pt]
			\toprule[1pt]
			\multicolumn{5}{c}{$ \sqrt{\frac{2}{3}}\left|\Sigma_c^{(*)+} {D}^{(*)-}\right\rangle+\sqrt{\frac{1}{3}}\left|\Sigma_c^{(*)0}\bar{D}^{(*)0}\right\rangle$}\\
			\toprule[1pt]
			$J^P$	& $^{2s+1}L_J$ & $J_{1}^{P_{1}}\otimes J_{2}^{P_{2}}\otimes J_{3}^{P_{3}}$   &Expressions & Results \\
			\hline
			{$\frac{1}{2}^{-}$}	&{{${^{2}S_{\frac{1}{2}}}$}}  &${\frac{1}{2}}^{+}\otimes0^{-}\otimes0^{+}$  &$\frac{1}{9} \left(4 \mu _{u}+8\mu _{d}-3\mu _{c}\right)$  &$-0.134$ \\

			&& ${\frac{1}{2}}^{+}\otimes1^{-}\otimes0^{+}$ & $\frac{1}{27} \left(2 \mu _{u}+4 \mu _{d}-15\mu _{c}\right)$ &$ -0.224$ \\
			
			&& ${\frac{3}{2}}^{+}\otimes1^{-}\otimes0^{+}$ & $\frac{1}{27} \left(7\mu _{u}+14 \mu _{d}+24\mu _{c}\right)$ &$ 0.359$ \\
			\hline
			
			{$\frac{3}{2}^{-}$}  &{{${^{4}S_{\frac{3}{2}}}$}} & ${\frac{1}{2}}^{+}\otimes1^{-}\otimes0^{+}$ &$\frac{1}{9} \left(7\mu _{u}+14 \mu _{d}-12\mu _{c}\right)$  &$ -0.538$ \\
			&& ${\frac{3}{2}}^{+}\otimes0^{-}\otimes0^{+}$ & $\frac{1}{3} \left(2\mu _{u}+4\mu _{d}+3\mu _{c}\right)$ &$ 0.403 $\\
			&& ${\frac{3}{2}}^{+}\otimes1^{-}\otimes0^{+}$ & $\frac{1}{45} \left(28\mu _{u}+56\mu _{d}+15\mu _{c}\right)$ &$ 0.134$ \\
			\hline
			{$\frac{5}{2}^{-}$}  &{{${^{6}S_{\frac{5}{2}}}$}} & ${\frac{3}{2}}^{+}\otimes1^{-}\otimes0^{+}$ &$\mu _{u}+2\mu _{d}$  & $0$ \\
			\hline
			{$\frac{1}{2}^{+}$}	&{{${^{2}P_{\frac{1}{2}}}$}}  &${\frac{1}{2}}^{+}\otimes0^{-}\otimes1^{-}$  &$\frac{1}{27} \left(-4\mu _{u}-8\mu _{d}+3\mu _{c}+12\mu_{  l(\Sigma_c^{+}/{D}^{-})  }+6\mu_{  l(\Sigma_c^{0}/\bar{D}^{0})  }\right)$  &$-0.009$  \\
			
			&& $({\frac{1}{2}}^{+}\otimes1^{-})_{\frac{1}{2}}\otimes1^{-}$ & $\frac{1}{81} \left(- 2\mu _{u}- 4\mu _{d}+15\mu _{c}+36\mu_{  l(\Sigma_c^{+}/{D}^{*-})  }+18\mu_{  l(\Sigma_c^{0}/\bar{D}^{*0})  }\right)$  &$0.037$  \\
			\cline{2-5}
			&{{${^{4}P_{\frac{1}{2}}}$}}  & $({\frac{1}{2}}^{+}\otimes1^{-})_{\frac{3}{2}}\otimes1^{-}$ &$\frac{1}{81} \left(35\mu _{u}+70\mu _{d}-60\mu _{c}-18\mu_{  l(\Sigma_c^{+}/{D}^{*-})  }-9\mu_{  l(\Sigma_c^{0}/\bar{D}^{*0})  }\right)$ & $-0.280$ \\
			&& ${\frac{3}{2}}^{+}\otimes0^{-}\otimes1^{-}$ & $\frac{1}{27} \left(10\mu _{u}+20\mu _{d}+15\mu _{c}-6\mu_{  l(\Sigma_c^{*+}/{D}^{-})  }-3\mu_{  l(\Sigma_c^{*0}/\bar{D}^{0})  }\right)$  &$0.251$  \\
			\cline{2-5}
			&{{${^{2}P_{\frac{1}{2}}}$}}  & $({\frac{3}{2}}^{+}\otimes1^{-})_{\frac{1}{2}}\otimes1^{-}$ &$\frac{1}{81} \left(-7\mu _{u}-14\mu _{d}-24\mu _{c}+36\mu_{  l(\Sigma_c^{*+}/{D}^{*-})  }+18\mu_{  l(\Sigma_c^{*0}/\bar{D}^{*0})  }\right)$ &$ -0.161$ \\
			&{{${^{4}P_{\frac{1}{2}}}$}}  & $({\frac{3}{2}}^{+}\otimes1^{-})_{\frac{3}{2}}\otimes1^{-}$ &$\frac{1}{81} \left(28\mu _{u}+56\mu _{d}+15\mu _{c}-18\mu_{  l(\Sigma_c^{*+}/{D}^{*-})  }-9\mu_{  l(\Sigma_c^{*0}/\bar{D}^{*0})  }\right)$ &$ 0.096$ \\
			\hline
			$ \frac{3}{2}^{+} $&{{${^{2}P_{\frac{3}{2}}}$}}  & ${\frac{1}{2}}^{+}\otimes0^{-}\otimes1^{-}$ &$\frac{1}{9} \left(4\mu _{u}+8\mu _{d}-3\mu _{c}+6\mu_{  l(\Sigma_c^{+}/{D}^{-})  }+3\mu_{  l(\Sigma_c^{0}/\bar{D}^{0})  }\right)$ &$ -0.214$ \\
			& & $({\frac{1}{2}}^{+}\otimes1^{-})_{\frac{1}{2}}\otimes1^{-}$ &$\frac{1}{27} \left(2\mu _{u}+4\mu _{d}-15\mu _{c}+18\mu_{  l(\Sigma_c^{+}/{D}^{*-})  }+9\mu_{  l(\Sigma_c^{0}/\bar{D}^{*0})  }\right)$ & $-0.281$ \\
			\cline{2-5}
			&{{${^{4}P_{\frac{3}{2}}}$}}  & $({\frac{1}{2}}^{+}\otimes1^{-})_{\frac{3}{2}}\otimes1^{-}$ &$\frac{1}{135} \left(77\mu _{u}+154\mu _{d}-132\mu _{c}+36\mu_{  l(\Sigma_c^{+}/{D}^{*-})  }+18\mu_{  l(\Sigma_c^{0}/\bar{D}^{*0})  }\right)$ &$ -0.417$ \\
			& & ${\frac{3}{2}}^{+}\otimes0^{-}\otimes1^{-}$ &$\frac{1}{45} \left(22\mu _{u}+44\mu _{d}+33\mu _{c}+12\mu_{  l(\Sigma_c^{*+}/{D}^{-})  }+6\mu_{  l(\Sigma_c^{*0}/\bar{D}^{0})  }\right)$ &$ 0.264$ \\
			\cline{2-5}
			&{{${^{2}P_{\frac{3}{2}}}$}}  & $({\frac{3}{2}}^{+}\otimes1^{-})_{\frac{1}{2}}\otimes1^{-}$ &$\frac{1}{27} \left(7\mu _{u}+14\mu _{d}+24\mu _{c}+18\mu_{  l(\Sigma_c^{*+}/{D}^{*-})  }+9\mu_{  l(\Sigma_c^{*0}/\bar{D}^{*0})  }\right)$ & $0.296$ \\
			&{{${^{4}P_{\frac{3}{2}}}$}}  & $({\frac{3}{2}}^{+}\otimes1^{-})_{\frac{3}{2}}\otimes1^{-}$ &$\frac{1}{675} \left(308\mu _{u}+616\mu _{d}+165\mu _{c}+180\mu_{  l(\Sigma_c^{*+}/{D}^{*-})  }+90\mu_{  l(\Sigma_c^{*0}/\bar{D}^{*0})  }\right)$ &$ 0.074$ \\
			&{{${^{6}P_{\frac{3}{2}}}$}}  & $({\frac{3}{2}}^{+}\otimes1^{-})_{\frac{5}{2}}\otimes1^{-}$ &$\frac{1}{75} \left(63\mu _{u}+126\mu _{d}-30\mu_{  l(\Sigma_c^{*+}/{D}^{*-})  }-15\mu_{  l(\Sigma_c^{*0}/\bar{D}^{*0})  }\right)$   &$ 0.038$ \\
			\hline
			$ \frac{5}{2}^{+} $&{{${^{4}P_{\frac{5}{2}}}$}}  & ${\frac{1}{2}}^{+}\otimes1^{-}\otimes1^{-}$ &$\frac{1}{9} \left(7\mu _{u}+14\mu _{d}-12\mu _{c}+6\mu_{  l(\Sigma_c^{+}/{D}^{*-})  }+3\mu_{  l(\Sigma_c^{0}/\bar{D}^{*0})  }\right)$ & $-0.532$ \\
			& & ${\frac{3}{2}}^{+}\otimes0^{-}\otimes1^{-}$ &$\frac{1}{3} \left(2\mu _{u}+4\mu _{d}+3\mu _{c}+2\mu_{  l(\Sigma_c^{*+}/{D}^{-})  }+\mu_{  l(\Sigma_c^{*0}/\bar{D}^{0})  }\right)$ &$ 0.323$ \\
			&& $({\frac{3}{2}}^{+}\otimes1^{-})_{\frac{3}{2}}\otimes1^{-}$ &$\frac{1}{45} \left(28\mu _{u}+56\mu _{d}+15\mu _{c}+30\mu_{  l(\Sigma_c^{*+}/{D}^{*-})  }+15\mu_{  l(\Sigma_c^{*0}/\bar{D}^{*0})  }\right)$   &$ 0.072$ \\
			\cline{2-5}
			&{{${^{6}P_{\frac{5}{2}}}$}}  &$({\frac{3}{2}}^{+}\otimes1^{-})_{\frac{5}{2}}\otimes1^{-}$  &$\frac{1}{105} \left(93\mu _{u}+186\mu _{d}+20\mu_{  l(\Sigma_c^{*+}/{D}^{*-})  }+10\mu_{  l(\Sigma_c^{*0}/\bar{D}^{*0})  }\right)$ &$ -0.018$ \\
			\hline
			$ \frac{7}{2}^{+} $&{{${^{6}P_{\frac{7}{2}}}$}}  & ${\frac{3}{2}}^{+}\otimes1^{-}\otimes1^{-}$ &$\frac{1}{3} \left(3\mu _{u}+6\mu _{d}+2\mu_{  l(\Sigma_c^{*+}/{D}^{*-})  }+\mu_{  l(\Sigma_c^{*0}/\bar{D}^{*0})  }\right)$ &$ -0.063$ \\
			\bottomrule[1pt]
			\bottomrule[1pt]
		\end{tabular}
	\end{table*}

	\section{Magnetic moments and transition magnetic moments of the $P^{N^{0}}_{\psi}  $ states and $P^{\Delta^{0}}_{\psi}$  states in the molecular model}
	\label{sec3}
	The magnetic moment of a   compound system is the sum of its constituent magnetic moments, including spin magnetic moment and orbital magnetic moment,
	\begin{align}
		\vec{\mu}&=\vec{\mu}_{spin}+\vec{\mu}_{orbital}
		=\sum_{i}(g_{i}{\vec{S}}_i+\vec{l_{i}})\mu_{i}, 
	\end{align}
	where $g _{i} $ is the g factor of $ i $-th constituent, $ \vec{S}_i $ is the spin of the $ i $-th constituent and $\mu _{i} $ is the
	magneton of the $ i $-th constituent,
	\begin{align}
	 \mu_{i}=\frac{q_{i}}{2m_{i}} .	 
	\end{align}
	
	In the above expression, $ q_{i} $ and $ m_{i} $ as the charge and mass of the $ i $-th
	 constituent, respectively. In the molecular model, the constituents of spin magnetic moment are baryons and mesons.
	
	The total magnetic moment of  hidden-charm pentaquark states   in the molecular model $(\bar{c}q_{3})(cq_{1}q_{2})  $  includes the sum of the meson spin magnetic moment, the baryon spin magnetic moment and the orbital magnetic moment
	\begin{eqnarray} \label{w1}
		\vec{\mu}&=&g_{m}{\mu_{m}}{\vec{S}_m}+g_{b}{\mu_b}{\vec{S}}_b+{\mu_l}\vec{l}, \\
		{\mu_l}&=&\frac{{m_b}\mu_{m}+{m_m}\mu_{b}}{m_m+m_b},
	\end{eqnarray}
	
	where $\mu_{l}  $ is the orbital magnetic moment between
	the two hadrons. The subscripts $ m $ and $ b $ represent meson and baryon, respectively. We use the masses of mesons and baryons from  Particle Data Group \cite{ParticleDataGroup:2022pth}. The magnetic moment formula of hidden-charm pentaquark  states $(\bar{c}q_{3})(cq_{1}q_{2})  $ in the molecular model is
\begin{align}
	\mu &=\left\langle \psi\left |g_{m}{\mu_{m}}{\vec{S}_m}+g_{b}{\mu_b}{\vec{S}}_b+{\mu_l}\vec{l}\,\,\right|\psi\right\rangle \nonumber\\
	&=\sum_{l_z, {S_z}}\left\langle l l_z,SS_z|JJ_{z}\right\rangle^2\bigg\{\mu_{l}l_z+\sum_{S'_b} \Big\langle S_bS'_b,S_m (S_z\nonumber\\
	&-S'_b)|SS_z\Big \rangle^2\Big[ (S_z-S'_b)(\mu_{\bar c}+\mu_{q_3})\nonumber\\
	&+\sum_{S'_{c},S'_{12}}\langle S_cS'_c,S_{12}S'_{12}|S_b
	S'_b\rangle^2\big(g_{c}\mu_{c}S'_{c}+S'_{12}(\mu_{q_1}\nonumber\\
&+\mu_{q_2})\big)\Big ]\bigg \},
\end{align}
	where $\psi$ represents the flavor wave function in molecular model in Table \ref{tab:k}, $ S_{12} $ and $ S'_{12} $ are the spin and its third componen of the diquark $ (q_{1}q_{2}) $ inside the baryon $(cq_{1}q_{2})  $. $ S $ is the spin of  pentaquark state. $J$ is the total angular momentum of pentaquark state.  $S_{c}$ and $S'_{c}$ are  spin and third component of quark $c $. $S_{z}, S' _{b} $ and $l _{z} $ are the spin third component of the  pentaquark state, the baryon, and the orbital excitation, respectively. In the generated table, $ \mu_{\bar{c}} $ did not appear in these pentaquark state magnetic moment expressions because the relationship $\mu_{c} =-\mu_{\bar{c}} $ can simplify these expressions.

	We	take $J^{P}=\frac{1}{2}^{-}({\frac{1}{2}}^{+}\otimes0^{-}\otimes0^{+})$ of the $\Sigma_c \bar{D}$ molecular state with the $8 _{1f} $ flavor representation  as an example. The  $J_{1}^{P_{1}}\otimes J_{2}^{P_{2}}\otimes J_{3}^{P_{3}}$ are corresponding to the angular momentum and parity of baryon, meson and orbital, respectively. According to  Table \ref{tab:k}, the flavor wave function can be written as 
	\begin{align}
		\psi_{\Sigma_c^{} \bar{D}}&=\sqrt{\frac{1}{3}}\left|\Sigma_c^{+} {D}^{-}\right\rangle-\sqrt{\frac{2}{3}}\left|\Sigma_c^{0}\bar{D}^{0}\right\rangle\nonumber\\
		&= \sqrt{\frac{1}{3}}(c\{ud\})({\bar c}d) -\sqrt{\frac{2}{3}}(c\{dd\})({\bar c}u).
	\end{align}
	
 We obtained the  magnetic moment expression  
	\begin{align}
		\mu_{\Sigma_c^{} \bar{D}} &=\left\langle \psi_{\Sigma_c^{} \bar{D}}\left |g_{m}{\mu_{m}}{\vec{S}_m}+g_{b}{\mu_b}{\vec{S}}_b+{\mu_l}\vec{l}\,\,\right|\psi_{\Sigma_c^{} \bar{D}}\right\rangle \nonumber\\
		& =
		\langle 00,\frac{1}{2}\frac{1}{2} |\frac{1}{2}\frac{1}{2}\rangle^{2}\Big [
		\langle \frac{1}{2}\frac{1}{2},0 0 |\frac{1}{2}\frac{1}{2}\rangle^{2}
		\big (\langle \frac{1}{2}\frac{1}{2},1 0 |\frac{1}{2}\frac{1}{2}\rangle^{2}		\mu_{c}
		\nonumber\\ 
		& \ +	\langle \frac{1}{2}-\frac{1}{2},1 1 |\frac{1}{2}\frac{1}{2}\rangle^{2} (-\mu_{c}+\frac{1}{3}\mu_{u}+\frac{5}{3}\mu_{d})\big)	\Big ]
		\nonumber\\ 
		& =
		\frac{1}{9} (2\mu_{u}+10\mu_{d}-3\mu_{c} ).
	\end{align}
The magnetic moments  of hidden-charm pentaquark states with isospin $(I,I_3)=(\frac{1}{2},-\frac{1}{2})$ and $(I,I_3)=(\frac{3}{2},-\frac{1}{2})$ in the molecular model are  collected in Table \ref{tab:mb} and Table \ref{tab:ms} respectively. 	In this work, we use the following constituent quark masses \cite{Wang:2016dzu},
	\begin{eqnarray}
		m_u \ =\ m_d \ =\  0.33\ \mbox{GeV},   m_c \ =\ 1.55\ \mbox{GeV}. \
	\end{eqnarray}

	When we express  transition magnetic moments of  hidden-charm pentaquark states in
	the molecular model, we need the magnetic moments and the transition magnetic moments of  baryon $\Sigma_c^{(*)} $  and meson $\bar{D}^{(*)}  $. Since the experimental values of the magnetic moments of  baryon $\Sigma_c^{(*)} $ and   meson $\bar{D}^{(*) } $ are still absent,  we  calculate their magnetic moments and transition magnetic moments. Our numerical results are consistent with the results of other theoretical works \cite{Wang:2022tib,Zhou:2022gra,Zhang:2021yul}. We take the baryon $ \Sigma_c^{+} $ as an example to illustrate how to obtain the magnetic moments of  conventional mesons and baryons. Similar to calculating the magnetic moments of pentaquarks, the magnetic moments of  baryon with the  configuration $(cq_{1}q_{2})  $ is  
	\begin{align}
		\mu_{b} =&\left\langle \psi_{b}\left |g_{b}{\mu_b}{\vec{S}}_b\,\right|\psi_{b}\right\rangle\nonumber \\
		=&\big\langle\frac{1}{2}\frac{1}{2},S_{12}(S'_b-\frac{1}{2})|S_b
		S'_b\big\rangle^2\big[\mu_{
			c}+(S'_b-\frac{1}{2})(\mu_{q_1}\nonumber\\
		+&\mu_{q_2})\big]
		+\big\langle\frac{1}{2}-\frac{1}{2},s_{12}(S'_b+\frac{1}{2})|S_b
		S'_b\big\rangle^2\big[-\mu_{c}\nonumber\\
    	+&(S'_b+\frac{1}{2})(\mu_{q_1}	+\mu_{q_2})\big],
		\end{align}
	where  $S_{b} $ is the spin of barons, $S' _{b} $ is the third component of   the spin of  baryon, $ S_{12} $  is the spin of the diquark $(q_{1}q_{2})  $ inside  baryon $(cq_{1}q_{2})  $.  $\psi_{b}$ represents  flavor wave function of  baryon.
	
	 We obtion the magnetic moment expression of the  baryon $ \Sigma_c^{+} $, 
	\begin{align}
		\mu_{ \Sigma_c^{+}} &=\left\langle \psi_{\Sigma_c^{+}}\left |g_{b}{\mu_b}{\vec{S}}_b\,\right|\psi_{\Sigma_c^{+}}\right\rangle \nonumber\\
		& =\langle \frac{1}{2}\frac{1}{2},1 0 |\frac{1}{2}\frac{1}{2}\rangle^{2}	\mu_{c}
		+	\langle \frac{1}{2}-\frac{1}{2},1 1 |\frac{1}{2}\frac{1}{2}\rangle^{2} (-\mu_{c}+\mu_{u}+\mu_{d})
		\nonumber\\ 
		& =
		\frac{2}{3}\mu_{u}+\frac{2}{3}\mu_{d}-\frac{1}{3}\mu_{c} .
	\end{align}

	Similarly to the solution procedure of the magnetic moment of the baryon $\Sigma_c^{+}  $, we obtion the expressions of  magnetic moments  of the $ S $-wave charmed baryon $ {\Sigma}_{c}^{(*)}$, $\Lambda_c^{} $ and the $ S $-wave anti-charmed meson $ \bar{D}^{(*)}  $ in Table \ref{tab:sl}.

	\renewcommand\tabcolsep{0.08cm}
	\renewcommand{\arraystretch}{1.50}
	\begin{table}[htbp]
		\caption{Magnetic moments of the $ S $-wave charmed baryon $ {\Sigma}_{c}^{(*)}$, $\Lambda_c^{} $ and the $ S $-wave anti-charmed meson ${\bar{D}^{*}} $.  Here, the magnetic moment of the $ S $-wave anti-charmed meson $ \bar{D} $ is zero. The unit is the nuclear magnetic moment $ \mu_{N} $.}
		\scriptsize
		\label{tab:sl} 
		\begin{tabular}{l|c|c|cccc}
			\toprule[1pt]
			\toprule[1pt]
			Hadrons	&$I (J^P)$ & Expressions  & Results &\cite{Wang:2022tib}&\cite{Zhou:2022gra}&\cite{Zhang:2021yul}  \\
			\hline
			${\mu}_{{\Sigma}_{c}^{++}}$	&$ 1(\frac{1}{2}^{+}) $   & $	\frac{4}{3}\mu_{u}-\frac{1}{3}\mu_{c}  $  &$2.392$&&$ 2.357 $&2.130 \\
			${\mu}_{{\Sigma}_{c}^{+}}$	&  & $	\frac{2}{3}\mu_{u}+\frac{2}{3}\mu_{d}-\frac{1}{3}\mu_{c}  $  &$0.497$&&$ 0.496 $&$ 0.410 $ \\
		
			${\mu}_{{\Sigma}_{c}^{0}}$	&  & $	\frac{4}{3}\mu_{d}-\frac{1}{3}\mu_{c}  $  
			&$-1.398$&&$-1.365$&$-1.310$ \\
				\hline
			${\mu}_{\Lambda_c^{+}}$	&$ 0(\frac{1}{2}^{+}) $   & $	\mu_{c} $  &$0.403$&&& \\
			
			\hline
			${\mu}_{{\Sigma}_{c}^{*++}}$	&$ 1(\frac{3}{2}^{+}) $   & $	2\mu_{u}+\mu_{c}  $  &$4.193$&&$ 4.094 $&$ 4.070 $ \\
			${\mu}_{{\Sigma}_{c}^{*+}}$	&  & $	\mu_{u}+\mu_{d}+\mu_{c}  $  &$1.351$&&$ 1.302 $&$ 1.390 $ \\
			${\mu}_{{\Sigma}_{c}^{*0}}$	&   & $	2\mu_{d}+\mu_{c}  $  &$-1.492$&&$ -1.490 $&$ -1.290 $ \\
			\hline
			${\mu}_{D^{*-}}$	&$ \frac{1}{2}(1^{-}) $   & $	\mu_{d}+\mu_{\bar{c}}  $  &$-1.351$&$ -1.303 $&& \\
			$ {\mu}_{\bar{D}^{*0}} $	&   & $	\mu_{u}+\mu_{\bar{c}}  $  &$1.492$&$ 1.489 $&& \\
			
			\bottomrule[1pt]
			\bottomrule[1pt]
		\end{tabular}
	\end{table}

We calculate the transition magnetic moments of   baryon $\Sigma_c^{(*)} $ and   meson $\bar{D}^{(*)} $.
The transition magnetic moment is obtained by the third component of the magnetic moment
operator $ \vec{\mu}_{z} $ acting on the hadron wave function \cite{Wang:2022ugk},
\begin{align}
	\mu_{H\rightarrow H'}=\left\langle J_{H'},J_{z} \left|\vec{\mu}_z\right| J_{H},J_{z}\right\rangle^{J_{z}=Min[J_{H},J_{H'}]}.
\end{align}
Here, $ H $ and $H'$ are the  corresponding wave functions of the initial and final states of the investigated hadronic state, respectively.  We use the the maximum spin third component of the lowest state to discuss the transition magnetic moment between pentaquark states.  For instance, we discuss the transition magnetic moment of the $\bar D^{*0}\rightarrow\bar D^0\gamma $ process.  We construct the flavor-spin wave functions of mesons $\bar D^{*0}  $ and $\bar D^0  $ as
	\begin{eqnarray}
		\phi_{\bar D^{*0}}^{S=1;\,S_3=0} &=\frac{1}{\sqrt{2}}|{\bar{c}u}\rangle|\uparrow\downarrow+\downarrow\uparrow\rangle,\\
		\phi_{\bar D^{0}}^{S=0;\,S_3=0} &=\frac{1}{\sqrt{2}}|{\bar{c}u}\rangle|\uparrow\downarrow-\downarrow\uparrow\rangle.
	\end{eqnarray}
	The transition magnetic moment of the $\bar D^{*0}\rightarrow \bar D^0\gamma $ process is	
	\begin{eqnarray}
		\mu_{\bar D^{*0}\rightarrow \bar D^0}&=&\left\langle\phi_{\bar D^{0}}^{S=0;\,S_3=0} \left|\hat{\mu}_z\right| \phi_{\bar D^{*0}}^{S=1;\,S_3=0}\right\rangle \notag\\
		&=& \left\langle \frac{\bar c u\uparrow \downarrow-\bar c u\downarrow\uparrow}{\sqrt{2}}\right|\hat{\mu}_z\left|\frac{\bar c u\uparrow\downarrow+\bar c u\downarrow\uparrow}{\sqrt{2}} \right\rangle \notag\\
		&=&\mu_{\bar c}-\mu_{u}.
	\end{eqnarray}
	
	In Table \ref{tab:ys}, we present the expressions and numerical results of the transition magnetic moments of the $ S $-wave charmed baryons ${\Sigma}_{c}^{(*)} $ and the $ S $-wave anti-charmed mesons ${\bar{D}^{(*)}} $.
	
	\renewcommand\tabcolsep{0.17cm}
	\renewcommand{\arraystretch}{1.5}
	\begin{table}[!htbp]
		\caption{ Transition magnetic moments of the $ S $-wave
			charmed baryons ${\Sigma}_{c}^{(*)} $ and the $ S $-wave anti-charmed mesons ${\bar{D}^{(*)}} $.
			The unit is the nuclear magnetic moment $ \mu_{N} $.}\label{}
		\scriptsize
		\label{tab:ys} 
		\begin{tabular}{l|c|ccc}
			\toprule[1pt]
			\toprule[1pt]
			Decay modes	& Expressions  & Results&\cite{Wang:2022tib}&\cite{Zhou:2022gra}  \\
			\hline
			$\Sigma^{*++}_{c} \to  \Sigma^{++}_{c} \gamma $	& $	\frac{2\sqrt{2}}{3}(\mu_{u}-\mu_{c})  $  &$1.406$&&$ 1.404 $ \\
			$\Sigma^{*+}_{c} \to  \Sigma^{+}_{c} \gamma $	& $	\frac{\sqrt{2}}{3}(\mu_{u}+\mu_{d}-2\mu_{c})  $  &$0.066$&&$ 0.088 $ \\
			$\Sigma^{*0}_{c} \to  \Sigma^{0}_{c} \gamma $& $	\frac{2\sqrt{2}}{3}(\mu_{d}-\mu_{c})  $  &$-1.274$&&$ -1.228 $ \\
			\hline
			$\bar{D}^{*0} \to \bar{D}^{0}\gamma	$	&$-\mu_{u}+\mu_{\bar{c}} $   &$-2.298$&$-2.234$& \\
			$D^{*-} \to  D^{-} \gamma$	& $-\mu_{d}+\mu_{\bar{c}} $  &$0.544$&$0.588  $& \\
			
			\bottomrule[1pt]
			\bottomrule[1pt]
		\end{tabular}
		
	\end{table}
	
		\renewcommand\tabcolsep{1.18cm}
	\renewcommand{\arraystretch}{1.5}
	\begin{table*}
		\caption{The  spin wave functions  of the $ S $-wave $\Sigma_c^{(*)} \bar{D}^{(*)}$ and $\Lambda_c^{} {D}^{(*)}$ systems.  Where $S$ and $S_z$ are the spin and its third component of the investigated system, respectively.}
		\label{tab:zx}
		\begin{tabular}{l|l|l}
			\toprule[1.0pt]
			\toprule[1.0pt]
			
			Systems&$\left|S,S_z\right\rangle$ & Spin wave functions \\
			\hline
			\multirow{1}{*}{$\Sigma_c^{} \bar{D}^{}$($\Lambda_c^{} {D}^{}$)}&$\left|\frac{1}{2}, \frac{1}{2}\right\rangle$ & $\left|\frac{1}{2},\frac{1}{2}\right\rangle\left|0,0\right\rangle$ \\
			\cline{1-3}
			
			\multirow{2}{*}	{$\Sigma_c^{*} \bar{D}^{}$}&$\left|\frac{3}{2}, \frac{3}{2}\right\rangle$ & $\left|\frac{3}{2},\frac{3}{2}\right\rangle\left|0,0\right\rangle$ \\
			&$\left|\frac{3}{2}, \frac{1}{2}\right\rangle$ & $\left|\frac{3}{2},\frac{1}{2}\right\rangle\left|0,0\right\rangle$ \\
			\cline{1-3}

			\multirow{3}{*}{$\Sigma_c^{} \bar{D}^{*}            (\Lambda_c^{} {D}^{*})$}&$\left|\frac{1}{2}, \frac{1}{2}\right\rangle$ &$\sqrt{\frac{1}{3}}\left|\frac{1}{2},\frac{1}{2}\right\rangle\left|1,0\right\rangle-\sqrt{\frac{2}{3}}\left|\frac{1}{2},-\frac{1}{2}\right\rangle\left|1,1\right\rangle$ \\
			\cline{2-3}
			&$\left|\frac{3}{2}, \frac{3}{2}\right\rangle$ & $\left|\frac{1}{2},\frac{1}{2}\right\rangle\left|1,1\right\rangle$ \\
			&$\left|\frac{3}{2}, \frac{1}{2}\right\rangle$ & $\sqrt{\frac{2}{3}}\left|\frac{1}{2},\frac{1}{2}\right\rangle\left|1,0\right\rangle+\sqrt{\frac{1}{3}}\left|\frac{1}{2},-\frac{1}{2}\right\rangle\left|1,1\right\rangle$ \\
			\cline{1-3}

			\multirow{6}{*}{$\Sigma_c^{*} \bar{D}^{*}$}&$\left|\frac{1}{2}, \frac{1}{2}\right\rangle$ & $\sqrt{\frac{1}{2}}\left|\frac{3}{2},\frac{3}{2}\right\rangle\left|1,-1\right\rangle-\sqrt{\frac{1}{3}}\left|\frac{3}{2},\frac{1}{2}\right\rangle\left|1,0\right\rangle+\sqrt{\frac{1}{6}}\left|\frac{3}{2},-\frac{1}{2}\right\rangle\left|1,1\right\rangle$ \\
			\cline{2-3}

			&$\left|\frac{3}{2}, \frac{3}{2}\right\rangle$ & $\sqrt{\frac{3}{5}}\left|\frac{3}{2},\frac{3}{2}\right\rangle\left|1,0\right\rangle-\sqrt{\frac{2}{5}}\left|\frac{3}{2},\frac{1}{2}\right\rangle\left|1,1\right\rangle$ \\
			
			&$\left|\frac{3}{2}, \frac{1}{2}\right\rangle$ & $\sqrt{\frac{2}{5}}\left|\frac{3}{2},\frac{3}{2}\right\rangle\left|1,-1\right\rangle+\sqrt{\frac{1}{15}}\left|\frac{3}{2},\frac{1}{2}\right\rangle\left|1,0\right\rangle-\sqrt{\frac{8}{15}}\left|\frac{3}{2},-\frac{1}{2}\right\rangle\left|1,1\right\rangle$ \\
			\cline{2-3}
			&$\left|\frac{5}{2}, \frac{5}{2}\right\rangle$ & $\left|\frac{3}{2},\frac{3}{2}\right\rangle\left|1,1\right\rangle$ \\
			&$\left|\frac{5}{2}, \frac{3}{2}\right\rangle$ & $\sqrt{\frac{2}{5}}\left|\frac{3}{2},\frac{3}{2}\right\rangle\left|1,0\right\rangle+\sqrt{\frac{3}{5}}\left|\frac{3}{2},\frac{1}{2}\right\rangle\left|1,1\right\rangle$ \\		
			
			&$\left|\frac{5}{2}, \frac{1}{2}\right\rangle$ & $\sqrt{\frac{1}{10}}\left|\frac{3}{2},\frac{3}{2}\right\rangle\left|1,-1\right\rangle+\sqrt{\frac{3}{5}}\left|\frac{3}{2},\frac{1}{2}\right\rangle\left|1,0\right\rangle+\sqrt{\frac{3}{10}}\left|\frac{3}{2},-\frac{1}{2}\right\rangle\left|1,1\right\rangle$ \\
			
			\bottomrule[1.0pt]
			\bottomrule[1.0pt]
		\end{tabular}
	\end{table*}
	Next, we calculate the transition magnetic moments of $ S $-wave  hidden-charm pentaquark $P^{N^{0}}_{\psi}  $ states and $P^{\Delta^{0}}_{\psi}$  states (including $\Sigma_c^{(*)} \bar{D}^{(*)}$ and $\Lambda_c^{} {D}^{(*)}$ systems) in the molecular model. For the convenience of calculation,  Table \ref{tab:zx}  lists the possible spin wave functions of the $ S $-wave $\Sigma_c^{(*)} \bar{D}^{(*)}$ and $\Lambda_c^{} {D}^{(*)}$ systems. 
	We take $\mu_{{\Sigma_c{\bar{D^{*}}}} |^{2} S_{1 / 2}\rangle \to {\Sigma_c{\bar{D}}} |^{2} S_{1 / 2}\rangle} $ 
	as an example to illustrate the procedure of getting the magnetic moments between the $ S $-wave $\Sigma_c^{(*)} \bar{D}^{(*)}$ -type doubly charmed molecular pentaquark states.  According to  Table \ref{tab:k} and Table \ref{tab:zx}, the flavor-spin wave functions of the $	{\Sigma_c{\bar{D^{*}}}} |^{2} S_{1 / 2}\rangle  $ and ${\Sigma_c{\bar{D}}} |^{2} S_{1 / 2}\rangle  $ states with $ (I,I_{3})=(\frac{1}{2},-\frac{1}{2})$ can be constructed as
	\begin{eqnarray}
		\psi_{\Sigma_c{\bar{D^{*}}}} |^{2} S_{1 / 2}\rangle&=&\Bigg[\sqrt{\frac{1}{3}}\left|\Sigma_c^{+} {D}^{*-}\right\rangle-\sqrt{\frac{2}{3}}\left|\Sigma_c^{0}\bar{D}^{*0}\right\rangle\Bigg]\otimes\nonumber\\ 
		&&\Bigg[\sqrt{\frac{1}{3}}\left|\frac{1}{2},\frac{1}{2}\right\rangle\left|1,0\right\rangle-\sqrt{\frac{2}{3}}\left|\frac{1}{2},-\frac{1}{2}\right\rangle\left|1,1\right\rangle\Bigg],\nonumber\\
		\psi_{\Sigma_c{\bar{D}}} |^{2} S_{1 / 2}\rangle&=&\Bigg[\sqrt{\frac{1}{3}}\left|\Sigma_c^{+}D^{-}\right\rangle-\sqrt{\frac{2}{3}}\left|\Sigma_c^{0}\bar{D}^{0}\right\rangle\Bigg]\otimes\nonumber\\ 
		&&\left|\frac{1}{2},\frac{1}{2}\right\rangle\left|0,0\right\rangle.
	\end{eqnarray}
	The transition magnetic moment of the ${\Sigma_c{\bar{D^{*}}}} |^{2} S_{1 / 2}\rangle \to{\Sigma_c{\bar{D}}} |^{2} S_{1 / 2}\rangle_{\gamma}  $ process is
	\begin{eqnarray}
		&&\mu_{{\Sigma_c{\bar{D^{*}}}} |^{2} S_{1 / 2}\rangle \to {\Sigma_c{\bar{D}}} |^{2} S_{1 / 2}\rangle}\nonumber\\
		&&=\left\langle \psi_{\Sigma_{c}  \bar{D}|^{2} S_{1/2}\rangle} \left|\hat{\mu}_z\right| \psi_{\Sigma_{c} \bar D^{*}|^{2} S_{1/2}\rangle}\right\rangle\nonumber\\
		&&=\frac{\sqrt{3}}{9}\mu_{D^{*-} \to  D^{-}}+\frac{2\sqrt{3}}{9}\mu_{\bar{D}^{*0} \to \bar{D}^{0}}.
	\end{eqnarray}

	\renewcommand\tabcolsep{1.2cm}
	\renewcommand{\arraystretch}{1.7}
	\begin{table*}[!htbp]
		\caption{Transition magnetic moments between the $S$-wave  $\Sigma_c^{(*)} \bar{D}^{(*)}$-type hidden-charm molecular pentaquarks  $P^{N^{0}}_{\psi}  $ with $8_{1f}$ flavor representation. The unit is the nuclear magnetic moment $ \mu_{N} $.}
		\label{tab:mby}
		\begin{tabular}{l|c|c}
			\toprule[1.0pt]
			\toprule[1.0pt]

			Decay modes & Expressions & Results \\
			\hline
			$\Sigma_c^{} \bar{D}^{*}|\frac{1}{2}^-\rangle \to\Sigma_c^{} \bar{D}^{}|\frac{1}{2}^-\rangle\gamma$ & $\frac{\sqrt{3}}{9}(\mu_{D^{*-} \to  D^{-}}+2\mu_{\bar{D}^{*0} \to \bar{D}^{0}})$ & $-0.780$ \\

			$\Sigma_c^{} \bar{D}^{*}|\frac{3}{2}^-\rangle \to\Sigma_c^{} \bar{D}^{}|\frac{1}{2}^-\rangle\gamma$ & $\frac{\sqrt{6}}{9}(\mu_{D^{*-} \to  D^{-}}+2\mu_{\bar{D}^{*0} \to \bar{D}^{0}})$ & $-1.103$ \\
			$\Sigma_c^{} \bar{D}^{*}|\frac{3}{2}^-\rangle \to\Sigma_c^{} \bar{D}^{*}|\frac{1}{2}^-\rangle\gamma$ & $\frac{\sqrt{2}}{9}({2\mu}_{{\Sigma}_{c}^{+}}+4{\mu}_{{\Sigma}_{c}^{0}})-\frac{\sqrt{2}}{9}({\mu}_{{D}^{*-}}+2{\mu}_{\bar{D}^{*0}})$ & $-0.979$  \\
			$\Sigma_c^{*} \bar{D}^{*}|\frac{1}{2}^-\rangle \to\Sigma_c^{} \bar{D}^{*}|\frac{3}{2}^-\rangle\gamma$  & $-\frac{\sqrt{2}}{6}(\mu_{\Sigma_{c}^{*+} \to  \Sigma_{c}^{+}}+2\mu_{\Sigma_{c}^{*0} \to  \Sigma_{c}^{0}})$ & $0.585$  \\
			$\Sigma_c^{*} \bar{D}^{}|\frac{3}{2}^-\rangle \to\Sigma_c^{} \bar{D}^{}|\frac{1}{2}^-\rangle\gamma$ & $\frac{1}{3}(\mu_{\Sigma_{c}^{*+} \to  \Sigma_{c}^{+}}+2\mu_{\Sigma_{c}^{*0} \to  \Sigma_{c}^{0}})$ & $-0.827$  \\
			
			$\Sigma_c^{*} \bar{D}^{*}|\frac{1}{2}^-\rangle \to\Sigma_c^{*} \bar{D}^{}|\frac{3}{2}^-\rangle\gamma$ & $-\frac{\sqrt{3}}{9}(\mu_{D^{*-} \to  D^{-}}+2\mu_{\bar{D}^{*0} \to \bar{D}^{0}})$ & $0.780$ \\
			
			$\Sigma_c^{*} \bar{D}^{*}|\frac{3}{2}^-\rangle \to\Sigma_c^{} \bar{D}^{*}|\frac{1}{2}^-\rangle\gamma$ & $-\frac{\sqrt{5}}{15}(\mu_{\Sigma_{c}^{*+} \to  \Sigma_{c}^{+}}+2\mu_{\Sigma_{c}^{*0} \to  \Sigma_{c}^{0}})$ & $0.370$  \\
			$\Sigma_c^{*} \bar{D}^{*}|\frac{3}{2}^-\rangle \to\Sigma_c^{*} \bar{D}^{*}|\frac{1}{2}^-\rangle\gamma$ & $\frac{\sqrt{5}}{75}({2\mu}_{{\Sigma}_{c}^{*+}}+4{\mu}_{{\Sigma}_{c}^{*0}})-\frac{\sqrt{5}}{75}(3{\mu}_{{D}^{*-}}+6{\mu}_{\bar{D}^{*0}})$ & $-0.676$  \\
			$\Sigma_c^{*} \bar{D}^{*}|\frac{3}{2}^-\rangle \to\Sigma_c^{} \bar{D}^{*}|\frac{3}{2}^-\rangle\gamma$ & $-\frac{\sqrt{10}}{15}(\mu_{\Sigma_{c}^{*+} \to  \Sigma_{c}^{+}}+2\mu_{\Sigma_{c}^{*0} \to  \Sigma_{c}^{0}})$ & $0.523$ \\
			$\Sigma_c^{*} \bar{D}^{*}|\frac{3}{2}^-\rangle \to\Sigma_c^{*} \bar{D}^{}|\frac{3}{2}^-\rangle\gamma$ & $\frac{\sqrt{15}}{15}(\mu_{D^{*-} \to  D^{-}}+2\mu_{\bar{D}^{*0} \to \bar{D}^{0}})$ & $-1.046$ \\

			$\Sigma_c^{*} \bar{D}^{*}|\frac{5}{2}^-\rangle \to\Sigma_c^{} \bar{D}^{*}|\frac{1}{2}^-\rangle\gamma$ & $\frac{2\sqrt{5}}{15}(\mu_{\Sigma_{c}^{*+} \to  \Sigma_{c}^{+}}+2\mu_{\Sigma_{c}^{*0} \to  \Sigma_{c}^{0}})$ & $-0.740$  \\

			$\Sigma_c^{*} \bar{D}^{*}|\frac{5}{2}^-\rangle \to\Sigma_c^{*} \bar{D}^{}|\frac{3}{2}^-\rangle\gamma$ & $\frac{\sqrt{10}}{15}(\mu_{D^{*-} \to  D^{-}}+2\mu_{\bar{D}^{*0} \to \bar{D}^{0}})$ & $-0.854$ \\
			
			$\Sigma_c^{*} \bar{D}^{*}|\frac{5}{2}^-\rangle \to\Sigma_c^{} \bar{D}^{*}|\frac{3}{2}^-\rangle\gamma$& $\frac{\sqrt{15}}{15}(\mu_{\Sigma_{c}^{*+} \to  \Sigma_{c}^{+}}+2\mu_{\Sigma_{c}^{*0} \to  \Sigma_{c}^{0}})$ & $-0.641$  \\
			$\Sigma_c^{*} \bar{D}^{*}|\frac{5}{2}^-\rangle \to\Sigma_c^{*} \bar{D}^{*}|\frac{3}{2}^-\rangle\gamma$ &$	\frac{\sqrt{6}}{45}(2{\mu}_{{\Sigma}_{c}^{*+}}+4{\mu}_{{\Sigma}_{c}^{*0}})-\frac{\sqrt{6}}{45}(3{\mu}_{{D}^{*-}}+6{\mu}_{{D}^{*0}}) $ & $-0.444$  \\
			\bottomrule[1.0pt]
			\bottomrule[1.0pt]
		\end{tabular}
		
	\end{table*}
	
	\renewcommand\tabcolsep{0.24cm}
	\renewcommand{\arraystretch}{1.7}
	\begin{table}[!htbp]
		\caption{ Transition magnetic moments between the $S$-wave  $\Lambda_c^{} {D}^{(*)}$-type hidden-charm molecular pentaquarks  $P^{N^{0}}_{\psi}  $ with $8_{2f}$ flavor representation.  The unit is the nuclear magnetic moment $ \mu_{N} $.}
		\label{tab:mbey}
		\begin{tabular}{l|c|c}
			\toprule[1.0pt]
			\toprule[1.0pt]
			Decay modes & Expressions & Results \\
			\hline
			$\Lambda_c^{} {D}^{*}|\frac{1}{2}^-\rangle \to\Lambda_c^{} {D}^{}|\frac{1}{2}^-\rangle\gamma$ & $\frac{\sqrt{3}}{3}\mu_{D^{*-} \to  D^{-}}$ & $0.314$ \\
			$\Lambda_c^{} {D}^{*}|\frac{3}{2}^-\rangle \to\Lambda_c^{} {D}^{*}|\frac{1}{2}^-\rangle\gamma$ & $\frac{\sqrt{2}}{3}(2\mu_{\Lambda_c^{+}}-{\mu}_{{D}^{*-}})$ & $1.017$ \\
			$\Lambda_c^{} {D}^{*}|\frac{3}{2}^-\rangle \to\Lambda_c^{} {D}^{}|\frac{1}{2}^-\rangle\gamma$ & $\frac{\sqrt{6}}{3}\mu_{D^{*-} \to  D^{-}}$ & $0.444$ \\
			\bottomrule[1.0pt]
			\bottomrule[1.0pt]
		\end{tabular}
	\end{table}
	
	Similarly, we obtain the transition magnetic moments of the $ S $-wave hidden-charm  pentaquark states in the molecular model. The transition magnetic moments between  the $S$-wave  $\Sigma_c^{(*)} \bar{D}^{(*)}$-type hidden-charm molecular pentaquarks with isospin $(I,I_3) = (\frac{1}{2},-\frac{1}{2})$ are collected in Table \ref{tab:mby}. The transition magnetic moments between  the $S$-wave  $\Sigma_c^{(*)} \bar{D}^{(*)}$-type hidden-charm molecular pentaquarks with isospin $(I,I_3) = (\frac{3}{2},-\frac{1}{2})$ are collected in Table \ref{tab:msy}. The transition magnetic moments between  the $S$-wave  $\Lambda_c^{} {D}^{(*)}$-type hidden-charm molecular pentaquarks with isospin $(I,I_3) = (\frac{1}{2},-\frac{1}{2})$ are collected in Table \ref{tab:mbey}. In the $S$-wave  $\Sigma_c^{(*)} \bar{D}^{(*)}$-type hidden-charm molecular pentaquarks, 
	the transition magnetic moments of the ${\Sigma_c^{*}{\bar{D^{*}}}} |^{2} S_{1 / 2}\rangle \to{\Sigma_c{\bar{D^{*}}}} |^{2} S_{1 / 2}\rangle_{\gamma}  $ and ${\Sigma_c^{*}{\bar{D^{*}}}} |^{6} S_{5 / 2}\rangle \to{\Sigma_c^{*}{\bar{D^{*}}}} |^{2} S_{1 / 2}\rangle_{\gamma}  $ processes are zero.

The transition magnetic moments we obtain are only prediction results at the quark level. Whether they correspond to the hadron-level results needs further studies. In Ref. \cite{Jiang:2023zqq}, it is found that a scale factor probably exists for the coupling constants between the quark-level calculation and the hadron-level measurement by noticing that spin-different baryons are involved. Since the transition magnetic moment is also a coupling parameter in the hadron-level Lagrangian, we speculate that similar factors are also needed in linking the experiment data and our results if the proposal is correct. For the transition magnetic moment $|\frac32\rangle\to|\frac12\rangle\gamma$, the factor is $\sqrt{\frac{3}{2}}$, but such factors are currently unknown for other cases

		\renewcommand\tabcolsep{1.2cm}
		\renewcommand{\arraystretch}{1.7}
		\begin{table*}[!htbp]
			\caption{Transition magnetic moments between the $S$-wave  $\Sigma_c^{(*)} \bar{D}^{(*)}$-type hidden-charm molecular pentaquarks    $P^{\Delta^{0}}_{\psi}$ with $10_{f}$ flavor representation. The unit is the nuclear magnetic moment $ \mu_{N} $.}
			\label{PTM}
			
			\label{tab:msy}
			\begin{tabular}{l|c|c}
				\toprule[1.0pt]
				\toprule[1.0pt]
				Decay modes & Expressions & Results \\
				\hline
				$\Sigma_c^{} \bar{D}^{*}|\frac{1}{2}^-\rangle \to\Sigma_c^{} \bar{D}^{}|\frac{1}{2}^-\rangle\gamma$ & $\frac{\sqrt{3}}{9}(2\mu_{D^{*-} \to  D^{-}}+\mu_{\bar{D}^{*0} \to \bar{D}^{0}})$ & $-0.233$ \\

				$\Sigma_c^{} \bar{D}^{*}|\frac{3}{2}^-\rangle \to\Sigma_c^{} \bar{D}^{}|\frac{1}{2}^-\rangle\gamma$ & $\frac{\sqrt{6}}{9}(2\mu_{D^{*-} \to  D^{-}}+\mu_{\bar{D}^{*0} \to \bar{D}^{0}})$ & $-0.329$ \\
				$\Sigma_c^{} \bar{D}^{*}|\frac{3}{2}^-\rangle \to\Sigma_c^{} \bar{D}^{*}|\frac{1}{2}^-\rangle\gamma$ & $\frac{\sqrt{2}}{9}(4{\mu}_{{\Sigma}_{c}^{+}}+2{\mu}_{{\Sigma}_{c}^{0}})-\frac{\sqrt{2}}{9}(2{\mu}_{{D}^{*-}}+{\mu}_{\bar{D}^{*0}})$ & $0.063$  \\
				$\Sigma_c^{*} \bar{D}^{*}|\frac{1}{2}^-\rangle \to\Sigma_c^{} \bar{D}^{*}|\frac{3}{2}^-\rangle\gamma$  & $-\frac{\sqrt{2}}{6}(2\mu_{\Sigma_{c}^{*+} \to  \Sigma_{c}^{+}}+\mu_{\Sigma_{c}^{*0} \to  \Sigma_{c}^{0}})$ & $0.269$  \\
				$\Sigma_c^{*} \bar{D}^{}|\frac{3}{2}^-\rangle \to\Sigma_c^{} \bar{D}^{}|\frac{1}{2}^-\rangle\gamma$  & $\frac{1}{3}(2\mu_{\Sigma_{c}^{*+} \to  \Sigma_{c}^{+}}+\mu_{\Sigma_{c}^{*0} \to  \Sigma_{c}^{0}})$ & $-0.380$  \\
				
				$\Sigma_c^{*} \bar{D}^{*}|\frac{1}{2}^-\rangle \to\Sigma_c^{*} \bar{D}^{}|\frac{3}{2}^-\rangle\gamma$ & $-\frac{\sqrt{3}}{9}(2\mu_{D^{*-} \to  D^{-}}+\mu_{\bar{D}^{*0} \to \bar{D}^{0}})$ & $0.233$ \\
				
				$\Sigma_c^{*} \bar{D}^{*}|\frac{3}{2}^-\rangle \to\Sigma_c^{} \bar{D}^{*}|\frac{1}{2}^-\rangle\gamma$ & $-\frac{\sqrt{5}}{15}(\mu_{\Sigma_{c}^{*+} \to  \Sigma_{c}^{+}}+\mu_{\Sigma_{c}^{*0} \to  \Sigma_{c}^{0}})$ & $0.170$  \\
				$\Sigma_c^{*} \bar{D}^{*}|\frac{3}{2}^-\rangle \to\Sigma_c^{*} \bar{D}^{*}|\frac{1}{2}^-\rangle\gamma$ & $\frac{\sqrt{5}}{75}(4{\mu}_{{\Sigma}_{c}^{*+}}+2{\mu}_{{\Sigma}_{c}^{*0}})-\frac{\sqrt{5}}{75}(6{\mu}_{{D}^{*-}}-3{\mu}_{\bar{D}^{*0}})$ & $0.501$  \\
				$\Sigma_c^{*} \bar{D}^{*}|\frac{3}{2}^-\rangle \to\Sigma_c^{} \bar{D}^{*}|\frac{3}{2}^-\rangle\gamma$& $-\frac{\sqrt{10}}{15}(2\mu_{\Sigma_{c}^{*+} \to  \Sigma_{c}^{+}}+\mu_{\Sigma_{c}^{*0} \to  \Sigma_{c}^{0}})$ & $0.241$ \\
				$\Sigma_c^{*} \bar{D}^{*}|\frac{3}{2}^-\rangle \to\Sigma_c^{*} \bar{D}^{}|\frac{3}{2}^-\rangle\gamma$ & $\frac{\sqrt{15}}{15}(2\mu_{D^{*-} \to  D^{-}}+\mu_{\bar{D}^{*0} \to \bar{D}^{0}})$ & $-0.313$ \\

				$\Sigma_c^{*} \bar{D}^{*}|\frac{5}{2}^-\rangle \to\Sigma_c^{} \bar{D}^{*}|\frac{1}{2}^-\rangle\gamma$& $\frac{2\sqrt{5}}{15}(2\mu_{\Sigma_{c}^{*+} \to  \Sigma_{c}^{+}}+\mu_{\Sigma_{c}^{*0} \to  \Sigma_{c}^{0}})$ & $-0.340$  \\

				$\Sigma_c^{*} \bar{D}^{*}|\frac{5}{2}^-\rangle \to\Sigma_c^{*} \bar{D}^{}|\frac{3}{2}^-\rangle\gamma$ & $\frac{\sqrt{10}}{15}(2\mu_{D^{*-} \to  D^{-}}+\mu_{\bar{D}^{*0} \to \bar{D}^{0}})$ & $-0.255$ \\
				$\Sigma_c^{*} \bar{D}^{*}|\frac{5}{2}^-\rangle \to\Sigma_c^{} \bar{D}^{*}|\frac{3}{2}^-\rangle\gamma$ & $\frac{\sqrt{15}}{15}(2\mu_{\Sigma_{c}^{*+} \to  \Sigma_{c}^{+}}+\mu_{\Sigma_{c}^{*0} \to  \Sigma_{c}^{0}})$ & $-0.295$  \\
				$\Sigma_c^{*} \bar{D}^{*}|\frac{5}{2}^-\rangle \to\Sigma_c^{*} \bar{D}^{*}|\frac{3}{2}^-\rangle\gamma$ &$	\frac{\sqrt{6}}{45}(4{\mu}_{{\Sigma}_{c}^{*+}}+2{\mu}_{{\Sigma}_{c}^{*0}})-\frac{\sqrt{6}}{45}(6{\mu}_{{D}^{*-}}+3{\mu}_{{D}^{*0}}) $& $0.329$  \\
				\bottomrule[1.0pt]
				\bottomrule[1.0pt]
			\end{tabular}
			
		\end{table*}

	\section{ magnetic moments of the  $P^{N^{0}}_{\psi}  $ states and $P^{\Delta^{0}}_{\psi}$  states in  the diquark- diquark-antiquark model }
	\label{sec4}
	The magnetic moment formula of hidden-charm pentaquark states $ (cq_{3})(q_{1}q_{2})\bar{c} $ in the diquark-diquark-antiquark model  is
\begin{align}
		\mu =&\left\langle \psi\left |g_{cq_{3}}{\mu_{cq_{3}}}{\vec{S}_{cq_{3}}}+g_{q_{1}q_{2}}{\mu_{q_{1}q_{2}}}{\vec{S}}_{q_{1}q_{2}}+g_{\bar{c}}{\mu_{\bar{c}}}{\vec{S}_{\bar{c}}}+{\mu_l}\vec{l}\,\,\right|\psi\right\rangle \nonumber\\
		=& \sum_{S_z,l_z}\ \langle\ SS_z,ll_z|JJ_z\ \rangle^{2}\bigg \{\mu_{l} l_z+  \sum_{S'_{1}}\ \big\langle\ S_{1} S'_{1},S_{\bar{c}}(S_{z}
		\nonumber\\    
		& -S'_{1}) |SS_z\ \big\rangle^{2}\bigg [2(S_{z}-S'_{1})\mu_{\bar{c}} +\sum_{S'_{cq_{3}}}\ \big\langle\ S_{cq_{3}} S'_{cq_{3}},S_{q_{1}q_{2}}(S'_{1} \nonumber\\
		&-S'{cq_{3}}) |S_{1}S'_{1}\ \big\rangle^{2} \big (S'_{cq_{3}}(\mu_{c}+\mu_{q_{3}})+(S'_{1}-S'_{cq_{3}})(\mu_{q_1}\nonumber\\
&+\mu_{q_2})\big )	\Big ] \bigg \},
	\end{align}
where the subscripts $ cq_{3} $ and $ q_{1}q_{2} $ represent   the diquark $ (cq_{3}) $  and $( q_{1}q_{2}) $, respectively. $S_{cq_{3}}  $ and $ S_{q_{1}q_{2}} $ couple into the spin $ S_{1} $, and the spin coupling of $ S_{1} $  and $S_{\bar{c}}  $ forms the total spin $ S $ of the pentaquark state.
	$ l $ is the orbital excitation.  $ S'_{1} $, $ S_{z} $ and $S'_{cq_{3}}  $ are spin third component of  $(cq_{3})(q_{1}q_{2})  $,  pentaquark state and  diquark $( cq_{3}) $. $ \psi $ is the flavor wave function in  diquark-diquark-antiquark model in Table \ref{tab:k}. 
	
\begin{figure}
	\centering
	\includegraphics[width=1\linewidth, height=0.28\textheight]{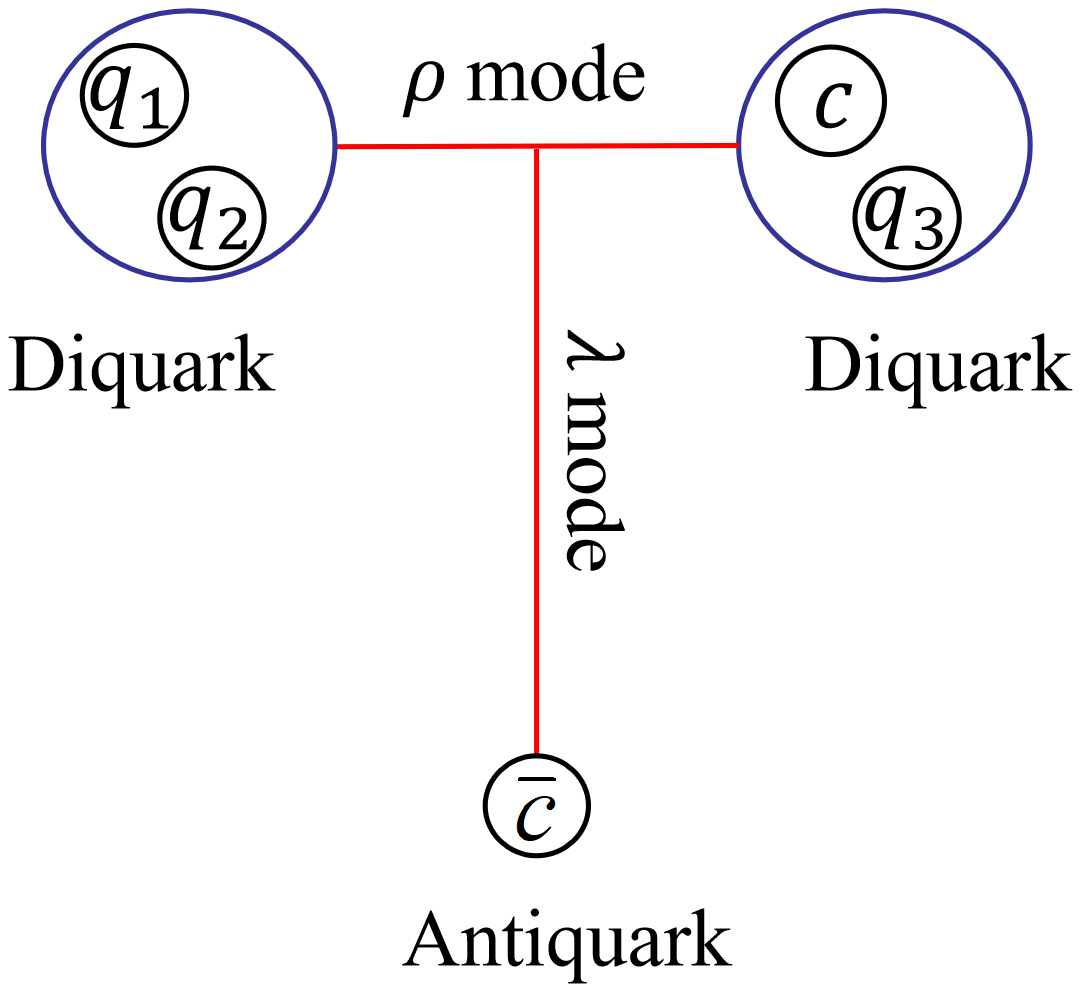}
	\caption{Two excitation modes of $ P $-wave in diquark-diquark-antiquark model.}
	\label{fig:x}
\end{figure}

	In the  diquark-diquark-antiquark model,  $ P $-wave orbital excitation includes $ \rho $ mode and $ \lambda $ mode. The $ P $-wave orbital excitation of the $ \rho $ mode lies between diquark $( cq_{3} )$ and diquark $( q_{1}q_{2}) $. The $ P $-wave orbital excitation of the $ \lambda $  mode lies between the $ \bar{c} $ and  the center of mass system of diquark  $( cq_{3}) $  and diquark  $( q_{1}q_{2}) $. We show these two excitation modes in the  Fig. \ref{fig:x}. The orbital magnetic moment $ \mu_{l} $ in $ \rho $  mode and $ \lambda $  mode is
	\begin{eqnarray} 
		\rho ~mode:~~~~~~	{\mu_l}&=&\frac{{m_{q_{1}q_{2}}}\mu_{cq_{3}}+{m_{cq_{3}}}\mu_{q_{1}q_{2}}}{m_{cq_{3}}+m_{q_{1}q_{2}}},\\
		\lambda ~mode:~~~~~~	{\mu_l}&=&\frac{{m_{\bar{c}}}\mu_{(cq_{3})(q_{1}q_{2})}+{m_{(cq_{3})(q_{1}q_{2})}}\mu_{\bar{c}}}{m_{(cq_{3})(q_{1}q_{2})}+m_{\bar{c}}},~~~~\\
		{m_{(cq_{3})(q_{1}q_{2})}}&=&\frac{{m_{q_{1}q_{2}}}{m_{cq_{3}}}}{m_{cq_{3}}+m_{q_{1}q_{2}}},
   	\end{eqnarray}
	where $ m $ and $ \mu $ represent the mass and magnetic moment of clusters represented by their subscripts. For example, $ m_{q_{1}q_{2}} $ represents the mass of diquark $ q_{1}q_{2} $.
	The mass of diquarks in diquark-diquark-antiquark model are \cite{Ebert:2010af}:
	\begin{align}
		&m_{[q,q]}= 0.710~\mbox{GeV}, ~~~~~~~m_{ \{q,q\}} =0.909~\mbox{GeV}, \nonumber\\
		&m_{[c,q]}= 1.937~\mbox{GeV}, ~~~~~~~m_{\{c,q\}} =2.036~\mbox{GeV}. \label{equ:op}
	\end{align}

The magnetic moments  of the $ S $-wave $P^{N^{0}}_{\psi} $ states and $P^{\Delta^{0}}_{\psi} $ states are presented in Table  \ref{tab:rb}. The magnetic moments  of the $ P $-wave excitation  $P^{N^{0}}_{\psi} $ states and $P^{\Delta^{0}}_{\psi} $ states in the $ \lambda $ mode and  $ \rho $ mode are presented in Table \ref{tab:lb} -Table \ref{tab:ls}.

	\renewcommand\tabcolsep{1cm}
	\renewcommand{\arraystretch}{1.7}
	\begin{table*}[!htbp]
		\caption{The  magnetic moments of the $ S $-wave $P^{N^{0}}_{\psi}  $ states    in the diquark-diquark-antiquark model with the $ 8_{2f} $ and $ 8_{1f} $ flavor representation, and the  magnetic moments of the $ S $-wave $P^{\Delta^{0}}_{\psi}  $ states in the diquark-diquark-antiquark model with the $ 10_{f} $ flavor representation.  The $J_{1}^{P_{1}}\otimes J_{2}^{P_{2}}\otimes J_{3}^{P_{3}}\otimes J_{4}^{P_{4}}$ are corresponding to the angular momentum and parity of $(cq_1)$, $(q_2q_3)$, $\bar{c}$ and orbital, respectively. The unit is the nuclear magnetic moment $ \mu_{N} $.}
		\scriptsize
		\label{tab:rb}
		\begin{tabular}{c|c|c|c|c}
			\toprule[1pt]
			\toprule[1pt]
			\multicolumn{5}{c}{ $ (cd)[ud]{\bar c} $}\\
			\toprule[1pt]
			$J^P$	& $^{2s+1}L_J$ &$J_{1}^{P_{1}}\otimes J_{2}^{P_{2}}\otimes J_{3}^{P_{3}}\otimes J_{4}^{P_{4}}$     &Expressions&Results \\
			\hline
			{$\frac{1}{2}^{-}$}	&{{${^{2}S_{\frac{1}{2}}}$}}  &$0^{+}\otimes0^{+}\otimes{\frac{1}{2}}^{-}\otimes0^{+}$   &$-\mu _{c}$&$-0.403$ \\

			&& $1^{+}\otimes0^{+}\otimes{\frac{1}{2}}^{-}\otimes0^{+}$ & $\frac{1}{3} \left(3 \mu _{c}+2 \mu _{d}\right)$  &$-0.228$ \\
			\hline
			{$\frac{3}{2}^{-}$}  &{{${^{4}S_{\frac{3}{2}}}$}} & $1^{+}\otimes0^{+}\otimes{\frac{1}{2}}^{-}\otimes0^{+}$ &$\mu_{d} $ &$-0.947$ \\
			
			\toprule[1pt]
			\multicolumn{5}{c}{$\sqrt{\frac{1}{3}}({c}d)\{ud\}{\bar c}-\sqrt{\frac{2}{3}}({c}u)\{dd\}{\bar c}$}\\
			\toprule[1pt]
			$J^P$	& $^{2s+1}L_J$ &$J_{1}^{P_{1}}\otimes J_{2}^{P_{2}}\otimes J_{3}^{P_{3}}\otimes J_{4}^{P_{4}}$     &Expressions&Results \\
			\hline
			{$\frac{1}{2}^{-}$}	&{{${^{2}S_{\frac{1}{2}}}$}}  &$0^{+}\otimes1^{+}\otimes{\frac{1}{2}}^{-}\otimes0^{+}$   &$\frac{1}{3} \left(2 \mu _{u}+10\mu _{d}+3\mu _{c}\right)$  &$-1.492$ \\

			&&$(1^{+}\otimes1^{+})_{0} \otimes {\frac{1}{2}}^{-}\otimes0^{+}$ & $-\mu _{c}$ &$-0.403$ \\
			
			&&$(1^{+}\otimes1^{+})_{1} \otimes {\frac{1}{2}}^{-}\otimes0^{+}$ & $\frac{1}{9} \left(3\mu _{u}+6 \mu _{d}+6\mu _{c}\right)$ &$0.269$ \\
			\hline
			
			{$\frac{3}{2}^{-}$}  &{{${^{4}S_{\frac{3}{2}}}$}} & $(0^{+}\otimes1^{+}) \otimes {\frac{1}{2}}^{-}\otimes0^{+}$  &$\frac{1}{3} \left(\mu _{u}+5\mu _{d}-3\mu _{c}\right)$&$-1.351$ \\
			&& $(1^{+}\otimes1^{+})_{1} \otimes {\frac{1}{2}}^{-}\otimes0^{+}$ & $\frac{1}{2} \left(\mu _{u}+2\mu _{d}-\mu _{c}\right)$ &$-0.202$  \\
			&& $(1^{+}\otimes1^{+})_{2} \otimes {\frac{1}{2}}^{-}\otimes0^{+}$ & $\frac{3}{10} \left(3\mu _{u}+6\mu _{d}+5\mu _{c}\right)$  &$0.605$ \\
			\hline
			{$\frac{5}{2}^{-}$}  &{{${^{6}S_{\frac{5}{2}}}$}}  &$1^{+}\otimes1^{+}\otimes{\frac{1}{2}}^{-}\otimes0^{+}$ &$\mu _{u}+2\mu _{d}$  &$0$ \\
			
			\toprule[1pt]
			\multicolumn{5}{c}{$\sqrt{\frac{2}{3}}({c}d)\{ud\}{\bar c}+\sqrt{\frac{1}{3}}({c}u)\{dd\}{\bar c}$}\\
			\toprule[1pt]
			$J^P$	& $^{2s+1}L_J$ &$J_{1}^{P_{1}}\otimes J_{2}^{P_{2}}\otimes J_{3}^{P_{3}}\otimes J_{4}^{P_{4}}$    & Expressions &Results   \\
			\hline
			{$\frac{1}{2}^{-}$}	&{{${^{2}S_{\frac{1}{2}}}$}}  &$0^{+}\otimes1^{+}\otimes{\frac{1}{2}}^{-}\otimes0^{+}$   &$\frac{1}{3} \left(4 \mu _{u}+8\mu _{d}+3\mu _{c}\right)$  &$0.403$ \\

			&&$(1^{+}\otimes1^{+})_{0} \otimes {\frac{1}{2}}^{-}\otimes0^{+}$ & $-\mu _{c}$&$-0.403$ \\
			
			&&$(1^{+}\otimes1^{+})_{1} \otimes {\frac{1}{2}}^{-}\otimes0^{+}$ & $\frac{1}{9} \left(3\mu _{u}+6 \mu _{d}+6\mu _{c}\right)$ &$0.269$ \\
			\hline
			
			{$\frac{3}{2}^{-}$}  &{{${^{4}S_{\frac{3}{2}}}$}} & $(0^{+}\otimes1^{+}) \otimes {\frac{1}{2}}^{-}\otimes0^{+}$  &$\frac{1}{3} \left(2\mu _{u}+4\mu _{d}-3\mu _{c}\right)$ & $-0.403$ \\
			&& $(1^{+}\otimes1^{+})_{1} \otimes {\frac{1}{2}}^{-}\otimes0^{+}$ & $\frac{1}{2} \left(\mu _{u}+2\mu _{d}-\mu _{c}\right)$ &$-0.202$  \\
			&& $(1^{+}\otimes1^{+})_{2} \otimes {\frac{1}{2}}^{-}\otimes0^{+}$ & $\frac{3}{10} \left(3\mu _{u}+6\mu _{d}+5\mu _{c}\right)$ &$0.605$ \\
			\hline
			{$\frac{5}{2}^{-}$}  &{{${^{6}S_{\frac{5}{2}}}$}}  &$1^{+}\otimes1^{+}\otimes{\frac{1}{2}}^{-}\otimes0^{+}$&$\mu _{u}+2\mu _{d}$ &$0$ \\
			
			\bottomrule[1pt]
			\bottomrule[1pt]
		\end{tabular}
		
	\end{table*}

	\renewcommand\tabcolsep{0.85cm}
	\renewcommand{\arraystretch}{1.65}
	\begin{table*}
		\caption{The  excited state magnetic moments of the $P^{N^{0}}_{\psi}  $ states in the diquark-diquark-antiquark model with the $ 8_{2f} $  flavor representation.  In the $\lambda$ mode $ P $-wave excitation, $ \mu _{l(S_{cd}\otimes S_{ud}/S_{\bar{c}})} $ represents the orbital magnetic moment between   $ \bar{c} $ and center of mass system of diquark $ (cd) $ and diquark $ (ud) $, $S_{cd}  $, $ S_{ud} $ and $ S_{\bar{c}} $ represent the spins of diquark $ (cd) $, diquark $ (ud) $ and $ \bar{c} $, respectively.  In the $\rho  $ mode $ P $-wave excitation, $ \mu _{l(S_{cd}/ S_{ud})} $ represents the orbital magnetic moment between   diquark $ (cd) $   and diquark $ (ud) $,  $S_{cd}  $ and $ S_{ud} $  represent the spins of diquark $ (cd) $ and diquark $ (ud) $, respectively.
		 The $J_{1}^{P_{1}}\otimes J_{2}^{P_{2}}\otimes J_{3}^{P_{3}}\otimes J_{4}^{P_{4}}$ are corresponding to the angular momentum and parity of $(cq_1)$, $(q_2q_3)$, $\bar{c}$ and orbital, respectively. The unit is the nuclear magnetic moment $ \mu_{N} $.}
		\scriptsize
		\label{tab:lb}
		\begin{tabular}{c|c|c|c|c}
			\toprule[1pt]
			\toprule[1pt]
			\multicolumn{5}{c}{ $ (cd)[ud]{\bar c} $}\\
			\toprule[1pt]
			$J^P$	& $^{2s+1}L_J$ &$J_{1}^{P_{1}}\otimes J_{2}^{P_{2}}\otimes J_{3}^{P_{3}}\otimes J_{4}^{P_{4}}$   &Expressions  & Results\\
			\hline
			
			{$\frac{1}{2}^{+}$}	&{{${^{2}P_{\frac{1}{2}}}$}}  &\multirow{2}{*}{$0^{+}\otimes0^{+}\otimes{\frac{1}{2}}^{-}\otimes1^{-}$ } &$\frac{1}{3} \left(\mu _{c}+2 \mu _{l(0\otimes0/\frac{1}{2})}\right)$   &$0.669_{(\lambda)}$  \\
			&&&$\frac{1}{3} \left(\mu _{c}+2 \mu _{l(0/0)}\right)$  &$0.378_{(\rho)}$\\
			\cline{3-5}

			&&\multirow{2}{*}{$(1^{+}\otimes0^{+}\otimes{\frac{1}{2}}^{-})_{\frac{1}{2}}\otimes1^{-}$ }  & $\frac{1}{9} \left(-3 \mu_{c}-2 \mu _{d}+6 \mu _{l(1\otimes0/\frac{1}{2})}\right)$   &$0.599_{(\lambda)}$  \\
			&& & $\frac{1}{9} \left(-3 \mu_{c}-2 \mu _{d}+6 \mu _{l(1/0)}\right)$ &$0.320_{(\rho)}$\\
			
			\cline{2-5}
			&{{${^{4}P_{\frac{1}{2}}}$}}  &\multirow{2}{*}{ $(1^{+}\otimes0^{+}\otimes{\frac{1}{2}}^{-})_{\frac{3}{2}}\otimes1^{-}$} &  $\frac{1}{9} \left(5 \mu _{d}-3 \mu_{l(1\otimes0/\frac{1}{2})}\right)$ &$ -0.788_{(\lambda)}$ \\
			&&&  $\frac{1}{9} \left(5 \mu _{d}-3 \mu_{l(1/0}\right)$&$-0.648_{(\rho)}$\\
			\hline
			
			{$\frac{3}{2}^{+}$}&{{${^{2}P_{\frac{3}{2}}}$}}  & \multirow{2}{*}{$0^{+}\otimes0^{+}\otimes{\frac{1}{2}}^{-}\otimes1^{-}$} &  $ - \mu _{c}+\mu_{l(0\otimes0/\frac{1}{2})}$ &$0.399_{(\lambda)}$\\
			&&&  $ - \mu _{c}+\mu_{l(0/0)}$&$-0.038_{(\rho)}$\\
			\cline{3-5}
			&&\multirow{2}{*}{$(1^{+}\otimes0^{+}\otimes{\frac{1}{2}}^{-})_{\frac{1}{2}}\otimes1^{-}$ } & $\frac{1}{3} \left(3 \mu _{c}+2 \mu _{d}+3 \mu_{l(1\otimes0/\frac{1}{2})}\right)$    & $0.556_{(\lambda)}$ \\
			&&& $\frac{1}{3} \left(3 \mu _{c}+2 \mu _{d}+3 \mu_{l(1/0)}\right)$  &$0.138_{(\rho)}$ \\
			\cline{2-5}
			&{{${^{4}P_{\frac{3}{2}}}$}}  &\multirow{2}{*}{$(1^{+}\otimes0^{+}\otimes{\frac{1}{2}}^{-})_{\frac{3}{2}}\otimes1^{-}$} &$\frac{1}{15} \left(11 \mu _{d}+6 \mu_{l(1\otimes0/\frac{1}{2})}\right)$   & $-0.381_{(\lambda)}$ \\
			&&&$\frac{1}{15} \left(11 \mu _{d}+6 \mu_{l(1/0)}\right)$  &$-0.548_{(\rho)}$\\
			
			\hline
			{$\frac{5}{2}^{+}$}	&{{${^{4}P_{\frac{5}{2}}}$}}   &\multirow{2}{*}{$1^{+}\otimes0^{+}\otimes{\frac{1}{2}}^{-}\otimes1^{-}$}  &$  \mu _{d}+\mu_{l(1\otimes0/\frac{1}{2})}$    &$ -0.163_{(\lambda)}$ \\
			&&&$  \mu _{d}+\mu_{l(1/0)}$  &$-0.581_{(\rho)}$ \\

			\bottomrule[1pt]
			\bottomrule[1pt]
		\end{tabular}
		
	\end{table*}

	\renewcommand\tabcolsep{0.485cm}
	\renewcommand{\arraystretch}{1.6}
	\begin{table*}
		\caption{The  excited state magnetic moments of the $P^{N^{0}}_{\psi}  $ states in the diquark-diquark-antiquark model with the $ 8_{1f} $  flavor representation.  In the $\lambda$ mode $ P $-wave excitation, $ \mu _{l(S_{cd}\otimes S_{ud}/S_{\bar{c}})} $ represents the orbital magnetic moment between   $ \bar{c} $ and center of mass system of diquark $ (cd) $ and diquark $ (ud) $, $S_{cd}  $, $ S_{ud} $ and $ S_{\bar{c}} $ represent the spins of diquark $ (cd) $, diquark $ (ud) $ and $ \bar{c} $, respectively. $ \mu _{l'(S_{cu}\otimes S_{dd}/S_{\bar{c}})} $ represents the orbital magnetic moment between   $ \bar{c} $ and center of mass system of diquark $ (cu) $ and diquark $ (dd) $, and $S_{cu}  $, $ S_{dd} $ and $ S_{\bar{c}} $ represent the spins of diquark $ (cu) $, diquark $ (dd) $ and $ \bar{c} $, respectively.  In the $\rho  $ mode $ P $-wave excitation, $ \mu _{l(S_{cd}/ S_{ud})} $ represents the orbital magnetic moment between   diquark $ (cd) $   and diquark $ (ud) $,  $S_{cd}  $ and $ S_{ud} $  represent the spins of diquark $ (cd) $ and diquark $ (ud) $, respectively.
			$ \mu _{l'(S_{cu}/ S_{dd})} $ represents the orbital magnetic moment between  diquark $ (cu) $ and diquark $ (dd) $,  $S_{cu}  $ and $ S_{dd} $  represent the spins of diquark $ (cu) $, diquark $ (dd) $, respectively. The $J_{1}^{P_{1}}\otimes J_{2}^{P_{2}}\otimes J_{3}^{P_{3}}\otimes J_{4}^{P_{4}}$ are corresponding to the angular momentum and parity of $(cq_1)$, $(q_2q_3)$, $\bar{c}$ and orbital, respectively. The unit is the nuclear magnetic moment $ \mu_{N} $. }
		\scriptsize
		\label{tab:lbb}
		\begin{tabular}{c|c|c|c|c}
			\toprule[1pt]
			\toprule[1pt]
			\multicolumn{5}{c}{$\sqrt{\frac{1}{3}}({c}d)\{ud\}{\bar c}-\sqrt{\frac{2}{3}}({c}u)\{dd\}{\bar c}$}\\
			\toprule[1pt]
			$J^P$	& $^{2s+1}L_J$ &$J_{1}^{P_{1}}\otimes J_{2}^{P_{2}}\otimes J_{3}^{P_{3}}\otimes J_{4}^{P_{4}}$   &Expressions  & Results\\
		\hline
			{$\frac{1}{2}^{+}$}	&{{${^{2}P_{\frac{1}{2}}}$}}  &\multirow{2}{*}{$(0^{+}\otimes1^{+}\otimes {\frac{1}{2}}^{-})_{\frac{1}{2}}\otimes1^{-}$   }&$\frac{1}{27} \left(-2\mu _{u}-10\mu _{d}-3\mu _{c}+6\mu _{l(0\otimes1/\frac{1}{2})}+12\mu _{l'(0\otimes1/\frac{1}{2})}\right)$  &$0.570_{(\lambda)}$  \\
			&&&$\frac{1}{27} \left(-2\mu _{u}-10\mu _{d}-3\mu _{c}+6\mu _{l(0/1)}+12\mu _{l'(0/1}\right)$ &$0.112_{(\rho)}$  \\
			\cline{3-5}
			
			&& \multirow{2}{*}{$((1^{+}\otimes1^{+})_{0}\otimes{\frac{1}{2}}^{-})_{\frac{1}{2}}\otimes1^{-}$ }& $\frac{1}{9} \left(3\mu _{c}+2\mu _{l(1\otimes1/\frac{1}{2})}+4\mu _{l'(1\otimes1/\frac{1}{2})}\right)$  &$0.529_{(\lambda)}$  \\
			&&& $\frac{1}{9} \left(3\mu _{c}+2\mu _{l(1/1)}+4\mu _{l'(1/1)}\right)$ &$0.071_{(\rho)}$  \\
			\cline{3-5}
			
			&&\multirow{2}{*}{$((1^{+}\otimes1^{+})_{1}\otimes{\frac{1}{2}}^{-})_{\frac{1}{2}}\otimes1^{-}$ }&$\frac{1}{27} \left(-3\mu _{u}-6\mu _{d}-6\mu _{c}+6\mu _{l(1\otimes1/\frac{1}{2})}+12\mu _{l'(1\otimes1/\frac{1}{2})}\right)$ & $0.305_{(\lambda)}$ \\
			&&&$\frac{1}{27} \left(-3\mu _{u}-6\mu _{d}-6\mu _{c}+6\mu _{l(1/1)}+12\mu _{l'(1/1)}\right)$&$-0.153_{(\rho)}$ \\
			\cline{2-5}
			&{{${^{4}P_{\frac{1}{2}}}$}}&\multirow{2}{*}{ $(0^{+}\otimes1^{+}\otimes {\frac{1}{2}}^{-})_{\frac{3}{2}}\otimes1^{-}$} & $\frac{1}{27} \left(5\mu _{u}+25\mu _{d}-15\mu _{c}-3\mu _{l(0\otimes1/\frac{1}{2})}-6\mu _{l'(0\otimes1/\frac{1}{2})}\right)$   &$-0.953_{(\lambda)}$  \\
			&& & $\frac{1}{27} \left(5\mu _{u}+25\mu _{d}-15\mu _{c}-3\mu _{l(0/1)}-6\mu _{l'(0/1)}\right)$  &$-0.724_{(\rho)}$  \\
			\cline{3-5}
			&&\multirow{2}{*}{$((1^{+}\otimes1^{+})_{1}\otimes{\frac{1}{2}}^{-})_{\frac{3}{2}}\otimes1^{-}$} &$\frac{1}{54} \left(15\mu _{u}+30\mu _{d}-15\mu _{c}-6\mu _{l(1\otimes1/\frac{1}{2})}-12\mu _{l'(1\otimes1/\frac{1}{2})}\right)$ &$ -0.309_{(\lambda)}$ \\
			&&&$\frac{1}{54} \left(15\mu _{u}+30\mu _{d}-15\mu _{c}-6\mu _{l(1/1)}-12\mu _{l'(1/1)}\right)$ &$-0.080_{(\rho)}$ \\
			\cline{3-5}
			&  &\multirow{2}{*}{ $((1^{+}\otimes1^{+})_{2}\otimes{\frac{1}{2}}^{-})_{\frac{3}{2}}\otimes1^{-}$} &$\frac{1}{18} \left(15\mu _{u}+18\mu _{d}+15\mu _{c}-2\mu _{l(1\otimes1/\frac{1}{2})}-4\mu _{l'(1\otimes1/\frac{1}{2})}\right)$ &$ 0.140_{(\lambda)}$ 
			\\
			&&·&$\frac{1}{18} \left(15\mu _{u}+18\mu _{d}+15\mu _{c}-2\mu _{l(1/1)}-4\mu _{l'(1/1)}\right)$&$0.224_{(\rho)}$ \\
			\hline
			$ \frac{3}{2}^{+} $&{{${^{2}P_{\frac{3}{2}}}$}}  &\multirow{2}{*}{ $(0^{+}\otimes1^{+}\otimes {\frac{1}{2}}^{-})_{\frac{1}{2}}\otimes1^{-}$} &$\frac{1}{9} \left(2\mu _{u}+10\mu _{d}+3\mu _{c}+3\mu _{l(0\otimes1/\frac{1}{2})}+6\mu _{l'(0\otimes1/\frac{1}{2})}\right)$ &$ 0.110_{(\lambda)}$ \\
			&&	&$\frac{1}{9} \left(2\mu _{u}+10\mu _{d}+3\mu _{c}+3\mu _{l(0/1)}+6\mu _{l'(0/1)}\right)$&$-0.577_{(\rho)}$ \\
			\cline{3-5}
			&&\multirow{2}{*}{$((1^{+}\otimes1^{+})_{0}\otimes{\frac{1}{2}}^{-})_{\frac{1}{2}}\otimes1^{-}$} &$\frac{1}{3} \left(-3\mu _{c}+\mu _{l(1\otimes1/\frac{1}{2})}+2\mu _{l'(1\otimes1/\frac{1}{2})}\right)$ &$ 0.186_{(\lambda)}$ \\
			&&&$\frac{1}{3} \left(-3\mu _{c}+\mu _{l(1/1)}+2\mu _{l'(1/1}\right)$ &$-0.499_{(\rho)}$ \\
			\cline{3-5}
			&&\multirow{2}{*}{$((1^{+}\otimes1^{+})_{1}\otimes{\frac{1}{2}}^{-})_{\frac{1}{2}}\otimes1^{-}$} &$\frac{1}{9} \left(3\mu _{u}+6\mu _{d}+6\mu _{c}+3\mu _{l(1\otimes1/\frac{1}{2})}+6\mu _{l'(1\otimes1/\frac{1}{2})}\right)$ &$ 0.858_{(\lambda)}$ \\
			&&&$\frac{1}{9} \left(3\mu _{u}+6\mu _{d}+6\mu _{c}+3\mu _{l(1/1)}+6\mu _{l'(1/1)}\right)$ &$0.174_{(\rho)}$ \\
			\cline{2-5}
			&{{${^{4}P_{\frac{3}{2}}}$}} &\multirow{2}{*}{$(0^{+}\otimes1^{+}\otimes {\frac{1}{2}}^{-})_{\frac{3}{2}}\otimes1^{-}$ } &$\frac{1}{45} \left(11\mu _{u}+55\mu _{d}-33\mu _{c}+6\mu _{l(0\otimes1/\frac{1}{2})}+12\mu _{l'(0\otimes1/\frac{1}{2})}\right)$  &$ -0.748_{(\lambda)}$ \\
			&&&$\frac{1}{45} \left(11\mu _{u}+55\mu _{d}-33\mu _{c}+6\mu _{l(0/1)}+12\mu _{l'(0/1)}\right)$ &$-1.022_{(\rho)}$ \\
			\cline{3-5}
			&&\multirow{2}{*}{$((1^{+}\otimes1^{+})_{1}\otimes{\frac{1}{2}}^{-})_{\frac{3}{2}}\otimes1^{-}$} &$\frac{1}{90} \left(33\mu _{u}+66\mu _{d}-33\mu _{c}+12\mu _{l(1\otimes1/\frac{1}{2})}+24\mu _{l'(1\otimes1/\frac{1}{2})}\right)$&$ 0.088_{(\lambda)}$ \\
			&&&$\frac{1}{90} \left(33\mu _{u}+66\mu _{d}-33\mu _{c}+12\mu _{l(1/1)}+24\mu _{l'(1/1)}\right)$ &$-0.186_{(\rho)}$ \\
			\cline{3-5}
			&&\multirow{2}{*}{$((1^{+}\otimes1^{+})_{2}\otimes{\frac{1}{2}}^{-})_{\frac{3}{2}}\otimes1^{-}$} &$\frac{1}{150} \left(99\mu _{u}+198\mu _{d}+165\mu _{c}+20\mu _{l(1\otimes1/\frac{1}{2})}+40\mu _{l'(1\otimes1/\frac{1}{2})}\right)$ &$ 0.679_{(\lambda)}$ \\
			&&&$\frac{1}{150} \left(99\mu _{u}+198\mu _{d}+165\mu _{c}+20\mu _{l(1/1)}+40\mu _{l'(1/1)}\right)$&$0.406_{(\rho)}$ \\
			\cline{2-5}
			&{{${^{6}P_{\frac{3}{2}}}$}}  &\multirow{2}{*}{ $((1^{+}\otimes1^{+})_{2}\otimes{\frac{1}{2}}^{-})_{\frac{5}{2}}\otimes1^{-}$ }&$\frac{1}{25} \left(14\mu _{u}+42\mu _{d}-5\mu _{l(1\otimes1/\frac{1}{2})}-10\mu _{l'(1\otimes1/\frac{1}{2})}\right)$   &$ -0.353_{(\lambda)}$ \\
			&&&$\frac{1}{25} \left(14\mu _{u}+42\mu _{d}-5\mu _{l(1/1)}-10\mu _{l'(1/1)}\right)$  &$0.057_{(\rho)}$ \\
			\hline
			$ \frac{5}{2}^{+} $&{{${^{4}P_{\frac{5}{2}}}$}}  & \multirow{2}{*}{$(0^{+}\otimes1^{+}\otimes {\frac{1}{2}}^{-})\otimes1^{-}$ }&$\frac{1}{3} \left(\mu _{u}+5\mu _{d}-3\mu _{c}+\mu _{l(0\otimes1/\frac{1}{2})}+2\mu _{l'(0\otimes1/\frac{1}{2})}\right)$& $-0.744_{(\lambda)}$ \\
			&&&$\frac{1}{3} \left(\mu _{u}+5\mu _{d}-3\mu _{c}+\mu _{l(0/1)}+2\mu _{l'(0/1)}\right)$ &$-1.430_{(\rho)}$ \\
			\cline{3-5}
			&&\multirow{2}{*}{$((1^{+}\otimes1^{+})_{1}\otimes{\frac{1}{2}}^{-})_{\frac{3}{2}}\otimes1^{-}$} &$\frac{1}{6} \left(3\mu _{u}+6\mu _{d}-3\mu _{c}+2\mu _{l(1\otimes1/\frac{1}{2})}+4\mu _{l'(1\otimes1/\frac{1}{2})}\right)$ &$ 0.387_{(\lambda)}$ \\
			&&&$\frac{1}{6} \left(3\mu _{u}+6\mu _{d}-3\mu _{c}+2\mu _{l(1/1)}+4\mu _{l'(1/1)}\right)$ &$-0.297_{(\rho)}$ \\
			\cline{3-5}
			&&\multirow{2}{*}{$((1^{+}\otimes1^{+})_{2}\otimes{\frac{1}{2}}^{-})_{\frac{3}{2}}\otimes1^{-}$} &$\frac{1}{30} \left(27\mu _{u}+54\mu _{d}+15\mu _{c}+10\mu _{l(1\otimes1/\frac{1}{2})}+20\mu _{l'(1\otimes1/\frac{1}{2})}\right)$   &$ 1.194_{(\lambda)}$ \\
			&&&$\frac{1}{30} \left(27\mu _{u}+54\mu _{d}+15\mu _{c}+10\mu _{l(1/1)}+20\mu _{l'(1/1)}\right)$  &$0.509_{(\rho)}$ \\
			\cline{2-5}
			&{{${^{6}P_{\frac{5}{2}}}$}}  &\multirow{2}{*}{$((1^{+}\otimes1^{+})_{2}\otimes{\frac{1}{2}}^{-})_{\frac{5}{2}}\otimes1^{-}$} &$\frac{1}{105} \left(93\mu _{u}+186\mu _{d}+10\mu _{l(1\otimes1/\frac{1}{2})}+20\mu _{l'(1\otimes1/\frac{1}{2})}\right)$ &$ 0.168_{(\lambda)}$ \\
			&&&$\frac{1}{105} \left(93\mu _{u}+186\mu _{d}+10\mu _{l(1/1)}+20\mu _{l'(1/1)}\right)$ &$-0.027_{(\rho)}$ \\
			\hline
			$ \frac{7}{2}^{+} $&{{${^{6}P_{\frac{7}{2}}}$}}  &\multirow{2}{*}{ $1^{+}\otimes1^{+}\otimes{\frac{1}{2}}^{-}\otimes1^{-}$} &$\frac{1}{3} \left(3\mu _{u}+6\mu _{d}+\mu _{l(1\otimes1/\frac{1}{2})}+2\mu _{l'(1\otimes1/\frac{1}{2})}\right)$ & $0.589_{(\lambda)}$ \\
			&&&$\frac{1}{3} \left(3\mu _{u}+6\mu _{d}+\mu _{l(1/1)}+2\mu _{l'(1/1)}\right)$&$-0.095_{(\rho)}$ \\
			\bottomrule[1pt]
			\bottomrule[1pt]
		\end{tabular}
		
	\end{table*}

	\renewcommand\tabcolsep{0.46cm}
	\renewcommand{\arraystretch}{1.6}
	\begin{table*}
		\caption{The  excited state magnetic moments of the $P^{\Delta^{0}}_{\psi}  $ states in the diquark-diquark-antiquark model with the $ 10_{f} $  flavor representation. In the $\lambda$ mode $ P $-wave excitation, $ \mu _{l(S_{cd}\otimes S_{ud}/S_{\bar{c}})} $ represents the orbital magnetic moment between   $ \bar{c} $ and center of mass system of diquark $ (cd) $ and diquark $ (ud) $, $S_{cd}  $, $ S_{ud} $ and $ S_{\bar{c}} $ represent the spins of diquark $ (cd) $, diquark $ (ud) $ and $ \bar{c} $, respectively. $ \mu _{l'(S_{cu}\otimes S_{dd}/S_{\bar{c}})} $ represents the orbital magnetic moment between   $ \bar{c} $ and center of mass system of diquark $ (cu) $ and diquark $ (dd) $, and $S_{cu}  $, $ S_{dd} $ and $ S_{\bar{c}} $ represent the spins of diquark $ (cu) $, diquark $ (dd) $ and $ \bar{c} $, respectively.  In the $\rho  $ mode $ P $-wave excitation, $ \mu _{l(S_{cd}/ S_{ud})} $ represents the orbital magnetic moment between   diquark $ (cd) $   and diquark $ (ud) $,  $S_{cd}  $ and $ S_{ud} $  represent the spins of diquark $ (cd) $ and diquark $ (ud) $, respectively.
			$ \mu _{l'(S_{cu}/ S_{dd})} $ represents the orbital magnetic moment between  diquark $ (cu) $ and diquark $ (dd) $,  $S_{cu}  $ and $ S_{dd} $  represent the spins of diquark $ (cu) $, diquark $ (dd) $, respectively. The $J_{1}^{P_{1}}\otimes J_{2}^{P_{2}}\otimes J_{3}^{P_{3}}\otimes J_{4}^{P_{4}}$ are corresponding to the angular momentum and parity of $(cq_1)$, $(q_2q_3)$, $\bar{c}$ and orbital, respectively. The unit is the nuclear magnetic moment $ \mu_{N} $.}
		\scriptsize
		\label{tab:ls}
		\begin{tabular}{c|c|c|c|c}
			\toprule[1pt]
			\toprule[1pt]
			\multicolumn{5}{c}{$\sqrt{\frac{2}{3}}({c}d)\{ud\}{\bar c}+\sqrt{\frac{1}{3}}({c}u)\{dd\}{\bar c}$ } \\
			\toprule[1pt]
			$J^P$	& $^{2s+1}L_J$ &$J_{1}^{P_{1}}\otimes J_{2}^{P_{2}}\otimes J_{3}^{P_{3}}\otimes J_{4}^{P_{4}}$    &Expressions &Results   \\
			\hline
			
			{$\frac{1}{2}^{+}$}	&{{${^{2}P_{\frac{1}{2}}}$}}  &\multirow{2}{*}{$(0^{+}\otimes1^{+}\otimes {\frac{1}{2}}^{-})_{\frac{1}{2}}\otimes1^{-}$ }  &$\frac{1}{27} \left(-4\mu _{u}-8\mu _{d}-3\mu _{c}+12\mu _{l(0\otimes1/\frac{1}{2})}+6\mu _{l'(0\otimes1/\frac{1}{2})}\right)$   &$0.360_{(\lambda)}$  \\
			&&&$\frac{1}{27} \left(-4\mu _{u}-8\mu _{d}-3\mu _{c}+12\mu _{l(0/1)}+6\mu _{l'(0/1}\right)$  &$0.024_{(\rho)}$  \\
			\cline{3-5}
			&&\multirow{2}{*}{$((1^{+}\otimes1^{+})_{0}\otimes{\frac{1}{2}}^{-})_{\frac{1}{2}}\otimes1^{-}$} & $\frac{1}{9} \left(3\mu _{c}+4\mu _{l(1\otimes1/\frac{1}{2})}+2\mu _{l'(1\otimes1/\frac{1}{2})}\right)$  &$0.529_{(\lambda)}$ \\
			&&& $\frac{1}{9} \left(3\mu _{c}+4\mu _{l(1/1)}+2\mu _{l'(1/1)}\right)$&$0.198_{(\rho)}$  \\
			\cline{3-5}
			&&\multirow{2}{*}{$((1^{+}\otimes1^{+})_{1}\otimes{\frac{1}{2}}^{-})_{\frac{1}{2}}\otimes1^{-}$} &$\frac{1}{27} \left(-3\mu _{u}-6\mu _{d}-6\mu _{c}+12\mu _{l(1\otimes1/\frac{1}{2})}+6\mu _{l'(1\otimes1/\frac{1}{2})}\right)$  & $0.305_{(\lambda)}$ \\
			&&&$\frac{1}{27} \left(-3\mu _{u}-6\mu _{d}-6\mu _{c}+12\mu _{l(1/1)}+6\mu _{l'(1/1)}\right)$&$-0.027_{(\rho)}$ \\
			\cline{2-5}
			&{{${^{4}P_{\frac{1}{2}}}$}}&\multirow{2}{*}{ $(0^{+}\otimes1^{+}\otimes {\frac{1}{2}}^{-})_{\frac{3}{2}}\otimes1^{-}$} & $\frac{1}{27} \left(10\mu _{u}+20\mu _{d}-15\mu _{c}-6\mu _{l(0\otimes1/\frac{1}{2})}-3\mu _{l'(0\otimes1/\frac{1}{2})}\right)$  &$-0.426_{(\lambda)}$  \\
			&&& $\frac{1}{27} \left(10\mu _{u}+20\mu _{d}-15\mu _{c}-6\mu _{l(0/1)}-3\mu _{l'(0/1)}\right)$  &$-0.259_{(\rho)}$  \\
			\cline{3-5}
			&&\multirow{2}{*}{$((1^{+}\otimes1^{+})_{1}\otimes{\frac{1}{2}}^{-})_{\frac{3}{2}}\otimes1^{-}$} &$\frac{1}{54} \left(15\mu _{u}+30\mu _{d}-15\mu _{c}-12\mu _{l(1\otimes1/\frac{1}{2})}-6\mu _{l'(1\otimes1/\frac{1}{2})}\right)$&$ -0.309_{(\lambda)}$\\
			&&&$\frac{1}{54} \left(15\mu _{u}+30\mu _{d}-15\mu _{c}-12\mu _{l(1/1)}-6\mu _{l^{'}(1/1)}\right)$ &$-0.144_{(\rho)}$ \\
			\cline{3-5}
			&&\multirow{2}{*}{$((1^{+}\otimes1^{+})_{2}\otimes{\frac{1}{2}}^{-})_{\frac{3}{2}}\otimes1^{-}$} &$\frac{1}{18} \left(15\mu _{u}+18\mu _{d}+15\mu _{c}-4\mu _{l(1\otimes1/\frac{1}{2})}-2\mu _{l'(1\otimes1/\frac{1}{2})}\right)$ &$ 0.140_{(\lambda)}$ \\
			&&&$\frac{1}{18} \left(15\mu _{u}+18\mu _{d}+15\mu _{c}-4\mu _{l(1/1)}-2\mu _{l'(1/1)}\right)$&$0.305_{(\rho)}$ \\
			\hline
			$ \frac{3}{2}^{+} $&{{${^{2}P_{\frac{3}{2}}}$}}  &\multirow{2}{*}{ $(0^{+}\otimes1^{+}\otimes {\frac{1}{2}}^{-})_{\frac{1}{2}}\otimes1^{-}$} &$\frac{1}{9} \left(4\mu _{u}+8\mu _{d}+3\mu _{c}+6\mu _{l(0\otimes1/\frac{1}{2})}+3\mu _{l'(0\otimes1/\frac{1}{2})}\right)$ &$ 0.741_{(\lambda)}$ \\
			&&&$\frac{1}{9} \left(4\mu _{u}+8\mu _{d}+3\mu _{c}+6\mu _{l(0/1)}+3\mu _{l'(0/1)}\right)$&$0.238_{(\rho)}$ \\
			\cline{3-5}
			&&\multirow{2}{*}{$((1^{+}\otimes1^{+})_{0}\otimes{\frac{1}{2}}^{-})_{\frac{1}{2}}\otimes1^{-}$} &$\frac{1}{3} \left(-3\mu _{c}+2\mu _{l(1\otimes1/\frac{1}{2})}+\mu _{l'(1\otimes1/\frac{1}{2})}\right)$ &$ 0.186_{(\lambda)}$ \\
			&&&$\frac{1}{3} \left(-3\mu _{c}+2\mu _{l(1/1)}+1\mu _{l'(1/1}\right)$ &$-0.309_{(\rho)}$ \\
			\cline{3-5}
			
			&&\multirow{2}{*}{$((1^{+}\otimes1^{+})_{1}\otimes{\frac{1}{2}}^{-})_{\frac{1}{2}}\otimes1^{-}$} &$\frac{1}{9} \left(3\mu _{u}+6\mu _{d}+6\mu _{c}+6\mu _{l(1\otimes1/\frac{1}{2})}+3\mu _{l'(1\otimes1/\frac{1}{2})}\right)$&$ 0.858_{(\lambda)}$ \\
			&&&$\frac{1}{9} \left(3\mu _{u}+6\mu _{d}+6\mu _{c}+6\mu _{l(1/1)}+3\mu _{l'(1/1)}\right)$&$0.364_{(\rho)}$ \\
			\cline{2-5}
			&{{${^{4}P_{\frac{3}{2}}}$}} &\multirow{2}{*}{$(0^{+}\otimes1^{+}\otimes {\frac{1}{2}}^{-})_{\frac{3}{2}}\otimes1^{-}$ } &$\frac{1}{45} \left(22\mu _{u}+44\mu _{d}-33\mu _{c}+12\mu _{l(0\otimes1/\frac{1}{2})}+6\mu _{l'(0\otimes1/\frac{1}{2})}\right)$  &$ -0.053_{(\lambda)}$ \\
			&&&$\frac{1}{45} \left(22\mu _{u}+44\mu _{d}-33\mu _{c}+12\mu _{l(0/1)}+6\mu _{l'(0/1)}\right)$ &$-0.255_{(\rho)}$ \\
			\cline{3-5}
			&&\multirow{2}{*}{$((1^{+}\otimes1^{+})_{1}\otimes{\frac{1}{2}}^{-})_{\frac{3}{2}}\otimes1^{-}$} &$\frac{1}{90} \left(33\mu _{u}+66\mu _{d}-33\mu _{c}+24\mu _{l(1\otimes1/\frac{1}{2})}+12\mu _{l'(1\otimes1/\frac{1}{2})}\right)$ &$ 0.088_{(\lambda)}$ \\
			&&&$\frac{1}{90} \left(33\mu _{u}+66\mu _{d}-33\mu _{c}+24\mu _{l(1/1)}+12\mu _{l'(1/1)}\right)$&$-0.110_{(\rho)}$ \\
			\cline{3-5}
			
			&&\multirow{2}{*}{$((1^{+}\otimes1^{+})_{2}\otimes{\frac{1}{2}}^{-})_{\frac{3}{2}}\otimes1^{-}$} &$\frac{1}{150} \left(99\mu _{u}+198\mu _{d}+165\mu _{c}+40\mu _{l(1\otimes1/\frac{1}{2})}+20\mu _{l'(1\otimes1/\frac{1}{2})}\right)$ &$ 0.679_{(\lambda)}$ \\
			&& &$\frac{1}{150} \left(99\mu _{u}+198\mu _{d}+165\mu _{c}+40\mu _{l(1/1)}+20\mu _{l'(1/1)}\right)$&$0.482_{(\rho)}$ \\
			
			\cline{2-5}
			&{{${^{6}P_{\frac{3}{2}}}$}}  &\multirow{2}{*}{ $((1^{+}\otimes1^{+})_{2}\otimes{\frac{1}{2}}^{-})_{\frac{5}{2}}\otimes1^{-}$} &$\frac{1}{25} \left(14\mu _{u}+42\mu _{d}-10\mu _{l(1\otimes1/\frac{1}{2})}-5\mu _{l'(1\otimes1/\frac{1}{2})}\right)$ &$ -0.353_{(\lambda)}$ \\
			&&&$\frac{1}{25} \left(14\mu _{u}+42\mu _{d}-10\mu _{l(1/1)}-5\mu _{l'(1/1)}\right)$  &$-0.057_{(\rho)}$ \\
			\hline
			$ \frac{5}{2}^{+} $&{{${^{4}P_{\frac{5}{2}}}$}}  &\multirow{2}{*}{ $(0^{+}\otimes1^{+}\otimes {\frac{1}{2}}^{-})\otimes1^{-}$} &$\frac{1}{3} \left(2\mu _{u}+4\mu _{d}-3\mu _{c}+2\mu _{l(0\otimes1/\frac{1}{2})}+\mu _{l'(0\otimes1/\frac{1}{2})}\right)$ & $0.204_{(\lambda)}$ \\
			&&&$\frac{1}{3} \left(2\mu _{u}+4\mu _{d}-3\mu _{c}+2\mu _{l(0/1)}+\mu _{l'(0/1)}\right)$ &$-0.300_{(\rho)}$ \\
			\cline{3-5}
			&&\multirow{2}{*}{$((1^{+}\otimes1^{+})_{1}\otimes{\frac{1}{2}}^{-})_{\frac{3}{2}}\otimes1^{-}$} &$\frac{1}{6} \left(3\mu _{u}+6\mu _{d}-3\mu _{c}+4\mu _{l(1\otimes1/\frac{1}{2})}+2\mu _{l'(1\otimes1/\frac{1}{2})}\right)$ &$ 0.387_{(\lambda)}$ \\
			&&&$\frac{1}{6} \left(3\mu _{u}+6\mu _{d}-3\mu _{c}+4\mu _{l(1/1)}+2\mu _{l'(1/1)}\right)$&$-0.107_{(\rho)}$ \\
			\cline{3-5}
			&&\multirow{2}{*}{$((1^{+}\otimes1^{+})_{2}\otimes{\frac{1}{2}}^{-})_{\frac{3}{2}}\otimes1^{-}$} &$\frac{1}{30} \left(27\mu _{u}+54\mu _{d}+15\mu _{c}+20\mu _{l(1\otimes1/\frac{1}{2})}+10\mu _{l'(1\otimes1/\frac{1}{2})}\right)$   &$ 1.194_{(\lambda)}$ \\
			&&&$\frac{1}{30} \left(27\mu _{u}+54\mu _{d}+15\mu _{c}+20\mu _{l(1/1)}+10\mu _{l'(1/1)}\right)$  &$0.701_{(\rho)}$ \\
			\cline{2-5}
			&{{${^{6}P_{\frac{5}{2}}}$}}  &\multirow{2}{*}{$((1^{+}\otimes1^{+})_{2}\otimes{\frac{1}{2}}^{-})_{\frac{5}{2}}\otimes1^{-}$} &$\frac{1}{105} \left(93\mu _{u}+186\mu _{d}+20\mu _{l(1\otimes1/\frac{1}{2})}+10\mu _{l'(1\otimes1/\frac{1}{2})}\right)$ &$ 0.168_{(\lambda)}$ \\
			&&&$\frac{1}{105} \left(93\mu _{u}+186\mu _{d}+20\mu _{l(1/1)}+10\mu _{l'(1/1)}\right)$ &$0.027_{(\rho)}$ \\
			\hline
			$ \frac{7}{2}^{+} $&{{${^{6}P_{\frac{7}{2}}}$}}  &\multirow{2}{*}{ $1^{+}\otimes1^{+}\otimes{\frac{1}{2}}^{-}\otimes1^{-}$} &$\frac{1}{3} \left(3\mu _{u}+6\mu _{d}+2\mu _{l(1\otimes1/\frac{1}{2})}+\mu _{l'(1\otimes1/\frac{1}{2})}\right)$ & $0.589_{(\lambda)}$\\
			&&&$\frac{1}{3} \left(3\mu _{u}+6\mu _{d}+2\mu _{l(1/1)}+\mu _{l'(1/1)}\right)$ &$0.095_{(\rho)}$ \\
			\bottomrule[1pt]
			\bottomrule[1pt]
		\end{tabular}
		
	\end{table*}

	\section{ magnetic moments of the  $P^{N^{0}}_{\psi}  $ states and $P^{\Delta^{0}}_{\psi}$  states in the Diquark-triquark model }
	\label{sec5}

	We calculated the magnetic moments in the diquark-triquark model with configurations $ (cq_{3})(\bar{c}q_{1}q_{2}) $. Similar to the magnetic moments expression in the molecular model, the magnetic moments in diquark-triquark model is
	\begin{align}
		\mu &=\left\langle \psi\left |g_{cq_{3}}{\mu_{cq_{3}}}{\vec{S}_{cq_{3}}}+g_{\bar{c}q_{1}q_{2}}{\mu_{\bar{c}q_{1}q_{2}}}{\vec{S}}_{\bar{c}q_{1}q_{2}}+{\mu_l}\vec{l}\,\,\right|\psi\right\rangle \nonumber\\
		&=\sum_{l_z, {S_z}}\left\langle l l_z,SS_z|JJ_{z}\right\rangle^2
		\Big\{\sum_{S'_{\bar{c}q_{1}q_{2}}}\big\langle {S_{\bar{c}q_{1}q_{2}}}S'_{\bar{c}q_{1}q_{2}},
		S_{cq_{3}} (S_z\nonumber\\
		&-S'_{\bar{c}q_{1}q_{2}})|SS_z\big \rangle^2\Big[(S_z-S'_{\bar{c}q_{1}q_{2}})(\mu_{c}+\mu_{q_3})\nonumber\\
		&+\sum_{S'_{\bar{c}},S'_{q_{1}q_{2}}}\big\langle S_{\bar{c}}S'_{\bar{c}},S_{q_{1}q_{2}}S'_{q_{1}q_{2}}|S_{cq_{1}q_{2}}
		S'_{cq_{1}q_{2}}\big\rangle^2\big(g_{\bar{c}}\mu_{\bar{c}}S'_{\bar{c}}\nonumber\\
		&+S'_{q_{1}q_{2}}(\mu_{q_1}+\mu_{q_2})\big)\Big ]+\mu_{l}l_z\bigg \},
	\end{align}
	where  $S_{cq_{3}}  $ and $S_{cq_{1}q_{2}}  $ represent the spin of  the diquark $({cq_{3}})$ and the triquark $( cq_{1}q_{2}) $, respectively. $ S_{q_{1}q_{2}} $ represent the spin of  the  diquark $(q_{1}q_{2}) $  inside the triquark $ (cq_{1}q_{2}) $. $ S'_{\bar{c}q_{1}q_{2}} $, $ S_{z} $ and $S'_{q_{1}q_{2}}  $ are  spin third component of  triquark $(\bar{c}q_{1}q_{2} ) $,  pentaquark state and  diquark $ (q_{1}q_{2}) $. $S_{\bar{c}}$ and $S'_{\bar{c}}$ are  spin and third component of antiquark $\bar{c}  $. $\psi$ represents the flavor wave function in diquark-triquark model in Table \ref{tab:k}, The orbital magnetic moment in diquark-triquark model  is
	\begin{eqnarray} \label{w1}
		{\mu_l}&=&\frac{{m_{cq_{3}}}\mu_{\bar{c}q_{1}q_{2} }+{m_{\bar{c}q_{1}q_{2} }}\mu_{cq_{3}}}{m_{\bar{c}q_{1}q_{2}}+m_{cq_{3}}},
	\end{eqnarray}
	where $ m $ and $ \mu $ are the mass and magnetic moment of the cluster represented by their subscripts. The mass of the triquarks is roughly equal to the sum of the mass of the corresponding diquarks and antiquarks \cite{Wang:2016dzu}. Here, we use the values of the  diquark  masses from Eq. (\ref{equ:op}). The magnetic moments  and corresponding expressions in the diquark-triquark mode  are presented in Table \ref{tab:dtb} and Table \ref{tab:dts}.

	\renewcommand\tabcolsep{0.63cm}
	\renewcommand{\arraystretch}{1.65}
	\begin{table*}[htbp]
		\caption{The magnetic moments of the $P^{N^{0}}_{\psi}  $ states in the diquark-triquark model with the $ 8_{1f} $ and $ 8_{2f} $ flavor representation. The  $J_{1}^{P_{1}}\otimes J_{2}^{P_{2}}\otimes J_{3}^{P_{3}}$ are corresponding to the angular momentum and parity of triquark, diquark and orbital, respectively. 
		Where the $ \mu _{l(S_{\bar{c}ud}/ S_{cd})} $ represents the orbital magnetic moment between   triquark $ (\bar{c}ud) $   and diquark $ (cd) $,  $S_{\bar{c}ud}  $ and $ S_{cd} $  represent the spins of triquark $ (\bar{c}ud) $ and diquark $ (cd) $, respectively.
		The $ \mu _{l'(S_{\bar{c}dd}/ S_{cu})} $ represents the orbital magnetic moment between   triquark $ (\bar{c}dd) $   and diquark $ (cu) $,  $S_{\bar{c}dd}  $ and $ S_{cu} $  represent the spins of triquark $ (\bar{c}dd) $ and diquark $ (cu) $, respectively.    The  unit is the nuclear magnetic moment $\mu_N$.}
		\scriptsize
		\label{tab:dtb}
		\begin{tabular}{c|c|c|c|c}
				\toprule[1pt]
			\toprule[1pt]
				\multicolumn{5}{c}{$({c}d)(\bar c[ud]) $}\\
			\toprule[1pt]
			$J^P$	& $^{2s+1}L_J$ & $J_{1}^{P_{1}}\otimes J_{2}^{P_{2}}\otimes J_{3}^{P_{3}}$   &Expressions  & Results \\
			\hline
			{$\frac{1}{2}^{-}$}	&{{${^{2}S_{\frac{1}{2}}}$}}  &${\frac{1}{2}}^{-}\otimes0^{+}\otimes0^{+}$  &$-\mu _{c}$  &$-0.403$ \\

			&& ${\frac{1}{2}}^{-}\otimes1^{+}\otimes0^{+}$ & $\frac{1}{3} \left(3 \mu _{c}+2 \mu _{d}\right)$ & $-0.228$ \\
			\hline
			{$\frac{3}{2}^{-}$}  &{{${^{4}S_{\frac{3}{2}}}$}} & ${\frac{1}{2}}^{-}\otimes1^{+}\otimes0^{+}$ &$\mu_{d} $  &$ -0.947$ \\
			\hline
			{$\frac{1}{2}^{+}$}	&{{${^{2}P_{\frac{1}{2}}}$}}  &${\frac{1}{2}}^{-}\otimes0^{+}\otimes1^{-}$  &$\frac{1}{3} \left(\mu _{c}+2 \mu _{l(\frac{1}{2}/0)}\right)$  &$0.150$  \\

			&& $({\frac{1}{2}}^{-}\otimes1^{+})_{\frac{1}{2}}\otimes1^{-}$ & $\frac{1}{9} \left(-3 \mu_{c}-2 \mu _{d}+6 \mu _{l(\frac{1}{2}/1)}\right)$  &$0.087$  \\
			\cline{2-5}
			&{{${^{4}P_{\frac{1}{2}}}$}}  & $({\frac{1}{2}}^{-}\otimes1^{+})_{\frac{3}{2}}\otimes1^{-}$ &  $\frac{1}{9} \left(5\mu _{d}-3 \mu _{l(\frac{1}{2}/1)}\right)$ &$ -0.532$ \\
			\hline
			{$\frac{3}{2}^{+}$}&{{${^{2}P_{\frac{3}{2}}}$}}  & ${\frac{1}{2}}^{-}\otimes0^{+}\otimes1^{-}$ &  $ - \mu _{c}+\mu _{l(\frac{1}{2}/0)}$ &$-0.380$  \\
			
			&&$({\frac{1}{2}}^{-}\otimes1^{+})_{\frac{1}{2}}\otimes1^{-}$  & $\frac{1}{3} \left(3 \mu _{c}+2 \mu _{d}+3\mu _{l(\frac{1}{2}/1)}\right)$   &$ -0.212$ \\
			\cline{2-5}
			&{{${^{4}P_{\frac{3}{2}}}$}}  &$({\frac{1}{2}}^{-}\otimes1^{+})_{\frac{3}{2}}\otimes1^{-}$  &$\frac{1}{15} \left(11 \mu _{d}+6 \mu _{l(\frac{1}{2}/1)}\right)$   &$ -0.688$ \\
			\hline
			{$\frac{5}{2}^{+}$}	&{{${^{4}P_{\frac{5}{2}}}$}}   &${\frac{1}{2}}^{-}\otimes1^{+}\otimes1^{-}$  &$  \mu _{d}+\mu _{l(\frac{1}{2}/1)}$  &$ -0.931 $\\
			\toprule[1pt]
				\multicolumn{5}{c}{$\sqrt{\frac{1}{3}}({c}d)(\bar c\{ud\})-\sqrt{\frac{2}{3}}({c}u)(\bar c\{dd\})$}\\
			\toprule[1pt]
			$J^P$	& $^{2s+1}L_J$ & $J_{1}^{P_{1}}\otimes J_{2}^{P_{2}}\otimes J_{3}^{P_{3}}$  &Expressions  & Results \\
			\hline
			{$\frac{1}{2}^{-}$}	&{{${^{2}S_{\frac{1}{2}}}$}}  &${\frac{1}{2}}^{-}\otimes0^{+}\otimes0^{+}$  &$\frac{1}{9} \left(2 \mu _{u}+10\mu _{d}+3\mu _{c}\right)$  &$-0.497$ \\

			&& ${\frac{1}{2}}^{-}\otimes1^{+}\otimes0^{+}$ & $\frac{1}{27} \left(10 \mu _{u}-4 \mu _{d}+15\mu _{c}\right)$ &$ 1.066$ \\
			
			&& ${\frac{3}{2}}^{-}\otimes1^{+}\otimes0^{+}$ & $\frac{1}{27} \left(-\mu _{u}+22 \mu _{d}-24\mu _{c}\right)$ & $-1.201$ \\
			\hline

			{$\frac{3}{2}^{-}$}  &{{${^{4}S_{\frac{3}{2}}}$}} & ${\frac{1}{2}}^{-}\otimes1^{-}\otimes0^{+}$ &$\frac{1}{9} \left(8\mu _{u}+13 \mu _{d}+12\mu _{c}\right)$  &$ 0.854$ \\
			&& ${\frac{3}{2}}^{-}\otimes0^{+}\otimes0^{+}$ & $\frac{1}{3} \left(\mu _{u}+5\mu _{d}-3\mu _{c}\right)$ & $-1.351$ \\
			&& ${\frac{3}{2}}^{-}\otimes1^{-}\otimes0^{+}$ & $\frac{1}{45} \left(23\mu _{u}+61\mu _{d}-15\mu _{c}\right)$ &$ -0.450$ \\
			\hline
			{$\frac{5}{2}^{-}$}  &{{${^{6}S_{\frac{5}{2}}}$}} & ${\frac{3}{2}}^{-}\otimes1^{+}\otimes0^{+}$ &$\mu _{u}+2\mu _{d}$  & $0$ \\
			\hline
			{$\frac{1}{2}^{+}$}	&{{${^{2}P_{\frac{1}{2}}}$}}  &${\frac{1}{2}}^{-}\otimes0^{+}\otimes1^{-}$  &$\frac{1}{27} \left(-2\mu _{u}-10\mu _{d}-3\mu _{c}+6\mu _{(\frac{1}{2}/0)}+12\mu _{l'(\frac{1}{2}/0)}\right)$  &$0.234 $ \\
			
			&& $({\frac{1}{2}}^{-}\otimes1^{+})_{\frac{1}{2}}\otimes1^{-}$ & $\frac{1}{81} \left(- 10\mu _{u}+4\mu _{d}-15\mu _{c}+18\mu _{l(\frac{1}{2}/1)}+36\mu _{l'(\frac{1}{2}/1)}\right)$  &$-0.377$  \\
			\cline{2-5}
			&{{${^{4}P_{\frac{1}{2}}}$}}  & $({\frac{1}{2}}^{-}\otimes1^{+})_{\frac{3}{2}}\otimes1^{-}$ &$\frac{1}{81} \left(40\mu _{u}+65\mu _{d}+60\mu _{c}-9\mu _{l(\frac{1}{2}/1)}-18\mu _{l'(\frac{1}{2}/1)}\right)$ &$ 0.286$\\
			&& ${\frac{3}{2}}^{-}\otimes0^{+}\otimes1^{-}$ & $\frac{1}{27} \left(5\mu _{u}+25\mu _{d}-15\mu _{c}-3\mu _{l(\frac{3}{2}/0)}-6\mu _{l'(\frac{3}{2}/0)}\right)$  &$-0.785$  \\
			\cline{2-5}
			&{{${^{2}P_{\frac{1}{2}}}$}}  & $({\frac{3}{2}}^{-}\otimes1^{+})_{\frac{1}{2}}\otimes1^{-}$ &$\frac{1}{81} \left(\mu _{u}-22\mu _{d}+24\mu _{c}+18\mu _{l(\frac{3}{2}/1)}+36\mu _{l'(\frac{3}{2}/1)}\right)$ &$ 0.379$ \\
			&{{${^{4}P_{\frac{1}{2}}}$}}  & $({\frac{3}{2}}^{-}\otimes1^{+})_{\frac{3}{2}}\otimes1^{-}$ &$\frac{1}{81} \left(23\mu _{u}+61\mu _{d}-15\mu _{c}-9\mu _{l(\frac{3}{2}/1)}-18\mu _{l'(\frac{3}{2}/1)}\right)$ &$ -0.240$ \\
			\hline
			$ \frac{3}{2}^{+} $&{{${^{2}P_{\frac{3}{2}}}$}}  & ${\frac{1}{2}}^{-}\otimes0^{+}\otimes1^{-}$ &$\frac{1}{9} \left(2\mu _{u}+10\mu _{d}+3\mu _{c}+3\mu _{l(\frac{1}{2}/0)}+6\mu _{l'(\frac{1}{2}/0)}\right)$ & $-0.395$ \\
			& & $({\frac{1}{2}}^{-}\otimes1^{+})_{\frac{1}{2}}\otimes1^{-}$ &$\frac{1}{27} \left(10\mu _{u}-4\mu _{d}+15\mu _{c}+9\mu _{l(\frac{1}{2}/1)}+18\mu _{l'(\frac{1}{2}/1)}\right)$ &$ 1.034$ \\
			\cline{2-5}
			&{{${^{4}P_{\frac{3}{2}}}$}}  & $({\frac{1}{2}}^{-}\otimes1^{+})_{\frac{3}{2}}\otimes1^{-}$ &$\frac{1}{135} \left(88\mu _{u}+143\mu _{d}+132\mu _{c}+18\mu _{l(\frac{1}{2}/1)}+36\mu _{l'(\frac{1}{2}/1)}\right)$ &$ 0.613$ \\
			& & ${\frac{3}{2}}^{-}\otimes0^{+}\otimes1^{-}$ &$\frac{1}{45} \left(11\mu _{u}+55\mu _{d}-33\mu _{c}+6\mu _{l(\frac{3}{2}/0)}+12\mu _{l'(\frac{3}{2}/0)}\right)$ &$ -0.950$ \\
			\cline{2-5}
			&{{${^{2}P_{\frac{3}{2}}}$}}  & $({\frac{3}{2}}^{-}\otimes1^{+})_{\frac{1}{2}}\otimes1^{-}$ &$\frac{1}{27} \left(-\mu _{u}+22\mu _{d}-24\mu _{c}+9\mu _{l(\frac{3}{2}/1)}+18\mu _{l'(\frac{3}{2}/1)}\right)$ &$ -1.233$ \\
			&{{${^{4}P_{\frac{3}{2}}}$}}  & $({\frac{3}{2}}^{-}\otimes1^{+})_{\frac{3}{2}}\otimes1^{-}$ &$\frac{1}{675} \left(253\mu _{u}+671\mu _{d}-165\mu _{c}+90\mu _{l(\frac{3}{2}/1)}+180\mu _{l'(\frac{3}{2}/1)}\right)$ &$-0.343$ \\
			&{{${^{6}P_{\frac{3}{2}}}$}}  & $({\frac{3}{2}}^{-}\otimes1^{+})_{\frac{5}{2}}\otimes1^{-}$ &$\frac{1}{75} \left(63\mu _{u}+126\mu _{d}-15\mu _{l(\frac{3}{2}/1)}-30\mu _{l'(\frac{3}{2}/1)}\right)$   &$ 0.019$ \\
			\hline
			$ \frac{5}{2}^{+} $&{{${^{4}P_{\frac{5}{2}}}$}}  & ${\frac{1}{2}}^{-}\otimes1^{+}\otimes1^{-}$ &$\frac{1}{9} \left(8\mu _{u}+13\mu _{d}+12\mu _{c}+3\mu _{l(\frac{1}{2}/1)}+6\mu _{l'(\frac{1}{2}/1)}\right)$ & $0.822$ \\
			
			& & ${\frac{3}{2}}^{-}\otimes0^{+}\otimes1^{-}$ &$\frac{1}{3} \left(\mu _{u}+5\mu _{d}-3\mu _{c}+\mu _{l(\frac{3}{2}/0)}+2\mu _{l'(\frac{3}{2}/0)}\right)$ & $-1.248$ \\
			
			&& $({\frac{3}{2}}^{-}\otimes1^{+})_{\frac{3}{2}}\otimes1^{-}$ &$\frac{1}{45} \left(23\mu _{u}+61\mu _{d}-15\mu _{c}+15\mu _{l(\frac{3}{2}/1)}+30\mu _{l'(\frac{3}{2}/1)}\right)$   &$ -0.482 $\\
			\cline{2-5}
			&{{${^{6}P_{\frac{5}{2}}}$}}  &$({\frac{3}{2}}^{-}\otimes1^{+})_{\frac{5}{2}}\otimes1^{-}$  &$\frac{1}{105} \left(93\mu _{u}+186\mu _{d}+10\mu _{l(\frac{3}{2}/1)}+20\mu _{l'(\frac{3}{2}/1)}\right)$ & $-0.009$ \\
			\hline
			$ \frac{7}{2}^{+} $&{{${^{6}P_{\frac{7}{2}}}$}}  & ${\frac{3}{2}}^{-}\otimes1^{+}\otimes1^{-}$ &$\frac{1}{3} \left(3\mu _{u}+6\mu _{d}+\mu _{l(\frac{3}{2}/1)}+2\mu _{l'(\frac{3}{2}/1)}\right)$ &$ -0.032$ \\
			\bottomrule[1pt]
			\bottomrule[1pt]
		\end{tabular}
		
	\end{table*}

	\renewcommand\tabcolsep{0.64cm}
	\renewcommand{\arraystretch}{1.65}
	\begin{table*}[htbp]
		\caption{The magnetic moments of the $P^{\Delta^{0}}_{\psi}  $ states in the diquark-triquark model with the $ 10_{f} $  flavor representation. The  $J_{1}^{P_{1}}\otimes J_{2}^{P_{2}}\otimes J_{3}^{P_{3}}$ are corresponding to the angular momentum and parity of triquark, diquark and orbital, respectively. 	Where the $ \mu _{l(S_{\bar{c}ud}/ S_{cd})} $ represents the orbital magnetic moment between   triquark $ (\bar{c}ud) $   and diquark $ (cd) $,  $S_{\bar{c}ud}  $ and $ S_{cd} $  represent the spins of triquark $ (\bar{c}ud) $ and diquark $ (cd) $, respectively.
		The $ \mu _{l'(S_{\bar{c}dd}/ S_{cu})} $ represents the orbital magnetic moment between   triquark $ (\bar{c}dd) $   and diquark $ (cu) $,  $S_{\bar{c}dd}  $ and $ S_{cu} $  represent the spins of triquark $ (\bar{c}dd) $ and diquark $ (cu) $, respectively.    The  unit is the nuclear magnetic moment $\mu_N$.  }
		\scriptsize
		\label{tab:dts}
		\begin{tabular}{c|c|c|c|c}
			\toprule[1pt]
			\toprule[1pt]
			\multicolumn{5}{c}{$\sqrt{\frac{2}{3}}({c}d)(\bar c\{ud\})+\sqrt{\frac{1}{3}}({c}u)(\bar c\{dd\})$}\\
			\toprule[1pt]
			$J^P$	& $^{2s+1}L_J$ &$J_{1}^{P_{1}}\otimes J_{2}^{P_{2}}\otimes J_{3}^{P_{3}}$   &Expressions & Results \\
			\hline
			{$\frac{1}{2}^{-}$}	&{{${^{2}S_{\frac{1}{2}}}$}}  &${\frac{1}{2}}^{-}\otimes0^{+}\otimes0^{+}$  &$\frac{1}{9} \left(4 \mu _{u}+8\mu _{d}+3\mu _{c}\right)$  &$0.134$ \\

			&& ${\frac{1}{2}}^{-}\otimes1^{+}\otimes0^{+}$ & $\frac{1}{27} \left(2 \mu _{u}+4 \mu _{d}+15\mu _{c}\right)$ &$ 0.224$ \\
			
			&& ${\frac{3}{2}}^{-}\otimes1^{+}\otimes0^{+}$ & $\frac{1}{27} \left(7\mu _{u}+14 \mu _{d}-24\mu _{c}\right)$ &$ -0.359$ \\
			\hline
			
			{$\frac{3}{2}^{-}$}  &{{${^{4}S_{\frac{3}{2}}}$}} & ${\frac{1}{2}}^{-}\otimes1^{-}\otimes0^{+}$ &$\frac{1}{9} \left(7\mu _{u}+14 \mu _{d}+12\mu _{c}\right)$  &$ 0.539$ \\
			
			&& ${\frac{3}{2}}^{-}\otimes0^{+}\otimes0^{+}$ & $\frac{1}{3} \left(2\mu _{u}+4\mu _{d}-3\mu _{c}\right)$ & $-0.403$ \\
			&& ${\frac{3}{2}}^{-}\otimes1^{-}\otimes0^{+}$ & $\frac{1}{45} \left(28\mu _{u}+56\mu _{d}-15\mu _{c}\right)$ &$ -0.134$ \\
			\hline
			{$\frac{5}{2}^{-}$}  &{{${^{6}S_{\frac{5}{2}}}$}} & ${\frac{3}{2}}^{-}\otimes1^{+}\otimes0^{+}$ &$\mu _{u}+2\mu _{d}$  & $0 $\\
			\hline
			{$\frac{1}{2}^{+}$}	&{{${^{2}P_{\frac{1}{2}}}$}}  &${\frac{1}{2}}^{-}\otimes0^{+}\otimes1^{-}$  &$\frac{1}{27} \left(-4\mu _{u}-8\mu _{d}-3\mu _{c}+12\mu _{l(\frac{1}{2}/0)}+6\mu _{l'(\frac{1}{2}/0)}\right)$  &$0.001$  \\
			
			&& $({\frac{1}{2}}^{-}\otimes1^{+})_{\frac{1}{2}}\otimes1^{-}$ & $\frac{1}{81} \left(- 2\mu _{u}- 4\mu _{d}-15\mu _{c}+36\mu _{l(\frac{1}{2}/1)}+18\mu _{l'(\frac{1}{2}/1)}\right)$  &$-0.059$  \\
			\cline{2-5}
			&{{${^{4}P_{\frac{1}{2}}}$}}  & $({\frac{1}{2}}^{-}\otimes1^{+})_{\frac{3}{2}}\otimes1^{-}$ &$\frac{1}{81} \left(35\mu _{u}+70\mu _{d}+60\mu _{c}-18\mu _{l(\frac{1}{2}/1)}-9\mu _{l'(\frac{1}{2}/1)}\right)$ &$ 0.191$ \\
			
			&& ${\frac{3}{2}}^{-}\otimes0^{+}\otimes1^{-}$ & $\frac{1}{27} \left(10\mu _{u}+20\mu _{d}-15\mu _{c}-6\mu _{l(\frac{3}{2}/0)}-3\mu _{l'(\frac{3}{2}/0)}\right)$  &$-0.247$  \\
			\cline{2-5}
			&{{${^{2}P_{\frac{1}{2}}}$}}  & $({\frac{3}{2}}^{-}\otimes1^{+})_{\frac{1}{2}}\otimes1^{-}$ &$\frac{1}{81} \left(-7\mu _{u}-14\mu _{d}+24\mu _{c}+36\mu _{l(\frac{3}{2}/1)}+18\mu _{l'(\frac{3}{2}/1)}\right)$ &$ 0.136$ \\
			&{{${^{4}P_{\frac{1}{2}}}$}}  & $({\frac{3}{2}}^{-}\otimes1^{+})_{\frac{3}{2}}\otimes1^{-}$ &$\frac{1}{81} \left(28\mu _{u}+56\mu _{d}-15\mu _{c}-18\mu _{l(\frac{3}{2}/1)}-9\mu _{l'(\frac{3}{2}/1)}\right)$ &$ -0.083$ \\
			\hline
			$ \frac{3}{2}^{+} $&{{${^{2}P_{\frac{3}{2}}}$}}  & ${\frac{1}{2}}^{-}\otimes0^{+}\otimes1^{-}$ &$\frac{1}{9} \left(4\mu _{u}+8\mu _{d}+3\mu _{c}+6\mu _{l(\frac{1}{2}/0)}+3\mu _{l'(\frac{1}{2}/0)}\right)$ &$ 0.162$ \\
			& & $({\frac{1}{2}}^{-}\otimes1^{+})_{\frac{1}{2}}\otimes1^{-}$ &$\frac{1}{27} \left(2\mu _{u}+4\mu _{d}+15\mu _{c}+18\mu _{l(\frac{1}{2}/1)}+9\mu _{l'(\frac{1}{2}/1)}\right)$ &$ 0.248$ \\
			\cline{2-5}
			&{{${^{4}P_{\frac{3}{2}}}$}}  & $({\frac{1}{2}}^{-}\otimes1^{+})_{\frac{3}{2}}\otimes1^{-}$ &$\frac{1}{135} \left(77\mu _{u}+154\mu _{d}+132\mu _{c}+36\mu _{l(\frac{1}{2}/1)}+18\mu _{l'(\frac{1}{2}/1)}\right)$ &$ 0.404$ \\
			& & ${\frac{3}{2}}^{-}\otimes0^{+}\otimes1^{-}$ &$\frac{1}{45} \left(22\mu _{u}+44\mu _{d}-33\mu _{c}+12\mu _{l(\frac{3}{2}/0)}+6\mu _{l'(\frac{3}{2}/0)}\right)$ & $-0.269$ \\
			\cline{2-5}
			&{{${^{2}P_{\frac{3}{2}}}$}}  & $({\frac{3}{2}}^{-}\otimes1^{+})_{\frac{1}{2}}\otimes1^{-}$ &$\frac{1}{27} \left(7\mu _{u}+14\mu _{d}-24\mu _{c}+18\mu _{l(\frac{3}{2}/1)}+9\mu _{l'(\frac{3}{2}/1)}\right)$ &$ -0.335$ \\
			
			&{{${^{4}P_{\frac{3}{2}}}$}}  & $({\frac{3}{2}}^{-}\otimes1^{+})_{\frac{3}{2}}\otimes1^{-}$ &$\frac{1}{675} \left(308\mu _{u}+616\mu _{d}-165\mu _{c}+180\mu _{l(\frac{3}{2}/1)}+90\mu _{l'(\frac{3}{2}/1)}\right)$ &$ -0.089$ \\
			&{{${^{6}P_{\frac{3}{2}}}$}}  & $({\frac{3}{2}}^{-}\otimes1^{+})_{\frac{5}{2}}\otimes1^{-}$ &$\frac{1}{75} \left(63\mu _{u}+126\mu _{d}-30\mu _{l(\frac{3}{2}/1)}-15\mu _{l'(\frac{3}{2}/1)}\right)$   &$ -0.014$ \\
			\hline
			$ \frac{5}{2}^{+} $&{{${^{4}P_{\frac{5}{2}}}$}}  & ${\frac{1}{2}}^{-}\otimes1^{+}\otimes1^{-}$ &$\frac{1}{9} \left(7\mu _{u}+14\mu _{d}+12\mu _{c}+6\mu _{l(\frac{1}{2}/1)}+3\mu _{l'(\frac{1}{2}/1)}\right)$ & $0.562$ \\
			
			& & ${\frac{3}{2}}^{-}\otimes0^{+}\otimes1^{-}$ &$\frac{1}{3} \left(2\mu _{u}+4\mu _{d}-3\mu _{c}+2\mu _{l(\frac{3}{2}/0)}+\mu _{l'(\frac{3}{2}/0)}\right)$ &$ -0.335$ \\
			&& $({\frac{3}{2}}^{-}\otimes1^{+})_{\frac{3}{2}}\otimes1^{-}$ &$\frac{1}{45} \left(28\mu _{u}+56\mu _{d}-15\mu _{c}+30\mu _{l(\frac{3}{2}/1)}+15\mu _{l'(\frac{3}{2}/1)}\right)$   &$ -0.110$ \\
			\cline{2-5}
			&{{${^{6}P_{\frac{5}{2}}}$}}  &$({\frac{3}{2}}^{-}\otimes1^{+})_{\frac{5}{2}}\otimes1^{-}$  &$\frac{1}{105} \left(93\mu _{u}+186\mu _{d}+20\mu _{l(\frac{3}{2}/1)}+10\mu _{l'(\frac{3}{2}/1)}\right)$ &$ 0.007$ \\
			\hline
			$ \frac{7}{2}^{+} $&{{${^{6}P_{\frac{7}{2}}}$}}  & ${\frac{3}{2}}^{-}\otimes1^{+}\otimes1^{-}$ &$\frac{1}{3} \left(3\mu _{u}+6\mu _{d}+2\mu _{l(\frac{3}{2}/1)}+\mu_{l'(\frac{3}{2}/1)}\right)$ &$0.024$ \\
			\bottomrule[1pt]
			\bottomrule[1pt]
		\end{tabular}
		
	\end{table*}

	\section{NUMERICAL ANALYSIS}
		\label{sec6}
The analysis of  magnetic moments and transition magnetic moments of pentaquark states is an effective method to explore their  inner structures. This is an important work for the later discovery of the $P^{N^{0}}_{\psi}  $ states and $P^{\Delta^{0}}_{\psi} $ states, and opens up another way for us to explore the exotic hadrons. In this work, we  systematically  calculate the magnetic moments of $P^{N^{0}}_{\psi}  $ states and $P^{\Delta^{0}}_{\psi} $ states in diquark-diquark-antiquark model and diquark-triquark model, and also calculate the magnetic moments and transition magnetic moments of $P^{N^{0}}_{\psi}  $  states and $P^{\Delta^{0}}_{\psi} $ states in molecular model. In this section, we will analyze the results of magnetic moments and transition magnetic moments respectively.

\subsection{Magnetic moments of the $P^{N^{0}}_{\psi}  $ and $P^{\Delta^{0}}_{\psi}$  states states in molecular model, diquark-diquark-antiquark model and diquark-triquark model}

 We calculate the magnetic moments of  the $P^{N^{0}}_{\psi}  $ states and $P^{\Delta^{0}}_{\psi} $ states in three models.   On the basis of the magnetic moment results obtained in this work,  we summarized the following key points.

\begin{figure*}
	\centering
	\includegraphics[width=1.03\linewidth, height=0.45\textheight]{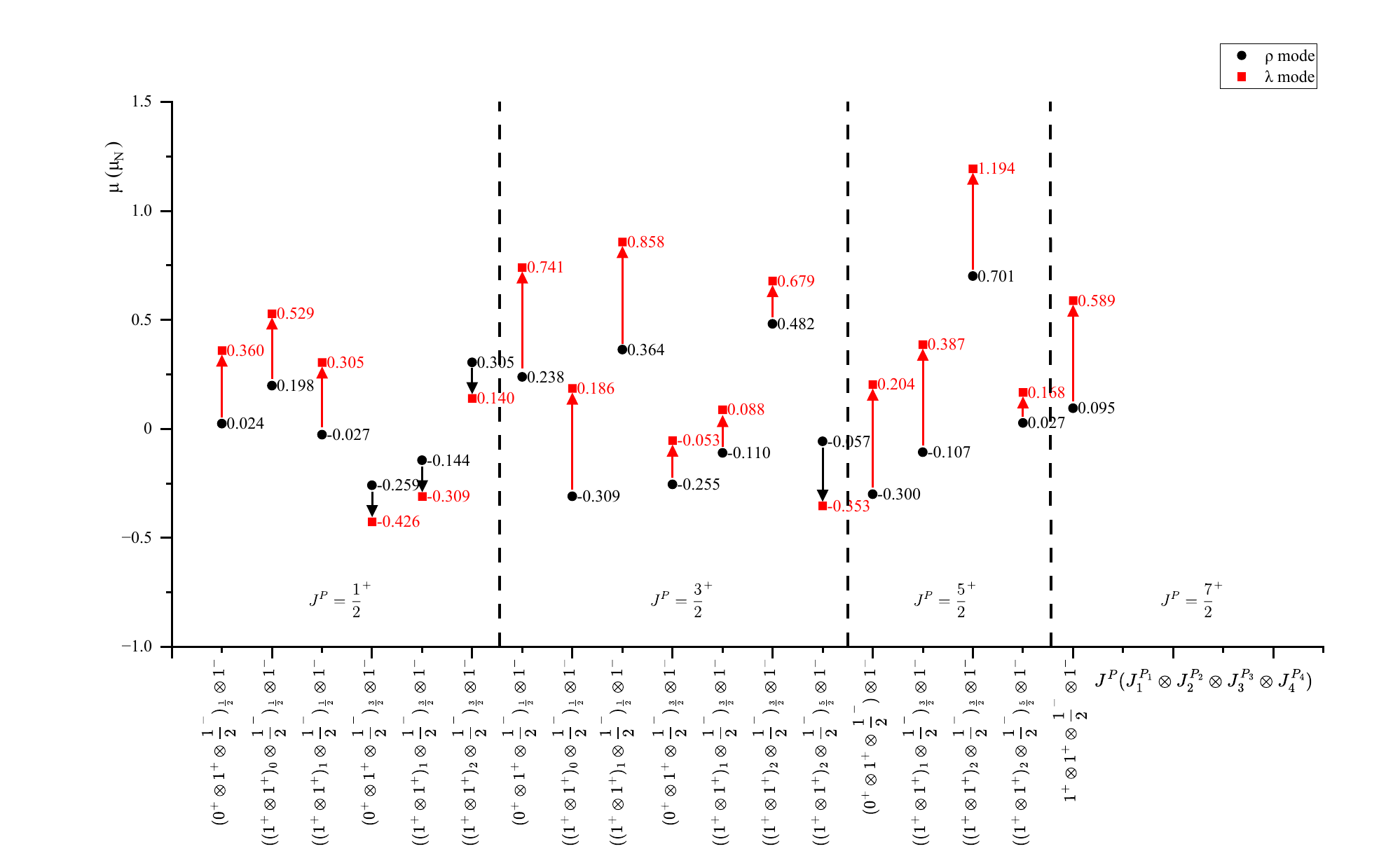}
	\caption{The excited state magnetic moments of the $P^{\Delta^{0}}_{\psi}  $ states in the diquark-diquark-antiquark model with the $ 10_{f} $ flavor representation. Here,  the red  arrow represents the magnetic moments of the $ \lambda $ excitation state are larger than the magnetic moments of the $ \rho $ excitation state, and the black  arrow represents the magnetic moments of the $ \rho $ excitation state are  larger than the magnetic moments of the $ \lambda $ excitation state.}
	\label{fig:10}
\end{figure*}

\begin{figure*}
	\centering
	\includegraphics[width=1.03\linewidth, height=0.45\textheight]{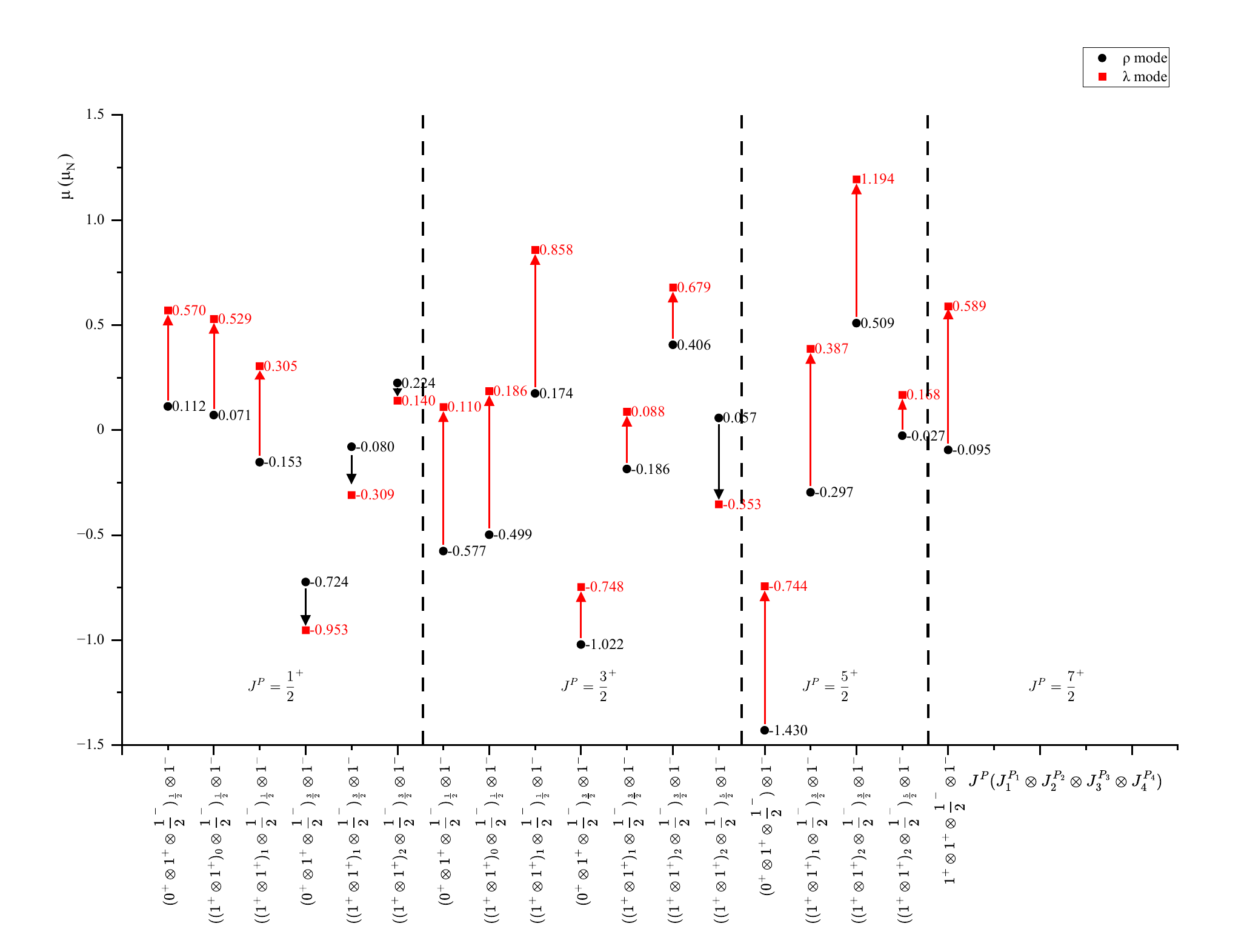}
	\caption{The excited state magnetic moments of the $P^{N^{0}}_{\psi} $ states in the diquark-diquark-antiquark model with the $ 8_{1f} $ flavor representation. Here,  the red  arrow represents the magnetic moments of the $ \lambda $ excitation state are larger than the magnetic moments of the $ \rho $ excitation state, and the black  arrow represents the magnetic moments of the $ \rho $ excitation state are  larger than the magnetic moments of the $ \lambda $ excitation state.}
	\label{fig:81}
\end{figure*}

\begin{figure*}
	\centering
	\includegraphics[width=1.03\linewidth, height=0.45\textheight]{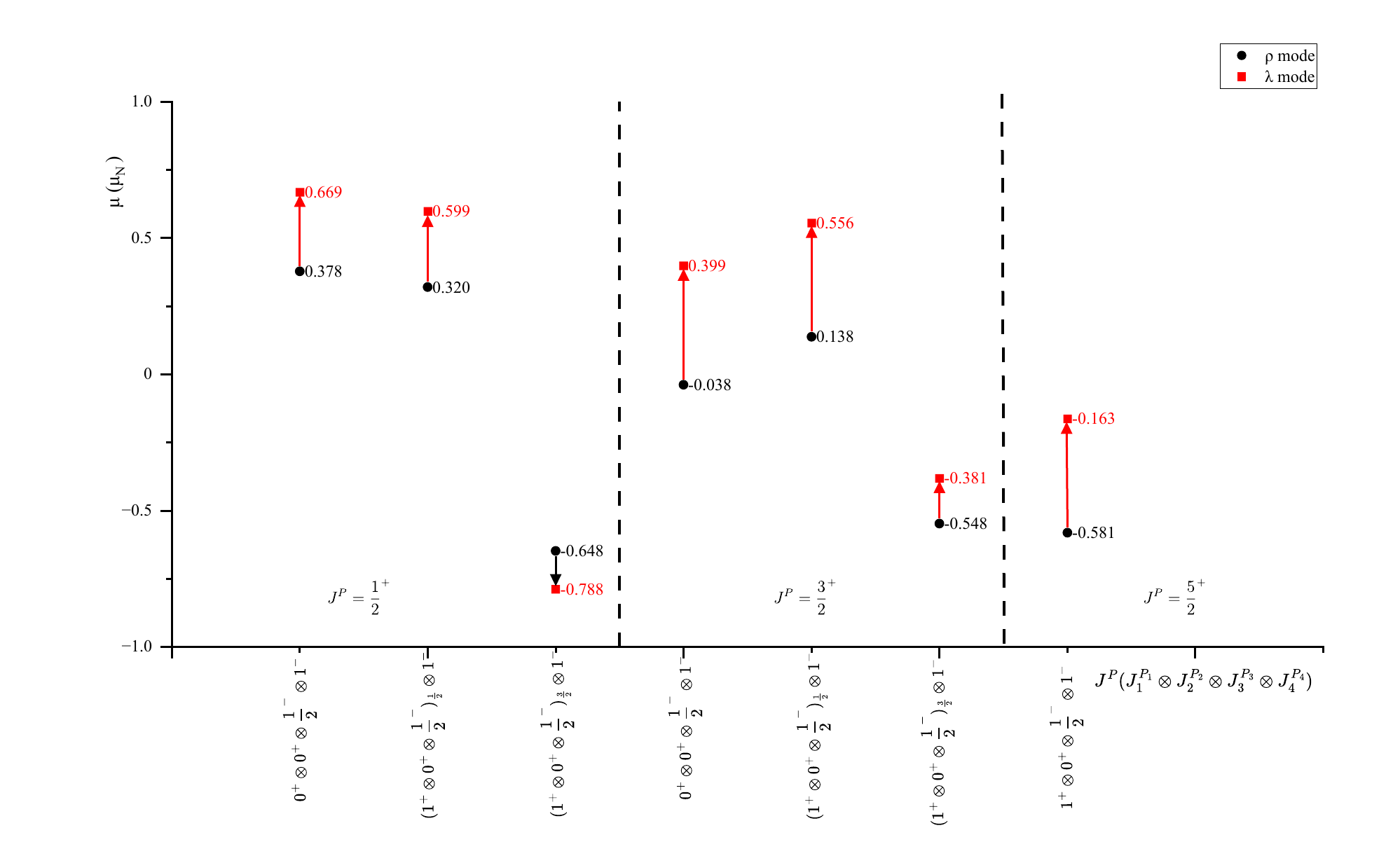}
	\caption{The excited state magnetic moments of the $P^{N^{0}}_{\psi} $ states in the diquark-diquark-antiquark model with the $ 8_{2f} $ flavor representation. Here,  the red  arrow represents the magnetic moments of the $ \lambda $ excitation state are larger than the magnetic moments of the $ \rho $ excitation state, and the black  arrow represents the magnetic moments of the $ \rho $ excitation state are  larger than the magnetic moments of the $ \lambda $ excitation state.}
	\label{fig:82}
\end{figure*}

• In the diquark-diquark-antiquark model, the $ \rho $ mode and the $ \lambda $ mode in the $ P $-wave excitation will lead to obvious difference in magnetic moment. The magnetic moments  in Table \ref{tab:lb} -Table \ref{tab:ls} indicate that the magnetic moments of the $ \lambda $ excitation state are usually larger than the magnetic moments of the $ \rho $ excitation state. This is caused by the orbital magnetic moment $ \mu_{l\lambda}>\mu_{l\rho} $. In order to facilitate analysis and comparison, we have shown the excited state magnetic moment results of these two excitation modes in Fig. \ref{fig:10} , Fig. \ref{fig:81} and Fig. \ref{fig:82}.

 In Fig. \ref{fig:10} and Fig. \ref{fig:81} ,we find that in the interval of quantum number $ J^{P}=\frac{1}{2}^{+} $ and $ J^{P}=\frac{3}{2}^{+} $, there are a few magnetic moments of the $ \rho $ excitation state are larger than magnetic moments  of the $ \lambda $ excitation state. Their spin configurations are 
	\begin{eqnarray} 
	J^{P}=\frac{1}{2}^{+} :	&&(0^{+}\otimes1^{+}\otimes {\frac{1}{2}}^{-})_{\frac{3}{2}}\otimes1^{-},\nonumber\\
	&&((1^{+}\otimes1^{+})_{1}\otimes{\frac{1}{2}}^{-})_{\frac{3}{2}}\otimes1^{-},\nonumber\\
&&((1^{+}\otimes1^{+})_{2}\otimes{\frac{1}{2}}^{-})_{\frac{3}{2}}\otimes1^{-}.\\
J^{P}=\frac{3}{2}^{+} :&&((1^{+}\otimes1^{+})_{2}\otimes{\frac{1}{2}}^{-})_{\frac{5}{2}}\otimes1^{-}.
\end{eqnarray}
In the interval of quantum number $ J^{P}=\frac{5}{2}^{+} $ and $ J^{P}=\frac{7}{2}^{+} $, the magnetic moments of the $ \lambda $ excitation state are all larger than the magnetic moments of the $ \rho $ excitation state. 

In Fig. \ref{fig:82}, we find that except for the interval of quantum number $ J^{P}=\frac{1}{2}^{+} $, there is a group of the magnetic moments of the $ \rho $  excitation state are  larger than the magnetic moments of the $ \lambda $ excitation state, other the magnetic moments of the $ \lambda $ excitation state are larger than the magnetic moments of the $ \rho $ excitation state. 
Its spin configuration is
\begin{eqnarray} 
	J^{P}=\frac{1}{2}^{+} :&&(1^{+}\otimes0^{+}\otimes{\frac{1}{2}}^{-})_{\frac{3}{2}}\otimes1^{-}.
\end{eqnarray}

In other words, with the increase of the quantum number $ J^{P} $, the phenomenon that the magnetic moments of the $ \lambda $ excitation state are larger than the magnetic moments of the $ \rho $ excitation state becomes more obvious.

•  In Table \ref{tab:rb}, the magnetic moment expressions of  $ S $-wave $P^{N^{0}}_{\psi} $ states with  $ 8_{1f} $ flavor representation  and $ S $-wave $P^{\Delta^{0}}_{\psi} $ states with $ 10_{f}$ flavor representation are equal whose spin configurations satisfing $J_{1}^{P_{1}}\otimes J_{2}^{P_{2}}\otimes J_{3}^{P_{3}}\otimes J_{4}^{P_{4}} = 1^{+}\otimes1^{+}\otimes{\frac{1}{2}}^{-}\otimes0^{+} $.
In Table \ref{tab:lbb} and Table \ref{tab:ls}, the magnetic moment expressions of  $ \lambda $ excitation $P^{N^{0}}_{\psi} $ states with $ 8_{1f} $ flavor representation  and $ \lambda $ excitation $P^{\Delta^{0}}_{\psi} $ states  with $ 10_{f} $  flavor representation are equal whose spin configurations satisfing $J_{1}^{P_{1}}\otimes J_{2}^{P_{2}}\otimes J_{3}^{P_{3}}\otimes J_{4}^{P_{4}} = 1^{+}\otimes1^{+}\otimes{\frac{1}{2}}^{-}\otimes1^{-} $. 
In the  magnetic moment expression of $ \lambda $ excitation $P^{N^{0}}_{\psi} $ states and $P^{\Delta^{0}}_{\psi} $ states, the orbital magnetic moment $\mu _{l(1\otimes1/\frac{1}{2})}=\mu _{l^{'}(1\otimes1/\frac{1}{2})}$, when we assume exact isospin symmetry with $ m_{u}=m_{d} $.
For example, in Table \ref{tab:rb}, the hidden-charm pentaquark states   wave function $\sqrt{\frac{1}{3}}({c}d)\{ud\}{\bar c}-\sqrt{\frac{2}{3}}({c}u)\{dd\}{\bar c}$ and $\sqrt{\frac{2}{3}}({c}d)\{ud\}{\bar c}+\sqrt{\frac{1}{3}}({c}u)\{dd\}{\bar c}$  with $J^{P}=\frac{3}{2}^{-}  $ has spin configurations $(1^{+}\otimes1^{+})_{2} \otimes {\frac{1}{2}}^{-}\otimes0^{+}$, and their magnetic moments are all $0.605\mu_{N}$.    In Table \ref{tab:lbb} and Table \ref{tab:ls},  the magnetic moments of the hidden-charm pentaquark states with $ J^{P}=\frac{7}{2}^{+} $ are both $0.589\mu_{N}$, and their corresponding spin configurations are $1^{+}\otimes1^{+}\otimes{\frac{1}{2}}^{-}\otimes1^{-}$.

 • In Table \ref{tab:ms},  the hidden-charm pentaquark states wave function  $ \sqrt{\frac{2}{3}}\left|\Sigma_c^{+} {D}^{*-}\right\rangle+\sqrt{\frac{1}{3}}\left|\Sigma_c^{0}\bar{D}^{*0}\right\rangle$ with $J^{P}=\frac{1}{2}^{+}  $ has spin configurations $({\frac{1}{2}}^{+}\otimes1^{-})_{\frac{1}{2}}\otimes1^{-}$ and $({\frac{1}{2}}^{+}\otimes1^{-})_{\frac{3}{2}}\otimes1^{-}$ in the $ P $-wave excitation. Their magnetic moments are  $ 0.037\mu_{N} $ and  $ -0.280\mu_{N} $, respectively. The magnetic moments of the same quantum numbers and quark configurations of the pentaquark states exist obvious differences, indicating that the magnetic moments can indeed distinguish the  inner structures of the pentaquark states.

  • When the $ S $-wave pentaquark has the highest spin $S= \frac{5}{2} $, the three models give the same magnetic moment $\mu_u+2\mu_d$. If we use $m_u=m_d$, one gets a vanishing number. The reason for the same result in different models is that  magnetic moment is only related with spin structure (of course it is affected by the flavor and color structures). The highest-spin case means that any two quarks are in the $ S=1 $ state. In the considered three models, $J^P=\frac52^-$ states exist in $8_f$ and $10_f$ flavor representations, which means that at least one $qq$ is in the $I=1$, $S=1$, and $\bar{3}_c$ state. To form a colorless pentaquark, $cq$ may be both $6_c$ and $\bar{3}_c$, but only $\bar{3}_c$ is possible from the model assumptions. Therefore, the color-spin structures for the highest-spin pentaquark in the three models are the same and thus the resulting magnetic moments are the same.

\subsection{Transition magnetic moments of the $P^{N^{0}}_{\psi}  $ states and $P^{\Delta^{0}}_{\psi}$  states in the molecular model}

We study the transition magnetic moment of the $ P_{\psi}^{N^{0}} $ states and $P^{\Delta^{0}}_{\psi}$  states in the molecular model.  Comparing the data, we obtain the following  interesting points.

•  There is an obvious rule between the expressions of transition magnetic moments with the $ 8_{1f} $ and $ 10_{f} $ flavor representation.
 In Table \ref{tab:mby} and Table \ref{tab:msy}, the  transition magnetic moment expression of $\Sigma_c^{} \bar{D}^{(*)}|\frac{1}{2}^-\rangle \to\Sigma_c^{} \bar{D}^{}|\frac{1}{2}^-\rangle\gamma$ process are in the form of  $k(\mu_{D^{*-} \to  D^{-}}+2\mu_{\bar{D}^{*0} \to \bar{D}^{0}})$ with $ 8_{1f}$ flavor representation and $k(2\mu_{D^{*-} \to  D^{-}}+\mu_{\bar{D}^{*0} \to \bar{D}^{0}})$ with $ 10_{f} $ flavor representation. Here, $ k $ is a real number.

• The transition magnetic moment of hidden-charm molecular pentaquark can be expressed as a linear combination of transition magnetic moments or magnetic moments.  In Table  \ref{tab:mby}, the transition magnetic moment of the $\Sigma_c^{} \bar{D}^{*}|\frac{1}{2}^-\rangle \to\Sigma_c^{} \bar{D}^{}|\frac{1}{2}^-\rangle\gamma$ process is related to the transition magnetic moment of the  $ \mu_{D^{*-} \to  D^{-}} $   and $ \mu_{\bar{D}^{*0} \to \bar{D}^{0}} $. The transition magnetic moment of the 	$\Sigma_c^{*} \bar{D}^{*}|\frac{5}{2}^-\rangle \to\Sigma_c^{*} \bar{D}^{*}|\frac{3}{2}^-\rangle\gamma$ process can be expressed as a linear combination of magnetic moments ${\mu}_{{\Sigma}_{c}^{*+}}  $, ${\mu}_{{\Sigma}_{c}^{*0}}  $, $ {\mu}_{{D}^{*-}} $ and ${\mu}_{\bar{D}^{*0}}  $.

• The transition process is different due to different hadron states corresponding to the flavor wave functions.  For example, Table  \ref{tab:mby} lists the transition magnetic moments of $S$-wave  $\Sigma_c^{(*)} \bar{D}^{(*)}$-type hidden-charm molecular pentaquarks  between $8_{1f}$ states, Table  \ref{tab:mbey} lists the transition magnetic moments of $S$-wave  $\Lambda_c^{} {D}^{(*)}$-type hidden-charm molecular pentaquarks between $8_{2f}$ states.

\section{summary}
\label{sec7}
In recent years, the study of pentaquark states has made continuous breakthroughs. Inspired by the discovery of the $P^{N^{+}}_{\psi}  $ states, we believe that the discovery of the $P^{N^{0}}_{\psi}  $ states and $P^{\Delta^{0}}_{\psi} $ states  is inevitable over time.  At present, the study of strange hadron states has attracted extensive attention in both experiment and theory, but the study of their magnetic moments and transition magnetic moments has not received enough attention. Magnetic moment is an inherent attribute of particles. The magnetic moment and transition magnetic moment can provide very useful clues for studying the internal structure of strange hadrons. 

In this work,  we construct the flavor wave functions of the $P^{N^{0}}_{\psi}  $ states and $P^{\Delta^{0}}_{\psi}$  states in the molecular model, the diquark-diquark-antiquark model and the diquark-triquark, and discuss their color configurations.  We  systematically calculate the magnetic moments and transition magnetic moments of the $P^{N^{0}}_{\psi}  $ states and $P^{\Delta^{0}}_{\psi}$  states, and simply calculated the
magnetic moments and transition magnetic moments of $P^{\Delta^{++}}_{\psi}  $ states and $P^{\Delta^{-}}_{\psi}  $ states  in the molecular model that have not yet been explored (see the Appendix \ref{1}).    The results clearly show that the flavor-spin wave function fundamentally determines the magnetic moment and the transition magnetic moment. The  flavor-spin wave functions of different initial and final states determines the difference of transition magnetic moment. The same spin configuration has different magnetic moments in different models, and the different spin configuration has different magnetic moments in same models. The different flavor-spin compositions of pentaquark states contain important information about their internal structure. Therefore, this work will provide important data support for us to explore the  inner structures of the hidden-charm pentaquark states. At present, for the calculation of transition magnetic moments, we only consider molecular model. The molecular model is the mainstream model for studying the transition magnetic moment of pentaquark states, which includes two parts: baryons and mesons, making it easy to visually analyze its results. Calculating the transition magnetic moments of other models requires recombining the relevant flavor-spin wave function, which will be discussed in our future work.

\section*{ACKNOWLEDGMENTS}
	This project is supported by the National Natural Science Foundation of China under Grants No. 11905171, No. 12047502 and No. 12247103. This work is also supported by the
	Natural Science Basic Research Plan in Shaanxi Province
	of China (Grant No. 2022JQ-025) and Shaanxi Fundamental Science Research Project for Mathematics and Physics (Grant No.22JSQ016).

	\appendix
	\section{ MAGNETIC MOMENTS AND TRANSITION MAGNETIC MOMENTS OF THE $P^{\Delta^{++}}_{\psi}  $ STATES AND THE $P^{\Delta^{-}}_{\psi}  $ STATES IN THE MOLECULAR MODEL }
	\label{1}
	In this section, we briefly supplement the unexplored magnetic moments and transition magnetic moments of  $P^{\Delta^{++}}_{\psi}  $  states and  $P^{\Delta^{-}}_{\psi}  $ states in the molecular model.
	The flavor  wave function of  $P^{\Delta^{++}}_{\psi}  $ states and  $P^{\Delta^{-}}_{\psi}  $ states in the molecular model is  $ 10_{f} $  flavor representation.
	
	\begin{eqnarray} 
		P^{\Delta^{++}}_{\psi} :	&&\left|\Sigma_c^{(*)++} \bar{D}^{(*)0}\right\rangle,\\
		P^{\Delta^{-}}_{\psi} :	&&\left|\Sigma_c^{(*)0} D^{(*)-}\right\rangle.
	\end{eqnarray}
The spin wave functions of $P^{\Delta^{++}}_{\psi}  $  states and  $P^{\Delta^{-}}_{\psi}  $ states with the same spin configuration are the same, and here we list the spin wave functions of $P^{\Delta^{++}}_{\psi}  $  states.
	\begin{align} 
	&\left|S,S_{z}\right\rangle=\nonumber\\
	&\sum_{S'_{\Sigma_c^{(*)}},S'_{\bar{D}^{(*)}}}C_{S_{\Sigma_c^{(*)}}S'_{\Sigma_c^{(*)}}
		,S_{\bar{D}^{(*)}}S'_{\bar{D}^{(*)}}}^{S,S_{z}}\left|S_{\Sigma_c^{(*)}},S'_{\Sigma_c^{(*)}}\right\rangle
	\left|S_{\bar{D}^{(*)}},S'_{\bar{D}^{(*)}}\right\rangle.\label{equ:qb3}
\end{align}
Here, $ S $ and $ S_{z} $ represent the spin and the spin third components of $P^{\Delta^{++}}_{\psi}  $  states. The $ S_{\Sigma_c^{(*)}} $, $ S_{\bar{D}^{(*)}} $ are spin of $ \Sigma_c^{(*)} $ and $\bar{D}^{(*)}  $, respectively. The $ S'_{\Sigma_c^{(*)}} $, $ S'_{\bar{D}^{(*)}} $ are spin third component of $ \Sigma_c^{(*)} $ and $\bar{D}^{(*)}  $, respectively.  

	The magnetic moments and transition magnetic moments of $P^{\Delta^{++}}_{\psi}  $ states in the molecular model are presented in Table \ref{tab:ncj} and Table \ref{tab:nycj}, respectively.  The transition magnetic moments of the ${\Sigma_c^{*}{\bar{D^{*}}}} |^{2} S_{1 / 2}\rangle \to{\Sigma_c{\bar{D^{*}}}} |^{2} S_{1 / 2}\rangle_{\gamma}  $ and ${\Sigma_c^{*}{\bar{D^{*}}}} |^{6} S_{5 / 2}\rangle \to{\Sigma_c^{*}{\bar{D^{*}}}} |^{2} S_{1 / 2}\rangle_{\gamma}  $ processes in $P^{\Delta^{++}}_{\psi}  $ states are zero. The magnetic moments and transition magnetic moments of $P^{\Delta^{-}}_{\psi}  $ states in the molecular model are presented in Table \ref{tab:j} and Table \ref{tab:yj}, respectively. The transition magnetic moments of the ${\Sigma_c^{*}{D^{*}}} |^{2} S_{1 / 2}\rangle \to{\Sigma_c{D^{*}}} |^{2} S_{1 / 2}\rangle_{\gamma}  $ and ${\Sigma_c^{*}{D^{*}}} |^{6} S_{5 / 2}\rangle \to{\Sigma_c^{*}{D^{*}}} |^{2} S_{1 / 2}\rangle_{\gamma}  $ processes in $P^{\Delta^{-}}_{\psi}  $ states are zero.

	\renewcommand\tabcolsep{0.88cm}
	\renewcommand{\arraystretch}{1.65}
	\begin{table*}[htbp]
		\caption{ The magnetic moments of the $P^{\Delta^{++}}_{\psi}  $ states in the molecular model with the $ 10_{f} $  flavor representation.   The  $J_{1}^{P_{1}}\otimes J_{2}^{P_{2}}\otimes J_{3}^{P_{3}}$ are corresponding to the angular momentum and parity of baryon, meson and orbital, respectively.  The $ \mu_{l(baryon/ meson)} $ represents the orbital excitation between corresponding hadrons. The unit is the nuclear magnetic moment $ \mu_{N} $.}
		\scriptsize
		\label{tab:ncj} 
		\begin{tabular}{c|c|c|c|c}
			\toprule[1pt]
			\toprule[1pt]

			\multicolumn{5}{c}{$ \left|\Sigma_c^{(*)++} \bar{D}^{(*)0}\right\rangle$}\\
			\toprule[1pt]
			$J^P$	& $^{2s+1}L_J$ &  $J_{1}^{P_{1}}\otimes J_{2}^{P_{2}}\otimes J_{3}^{P_{3}}$  &Expressions& Results \\
			\hline
			{$\frac{1}{2}^{-}$}	&{{${^{2}S_{\frac{1}{2}}}$}}  &${\frac{1}{2}}^{+}\otimes0^{-}\otimes0^{+}$  &$\frac{1}{3} \left(4 \mu _{u}-\mu _{c}\right)$  &$  2.392$ \\

			&& ${\frac{1}{2}}^{+}\otimes1^{-}\otimes0^{+}$ & $\frac{1}{9} \left(2 \mu _{u}-5\mu _{c}\right)$ &$ 0.197$ \\
			
			&& ${\frac{3}{2}}^{+}\otimes1^{-}\otimes0^{+}$ & $\frac{1}{9} \left(7 \mu _{u}+8\mu _{c}\right)$ & $1.832$ \\
			\hline

			{$\frac{3}{2}^{-}$}  &{{${^{4}S_{\frac{3}{2}}}$}} & ${\frac{1}{2}}^{+}\otimes1^{-}\otimes0^{+}$ &$\frac{1}{3} \left(7 \mu _{u}-4\mu _{c}\right)$  & $3.884$ \\
			&& ${\frac{3}{2}}^{+}\otimes0^{-}\otimes0^{+}$ & $2 \mu _{u}+\mu _{c}$ &$ 4.193$ \\
			&& ${\frac{3}{2}}^{+}\otimes1^{-}\otimes0^{+}$ & $\frac{1}{15} \left(28 \mu _{u}+5\mu _{c}\right)$ &$ 3.672$ \\
			\hline
			{$\frac{5}{2}^{-}$}  &{{${^{6}S_{\frac{5}{2}}}$}} &  ${\frac{3}{2}}^{+}\otimes1^{-}\otimes0^{+}$ &$3\mu _{u}$  &$ 5.685$ \\
			\hline

			{$\frac{1}{2}^{+}$}	&{{${^{2}P_{\frac{1}{2}}}$}}  &${\frac{1}{2}}^{+}\otimes0^{-}\otimes1^{-}$  &$\frac{1}{9} \left(-4\mu _{u}+\mu _{c}+6\mu_{  l(\Sigma_c^{++}/{\bar{D}}^{0})  }\right)$  &$-0.577$  \\
			
			&& $({\frac{1}{2}}^{+}\otimes1^{-})_{\frac{1}{2}}\otimes1^{-}$ & $\frac{1}{27} \left(- 2\mu _{u}+5\mu _{c}+18\mu_{  l(\Sigma_c^{++}/\bar{D}^{*0})  }\right)$  &$0.164$  \\
			\cline{2-5}
			&{{${^{4}P_{\frac{1}{2}}}$}}  & $({\frac{1}{2}}^{+}\otimes1^{-})_{\frac{3}{2}}\otimes1^{-}$ &$\frac{1}{27} \left(35\mu _{u}-20\mu _{c}-9\mu_{  l(\Sigma_c^{++}/\bar{D}^{*0})  }\right)$ & $2.043$ \\
			&& ${\frac{3}{2}}^{+}\otimes0^{-}\otimes1^{-}$ & $\frac{1}{9} \left(10\mu _{u}+5\mu _{c}-3\mu_{  l(\Sigma_c^{*++}/\bar{D}^{0})  }\right)$  &$2.224$  \\
			\cline{2-5}
			&{{${^{2}P_{\frac{1}{2}}}$}}  & $({\frac{3}{2}}^{+}\otimes1^{-})_{\frac{1}{2}}\otimes1^{-}$ &$\frac{1}{27} \left(-7\mu _{u}-8\mu _{c}+18\mu_{  l(\Sigma_c^{*++}/\bar{D}^{*0})  }\right)$ &$ -0.391$ \\
			&{{${^{4}P_{\frac{1}{2}}}$}}  & $({\frac{3}{2}}^{+}\otimes1^{-})_{\frac{3}{2}}\otimes1^{-}$ &$\frac{1}{27} \left(28\mu _{u}+5\mu _{c}-9\mu_{  l(\Sigma_c^{*++}/\bar{D}^{*0})  }\right)$ &$1.930$ \\
			\hline
			$ \frac{3}{2}^{+} $&{{${^{2}P_{\frac{3}{2}}}$}}  & ${\frac{1}{2}}^{+}\otimes0^{-}\otimes1^{-}$ &$\frac{1}{3} \left(4\mu _{u}-\mu _{c}+3\mu_{  l(\Sigma_c^{++}/\bar{D}^{0})  }\right)$ &$ 2.722$ \\
			& & $({\frac{1}{2}}^{+}\otimes1^{-})_{\frac{1}{2}}\otimes1^{-}$ &$\frac{1}{9} \left(2\mu _{u}-5\mu _{c}+9\mu_{  l(\Sigma_c^{++}/\bar{D}^{*0})  }\right)$ & $0.541$ \\
			\cline{2-5}
			&{{${^{4}P_{\frac{3}{2}}}$}}  & $({\frac{1}{2}}^{+}\otimes1^{-})_{\frac{3}{2}}\otimes1^{-}$ &$\frac{1}{45} \left(77\mu _{u}-44\mu _{c}+18\mu_{  l(\Sigma_c^{++}/\bar{D}^{*0})  }\right)$ &$ 2.986$ \\
			& & ${\frac{3}{2}}^{+}\otimes0^{-}\otimes1^{-}$ &$\frac{1}{15} \left(22\mu _{u}+11\mu _{c}+6\mu_{  l(\Sigma_c^{*++}/\bar{D}^{0})  }\right)$ &$ 3.202$ \\
			\cline{2-5}
			&{{${^{2}P_{\frac{3}{2}}}$}}  & $({\frac{3}{2}}^{+}\otimes1^{-})_{\frac{1}{2}}\otimes1^{-}$ &$\frac{1}{9} \left(7\mu _{u}+8\mu _{c}+9\mu_{  l(\Sigma_c^{*++}/\bar{D}^{*0})  }\right)$ & $2.163$ \\
			&{{${^{4}P_{\frac{3}{2}}}$}}  & $({\frac{3}{2}}^{+}\otimes1^{-})_{\frac{3}{2}}\otimes1^{-}$ &$\frac{1}{225} \left(308\mu _{u}+55\mu _{c}+90\mu_{  l(\Sigma_c^{*++}/\bar{D}^{*0})  }\right)$ &$ 2.825$ \\
			&{{${^{6}P_{\frac{3}{2}}}$}}  & $({\frac{3}{2}}^{+}\otimes1^{-})_{\frac{5}{2}}\otimes1^{-}$ &$\frac{1}{25} \left(63\mu _{u}-15\mu_{  l(\Sigma_c^{*++}/\bar{D}^{*0})  }\right)$   &$ 4.577$ \\
			\hline
			$ \frac{5}{2}^{+} $&{{${^{4}P_{\frac{5}{2}}}$}}  & ${\frac{1}{2}}^{+}\otimes1^{-}\otimes1^{-}$ &$\frac{1}{3} \left(7\mu _{u}-4\mu _{c}+3\mu_{  l(\Sigma_c^{++}/\bar{D}^{*0})  }\right)$ & $4.228$ \\
			& & ${\frac{3}{2}}^{+}\otimes0^{-}\otimes1^{-}$ &$2\mu _{u}+\mu _{c}+\mu_{  l(\Sigma_c^{*++}/\bar{D}^{0})  }$ &$4.510 $ \\
			&& $({\frac{3}{2}}^{+}\otimes1^{-})_{\frac{3}{2}}\otimes1^{-}$ &$\frac{1}{15} \left(28\mu _{u}+5\mu _{c}+15\mu_{  l(\Sigma_c^{*++}/\bar{D}^{*0})  }\right)$   &$ 4.002$ \\
			\cline{2-5}
			&{{${^{6}P_{\frac{5}{2}}}$}}  &$({\frac{3}{2}}^{+}\otimes1^{-})_{\frac{5}{2}}\otimes1^{-}$  &$\frac{1}{35} \left(93\mu _{u}+10\mu_{  l(\Sigma_c^{*++}/\bar{D}^{*0})  }\right)$ &$ 5.130$ \\
			\hline
			$ \frac{7}{2}^{+} $&{{${^{6}P_{\frac{7}{2}}}$}}  & ${\frac{3}{2}}^{+}\otimes1^{-}\otimes1^{-}$ &$3\mu _{u}+\mu_{  l(\Sigma_c^{*++}/\bar{D}^{*0})  }$ &$6.015$ \\
			\bottomrule[1pt]
			\bottomrule[1pt]
		\end{tabular}
	\end{table*}

	\renewcommand\tabcolsep{0.88cm}
	\renewcommand{\arraystretch}{1.65}
	\begin{table*}[htbp]
		\caption{ The magnetic moments of the $P^{\Delta^{-}}_{\psi}  $ states in the molecular model with the $ 10_{f} $  flavor representation.   The  $J_{1}^{P_{1}}\otimes J_{2}^{P_{2}}\otimes J_{3}^{P_{3}}$ are corresponding to the angular momentum and parity of baryon, meson and orbital, respectively.  The $ \mu_{l(baryon/ meson)} $ represents the orbital excitation between corresponding hadrons. The unit is the nuclear magnetic moment $ \mu_{N} $.}
		\scriptsize
		\label{tab:j} 
		\begin{tabular}{c|c|c|c|c}
			\toprule[1pt]
			\toprule[1pt]

			\multicolumn{5}{c}{$ \left|\Sigma_c^{(*)0} D^{(*)-}\right\rangle$}\\
			\toprule[1pt]
			$J^P$	& $^{2s+1}L_J$ &  $J_{1}^{P_{1}}\otimes J_{2}^{P_{2}}\otimes J_{3}^{P_{3}}$  &Expressions& Results \\
			\hline
			{$\frac{1}{2}^{-}$}	&{{${^{2}S_{\frac{1}{2}}}$}}  &${\frac{1}{2}}^{+}\otimes0^{-}\otimes0^{+}$  &$\frac{1}{3} \left(4 \mu _{d}-\mu _{c}\right)$  &$  -1.398$ \\

			&& ${\frac{1}{2}}^{+}\otimes1^{-}\otimes0^{+}$ & $\frac{1}{9} \left(2 \mu _{d}-5\mu _{c}\right)$ &$ -0.435$ \\
			
			&& ${\frac{3}{2}}^{+}\otimes1^{-}\otimes0^{+}$ & $\frac{1}{9} \left(7 \mu _{d}-8\mu _{c}\right)$ & $-0.378$ \\
			\hline

			{$\frac{3}{2}^{-}$}  &{{${^{4}S_{\frac{3}{2}}}$}} & ${\frac{1}{2}}^{+}\otimes1^{-}\otimes0^{+}$ &$\frac{1}{3} \left(7 \mu _{d}-4\mu _{c}\right)$  & $-2.749$ \\
			&& ${\frac{3}{2}}^{+}\otimes0^{-}\otimes0^{+}$ & $2 \mu _{d}+\mu _{c}$ &$-1.492 $ \\
			&& ${\frac{3}{2}}^{+}\otimes1^{-}\otimes0^{+}$ & $\frac{1}{15} \left(28 \mu _{d}+5\mu _{c}\right)$ &$-1.634 $ \\
			\hline
			{$\frac{5}{2}^{-}$}  &{{${^{6}S_{\frac{5}{2}}}$}}& ${\frac{3}{2}}^{+}\otimes1^{-}\otimes0^{+}$ & $3 \mu _{d}$   &$ -2.842$ \\
			\hline

			{$\frac{1}{2}^{+}$}	&{{${^{2}P_{\frac{1}{2}}}$}}  &${\frac{1}{2}}^{+}\otimes0^{-}\otimes1^{-}$  &$\frac{1}{9} \left(-4\mu _{d}+\mu _{c}+6\mu_{  l(\Sigma_c^{0}/{D}^{-})  }\right)$  &$0.276$  \\
			
			&& $({\frac{1}{2}}^{+}\otimes1^{-})_{\frac{1}{2}}\otimes1^{-}$ & $\frac{1}{27} \left(- 2\mu _{d}+5\mu _{c}+18\mu_{  l(\Sigma_c^{0}/D^{*-})  }\right)$  &$0.316$  \\
			\cline{2-5}
			&{{${^{4}P_{\frac{1}{2}}}$}}  & $({\frac{1}{2}}^{+}\otimes1^{-})_{\frac{3}{2}}\otimes1^{-}$ &$\frac{1}{27} \left(35\mu _{d}-20\mu _{c}-9\mu_{  l(\Sigma_c^{0}/D^{*-})  }\right)$ & $-1.442$ \\
			&& ${\frac{3}{2}}^{+}\otimes0^{-}\otimes1^{-}$ & $\frac{1}{9} \left(10\mu _{d}+5\mu _{c}-3\mu_{  l(\Sigma_c^{*0}/D^{-})  }\right)$  &$-0.733$  \\
			\cline{2-5}
			&{{${^{2}P_{\frac{1}{2}}}$}}  & $({\frac{3}{2}}^{+}\otimes1^{-})_{\frac{1}{2}}\otimes1^{-}$ &$\frac{1}{27} \left(-7\mu _{d}-8\mu _{c}+18\mu_{  l(\Sigma_c^{*0}/D^{*-})  }\right)$ &$ -0.047$ \\
			&{{${^{4}P_{\frac{1}{2}}}$}}  & $({\frac{3}{2}}^{+}\otimes1^{-})_{\frac{3}{2}}\otimes1^{-}$ &$\frac{1}{27} \left(28\mu _{d}+5\mu _{c}-9\mu_{  l(\Sigma_c^{*0}/D^{*-})  }\right)$ &$ -0.821$ \\
			\hline
			$ \frac{3}{2}^{+} $&{{${^{2}P_{\frac{3}{2}}}$}}  & ${\frac{1}{2}}^{+}\otimes0^{-}\otimes1^{-}$ &$\frac{1}{3} \left(4\mu _{d}-\mu _{c}+3\mu_{  l(\Sigma_c^{0}/D^{-})  }\right)$ &$ -1.682$ \\
			& & $({\frac{1}{2}}^{+}\otimes1^{-})_{\frac{1}{2}}\otimes1^{-}$ &$\frac{1}{9} \left(2\mu _{d}-5\mu _{c}+9\mu_{  l(\Sigma_c^{0}/D^{*-})  }\right)$ & $-0.691$ \\
			\cline{2-5}
			&{{${^{4}P_{\frac{3}{2}}}$}}  & $({\frac{1}{2}}^{+}\otimes1^{-})_{\frac{3}{2}}\otimes1^{-}$ &$\frac{1}{45} \left(77\mu _{d}-44\mu _{c}+18\mu_{  l(\Sigma_c^{0}/D^{*-})  }\right)$ &$ -2.118$ \\
			& & ${\frac{3}{2}}^{+}\otimes0^{-}\otimes1^{-}$ &$\frac{1}{15} \left(22\mu _{d}+11\mu _{c}+6\mu_{  l(\Sigma_c^{*0}/D^{-})  }\right)$ &$ -1.209$ \\
			\cline{2-5}
			&{{${^{2}P_{\frac{3}{2}}}$}}  & $({\frac{3}{2}}^{+}\otimes1^{-})_{\frac{1}{2}}\otimes1^{-}$ &$\frac{1}{9} \left(7\mu _{d}+8\mu _{c}+9\mu_{  l(\Sigma_c^{*0}/D^{*-})  }\right)$ & $-0.638$ \\
			&{{${^{4}P_{\frac{3}{2}}}$}}  & $({\frac{3}{2}}^{+}\otimes1^{-})_{\frac{3}{2}}\otimes1^{-}$ &$\frac{1}{225} \left(308\mu _{d}+55\mu _{c}+90\mu_{  l(\Sigma_c^{*0}/D^{*-})  }\right)$ &$ -1.302$ \\
			&{{${^{6}P_{\frac{3}{2}}}$}}  & $({\frac{3}{2}}^{+}\otimes1^{-})_{\frac{5}{2}}\otimes1^{-}$ &$\frac{1}{25} \left(63\mu _{d}-15\mu_{  l(\Sigma_c^{*0}/D^{*-})  }\right)$   &$ -2.232$ \\
			\hline
			$ \frac{5}{2}^{+} $&{{${^{4}P_{\frac{5}{2}}}$}}  & ${\frac{1}{2}}^{+}\otimes1^{-}\otimes1^{-}$ &$\frac{1}{3} \left(7\mu _{d}-4\mu _{c}+3\mu_{  l(\Sigma_c^{++}/D^{*0})  }\right)$ & $-3.005$ \\
			& & ${\frac{3}{2}}^{+}\otimes0^{-}\otimes1^{-}$ &$2\mu _{d}+\mu _{c}+\mu_{  l(\Sigma_c^{*0}/D^{-})  }$ &$-1.779 $ \\
			&& $({\frac{3}{2}}^{+}\otimes1^{-})_{\frac{3}{2}}\otimes1^{-}$ &$\frac{1}{15} \left(28\mu _{d}+5\mu _{c}+15\mu_{  l(\Sigma_c^{*0}/D^{*-})  }\right)$   &$ -1.894$ \\
			\cline{2-5}
			&{{${^{6}P_{\frac{5}{2}}}$}}  &$({\frac{3}{2}}^{+}\otimes1^{-})_{\frac{5}{2}}\otimes1^{-}$  &$\frac{1}{35} \left(93\mu _{d}+10\mu_{  l(\Sigma_c^{*0}/D^{*-})  }\right)$ &$ -2.592$ \\
			\hline
			$ \frac{7}{2}^{+} $&{{${^{6}P_{\frac{7}{2}}}$}}  & ${\frac{3}{2}}^{+}\otimes1^{-}\otimes1^{-}$ &$3\mu _{d}+\mu_{  l(\Sigma_c^{*0}/D^{*-})  }$ &$ -3.102$ \\
			\bottomrule[1pt]
			\bottomrule[1pt]
		\end{tabular}
	\end{table*}

	\renewcommand\tabcolsep{1.7cm}
	\renewcommand{\arraystretch}{1.75}
	\begin{table*}[!htbp]
		\caption{Transition magnetic moments between the $S$-wave  $\Sigma_c^{(*)} \bar{D}^{(*)}$-type hidden-charm molecular pentaquarks $P^{\Delta^{++}}_{\psi}$ with $10_{f}$ flavor representation. The unit is the nuclear magnetic moment $ \mu_{N} $.}
		\label{tab:nycj}
		\begin{tabular}{l|c|c}
			\toprule[1.0pt]
			\toprule[1.0pt]
			Decay modes & Expressions & Results \\
			\hline
			$\Sigma_c^{} \bar{D}^{*}|\frac{1}{2}^-\rangle \to\Sigma_c^{} \bar{D}^{}|\frac{1}{2}^-\rangle\gamma$ & $\frac{\sqrt{3}}{3}\mu_{\bar{D}^{*0} \to  \bar{D}^{0}}$ & $-1.327$ \\

			$\Sigma_c^{} \bar{D}^{*}|\frac{3}{2}^-\rangle \to\Sigma_c^{} \bar{D}^{}|\frac{1}{2}^-\rangle\gamma$ & $\frac{\sqrt{6}}{3}\mu_{\bar{D}^{*0} \to  \bar{D}^{0}}$ & $-1.877$ \\
			$\Sigma_c^{} \bar{D}^{*}|\frac{3}{2}^-\rangle \to\Sigma_c^{} \bar{D}^{*}|\frac{1}{2}^-\rangle\gamma$ & $\frac{\sqrt{2}}{3}({2\mu}_{{\Sigma}_{c}^{++}}-{\mu}_{{\bar{D}}^{*0}})$ & $1.552$  \\
			$\Sigma_c^{*} \bar{D}^{*}|\frac{1}{2}^-\rangle \to\Sigma_c^{} \bar{D}^{*}|\frac{3}{2}^-\rangle\gamma$  & $-\frac{\sqrt{2}}{2}\mu_{\Sigma_{c}^{*++} \to  \Sigma_{c}^{++}}$ & $-0.994$  \\
			$\Sigma_c^{*} \bar{D}^{}|\frac{3}{2}^-\rangle \to\Sigma_c^{} \bar{D}^{}|\frac{1}{2}^-\rangle\gamma$ & $\mu_{\Sigma_{c}^{*++} \to  \Sigma_{c}^{++}}$ & $1.406$  \\
			
			$\Sigma_c^{*} \bar{D}^{*}|\frac{1}{2}^-\rangle \to\Sigma_c^{*} \bar{D}^{}|\frac{3}{2}^-\rangle\gamma$ & $-\frac{\sqrt{3}}{3}\mu_{\bar{D}^{*0} \to  \bar{D}^{0}}$ & $1.327$ \\
			
			$\Sigma_c^{*} \bar{D}^{*}|\frac{3}{2}^-\rangle \to\Sigma_c^{} \bar{D}^{*}|\frac{1}{2}^-\rangle\gamma$ & $-\frac{\sqrt{5}}{5}\mu_{\Sigma_{c}^{*++} \to  \Sigma_{c}^{++}}$ & $-0.629$  \\
			$\Sigma_c^{*} \bar{D}^{*}|\frac{3}{2}^-\rangle \to\Sigma_c^{*} \bar{D}^{*}|\frac{1}{2}^-\rangle\gamma$ & $\frac{\sqrt{5}}{25}({2\mu}_{{\Sigma}_{c}^{*++}}-3{\mu}_{{\bar{D}}^{*0}})$ & $0.028$  \\
			
			$\Sigma_c^{*} \bar{D}^{*}|\frac{3}{2}^-\rangle \to\Sigma_c^{} \bar{D}^{*}|\frac{3}{2}^-\rangle\gamma$ & $-\frac{\sqrt{10}}{5}\mu_{\Sigma_{c}^{*++} \to  \Sigma_{c}^{++}}$ & $-0.889$ \\
			$\Sigma_c^{*} \bar{D}^{*}|\frac{3}{2}^-\rangle \to\Sigma_c^{*} \bar{D}^{}|\frac{3}{2}^-\rangle\gamma$ & $\frac{\sqrt{15}}{5}\mu_{\bar{D}^{*0} \to  \bar{D}^{0}}$ & $-1.780$ \\
			
			$\Sigma_c^{*} \bar{D}^{*}|\frac{5}{2}^-\rangle \to\Sigma_c^{} \bar{D}^{*}|\frac{1}{2}^-\rangle\gamma$ & $\frac{2\sqrt{5}}{5}\mu_{\Sigma_{c}^{*++} \to  \Sigma_{c}^{++}}$ & $1.258$  \\

			$\Sigma_c^{*} \bar{D}^{*}|\frac{5}{2}^-\rangle \to\Sigma_c^{*} \bar{D}^{}|\frac{3}{2}^-\rangle\gamma$ & $\frac{\sqrt{10}}{5}\mu_{\bar{D}^{*0} \to  \bar{D}^{0}}$ & $-1.454$ \\
			$\Sigma_c^{*} \bar{D}^{*}|\frac{5}{2}^-\rangle \to\Sigma_c^{} \bar{D}^{*}|\frac{3}{2}^-\rangle\gamma$& $\frac{\sqrt{15}}{5}\mu_{\Sigma_{c}^{*++} \to  \Sigma_{c}^{++}}$ & $1.089$  \\
			$\Sigma_c^{*} \bar{D}^{*}|\frac{5}{2}^-\rangle \to\Sigma_c^{*} \bar{D}^{*}|\frac{3}{2}^-\rangle\gamma$ &$	\frac{\sqrt{6}}{15}(2{\mu}_{{\Sigma}_{c}^{*++}}-3{\mu}_{{\bar{D}}^{*0}}) $ & $0.051$  \\
			\bottomrule[1.0pt]
			\bottomrule[1.0pt]
			
		\end{tabular}
		
	\end{table*}

	\renewcommand\tabcolsep{1.73cm}
	\renewcommand{\arraystretch}{1.75}
	\begin{table*}[!htbp]
		\caption{Transition magnetic moments between the $S$-wave  $\Sigma_c^{(*)} D^{(*)}$-type hidden-charm molecular pentaquarks $P^{\Delta^{-}}_{\psi}$ with $10_{f}$ flavor representation. The unit is the nuclear magnetic moment $ \mu_{N} $.}
		\label{tab:yj}
		\begin{tabular}{l|c|c}
			\toprule[1.0pt]
			\toprule[1.0pt]
			Decay modes & Expressions & Results \\
			\hline
			$\Sigma_c^{} D^{*}|\frac{1}{2}^-\rangle \to\Sigma_c^{} D^{}|\frac{1}{2}^-\rangle\gamma$ & $\frac{\sqrt{3}}{3}\mu_{D^{*-} \to  D^{-}}$ & $0.314$ \\

			$\Sigma_c^{} D^{*}|\frac{3}{2}^-\rangle \to\Sigma_c^{} D^{}|\frac{1}{2}^-\rangle\gamma$ & $\frac{\sqrt{6}}{3}\mu_{D^{*-} \to  D^{-}}$ & $0.444$ \\
			$\Sigma_c^{} D^{*}|\frac{3}{2}^-\rangle \to\Sigma_c^{}D^{*}|\frac{1}{2}^-\rangle\gamma$ & $\frac{\sqrt{2}}{3}({2\mu}_{{\Sigma}_{c}^{0}}-{\mu}_{{D}^{*-}})$ & $-0.681$  \\
			$\Sigma_c^{*} D^{*}|\frac{1}{2}^-\rangle \to\Sigma_c^{} D^{*}|\frac{3}{2}^-\rangle\gamma$  & $-\frac{\sqrt{2}}{2}\mu_{\Sigma_{c}^{*0} \to  \Sigma_{c}^{0}}$ & $0.901$  \\
			$\Sigma_c^{*}D^{}|\frac{3}{2}^-\rangle \to\Sigma_c^{} D^{}|\frac{1}{2}^-\rangle\gamma$ & $\mu_{\Sigma_{c}^{*0} \to  \Sigma_{c}^{0}}$ & $-1.273$  \\
		
			$\Sigma_c^{*} D^{*}|\frac{1}{2}^-\rangle \to\Sigma_c^{*} D^{}|\frac{3}{2}^-\rangle\gamma$ & $-\frac{\sqrt{3}}{3}\mu_{D^{*-} \to  D^{-}}$ & $-0.314$ \\
		
			$\Sigma_c^{*}D^{*}|\frac{3}{2}^-\rangle \to\Sigma_c^{} D^{*}|\frac{1}{2}^-\rangle\gamma$ & $-\frac{\sqrt{5}}{5}\mu_{\Sigma_{c}^{*0} \to  \Sigma_{c}^{0}}$ & $0.570$  \\
			$\Sigma_c^{*} D^{*}|\frac{3}{2}^-\rangle \to\Sigma_c^{*} D^{*}|\frac{1}{2}^-\rangle\gamma$ & $\frac{\sqrt{5}}{25}({2\mu}_{{\Sigma}_{c}^{*0}}-3{\mu}_{{D}^{*-}})$ & $0.112$  \\
			
		$\Sigma_c^{*} D^{*}|\frac{3}{2}^-\rangle \to\Sigma_c^{} D^{*}|\frac{3}{2}^-\rangle\gamma$ & $-\frac{\sqrt{10}}{5}\mu_{\Sigma_{c}^{*0} \to  \Sigma_{c}^{0}}$ & $0.806$ \\
		$\Sigma_c^{*} D^{*}|\frac{3}{2}^-\rangle \to\Sigma_c^{*} D^{}|\frac{3}{2}^-\rangle\gamma$ & $\frac{\sqrt{15}}{5}\mu_{D^{*-} \to  D^{-}}$ & $0.421$ \\

			$\Sigma_c^{*} D^{*}|\frac{5}{2}^-\rangle \to\Sigma_c^{} D^{*}|\frac{1}{2}^-\rangle\gamma$ & $\frac{2\sqrt{5}}{5}\mu_{\Sigma_{c}^{*0} \to  \Sigma_{c}^{0}}$ & $-1.139$  \\

			$\Sigma_c^{*} D^{*}|\frac{5}{2}^-\rangle \to\Sigma_c^{*} D^{}|\frac{3}{2}^-\rangle\gamma$ & $\frac{\sqrt{10}}{5}\mu_{D^{*-} \to  D^{-}}$ & $0.344$ \\
				$\Sigma_c^{*} D^{*}|\frac{5}{2}^-\rangle \to\Sigma_c^{} D^{*}|\frac{3}{2}^-\rangle\gamma$& $\frac{\sqrt{15}}{5}\mu_{\Sigma_{c}^{*0} \to  \Sigma_{c}^{0}}$ & $-0.987$  \\
			$\Sigma_c^{*} D^{*}|\frac{5}{2}^-\rangle \to\Sigma_c^{*} D^{*}|\frac{3}{2}^-\rangle\gamma$ &$	\frac{\sqrt{6}}{15}(2{\mu}_{{\Sigma}_{c}^{*0}}-3{\mu}_{{D}^{*-}}) $ & $0.205$  \\
			\bottomrule[1.0pt]
			\bottomrule[1.0pt]

		\end{tabular}
		
	\end{table*}

\end{document}